\pdfoutput=1
\documentclass{article}
\usepackage{natbib}
\usepackage{epubtk}    
\usepackage{booktabs}  
\usepackage{graphicx}  
\usepackage{esvect}
\usepackage{lscape}
\usepackage{array}
\usepackage{ragged2e} 

\newcommand{\nifs}{\ensuremath{^{56}\mathrm{Ni}}}
\newcommand{\Nifs}{\ensuremath{^{56}\mathrm{Ni}} $\;$}
\def\p{\partial}
\def\msun{$M_{\odot}$}
\def\Msun{$M_{\odot}$ }
\def\be{\begin{equation}}
\def\ee{\end{equation}}
\def\bea{\begin{eqnarray}}
\def\eea{\end{eqnarray}}
\def\gcc{gcm$^{-3}$}
\def\Gcc{gcm$^{-3} \;$}

\newcommand{\enth}{\mathcal{E}}
\def\divv{\nabla \cdot \vec{v}}

\def\M4{M$_4$}
\def\M6{M$_6$}
\def\lcm6{LCM$_6$}
\def\qcm6{QCM$_6$}
\def\whn{W$_{\rm H,n}$}
\def\wh3{W$_{\rm H,3}$}
\def\wh4{W$_{\rm H,4}$}
\def\wh5{W$_{\rm H,5}$}
\def\wh6{W$_{\rm H,6}$}
\def\wh7{W$_{\rm H,7}$}

\begin{document}

\title{SPH Methods in the Modelling of Compact Objects}

\author{\epubtkAuthorData{Stephan Rosswog}{%
The Oskar Klein Centre for Cosmoparticle Physics, \\
Department of Astronomy\\
AlbaNova University Centre,  106 91 Stockholm, Sweden}{%
stephan.rosswog@astro.su.se}{%
http://compact-merger.astro.su.se/}%
}

\date{}
\maketitle

\begin{abstract}
We review the current status of compact object simulations that are based on the 
Smooth Particle Hydrodynamics (SPH) method. The first main part of this review
is dedicated to SPH as a numerical method. We begin by discussing relevant
kernel approximation techniques and discuss the performance of different
kernel functions. Subsequently, we review a number of different SPH 
formulations of Newtonian, special- and general relativistic ideal fluid 
dynamics. We particularly point out recent developments that increase
the accuracy of SPH with respect to commonly used techniques.
The second main part of  the review is dedicated to the application
of SPH in compact object simulations. We 
discuss encounters between two white dwarfs, between two 
neutron stars and between a neutron star and a stellar-mass black hole. 
For each type of system, the main focus is on the more common, 
gravitational wave-driven binary mergers, but we also discuss 
dynamical collisions as they occur in dense stellar systems such as cores 
of globular clusters.
\end{abstract}

\epubtkKeywords{hydrodynamics, Smoothed Particle Hydrodynamics, binaries, white dwarfs, neutron stars, black holes}



\newpage


\section{Introduction}
\label{sec:intro}

\subsection{Relevance of compact object encounters}

The vast majority of stars in the Universe will finally become  a compact stellar object, either a 
white dwarf, a neutron star or a black hole. Our Galaxy therefore harbors large numbers of them,
probably $\sim 10^{10}$ white dwarfs, several $10^8$ neutron stars and tens of millions of 
stellar-mass black holes. These objects stretch the physics that is known 
from terrestrial laboratories to extreme limits. For example, the structure of white dwarfs 
is governed by electron degeneracy pressure, they are therefore Earth-sized manifestations 
of the quantum mechanical Pauli-principle. Neutron stars, on the other hand, reach in their 
cores multiples of the nuclear saturation density ($2.6 \times 10^{14}$ \gcc) which makes
them excellent probes for nuclear matter theories. The dimensionless compactness parameter
$\mathcal{C}= (G/c^2) (M/R)= R_{\rm S}/(2R)$, where $M$,$R$ and $R_{\rm S}$ are mass, radius 
and Schwarzschild radius of the object,
can be considered as a measure of the strength of  a gravitational field. It is proportional to
the Newtonian gravitational potential and directly related to the gravitational redshift. For
black holes, the parameter takes values of 0.5 at the Schwarzschild radius and for neutron stars
it is only moderately smaller, $\mathcal{C}\approx 0.2$, both are gigantic in comparison
to the solar value of $\sim 10^{-6}$. General relativity has so far passed all tests to high accuracy 
\citep{will14a}, but most of them have been performed in the limit of weak gravitational fields. Neutron 
stars and black holes, in contrast, offer the possibility to test gravity in the 
strong field regime \citep{psaltis08}.

Compact objects that had time to settle into an equilibrium state possess a high degree of symmetry 
and are essentially perfectly spherically symmetric. Moreover, they are cold enough to be excellently 
described in a $T=0$ approximation (since for all involved species $i$ the thermal energies are 
much smaller than the involved Fermi-energies, $k T_i \ll E_{{\rm F},i}$) and they are in chemical 
equilibrium. For such systems a number of interesting results can be obtained by (semi-) analytical 
methods. It is precisely because they are in a ``minimum energy state'' that such systems are 
hardly detectable in isolation, and certainly not out to large distances.

Compact objects, however, still possess -- at least in principle -- very large energy reservoirs and in
cases where these reservoirs can be tapped, they can produce electromagnetic emission that is so
bright that it can serve as cosmic beacons. For example, each nucleon inside of a carbon-oxygen 
white dwarf  can potentially still release $\approx$~0.9~MeV via nuclear burning to the most stable 
elements, or approximately $1.7 \times 10^{51} $ erg per solar mass. The gravitational binding energy 
of a neutron star or black hole is even  larger, $E_{\rm grav} \sim G M^2/R = \mathcal{C} M c^2 = 3.6 \times 10^{53} \; 
{\rm erg} \; (\mathcal{C}/0.2)  (M/$\msun). Tapping these gigantic energy reservoirs, however, usually 
requires special, often catastrophic circumstances, such as the collision or merger with yet another compact 
object. For example, the merger of a neutron star with another neutron star or with a black hole likely
powers the sub-class of short Gamma-ray bursts (GRBs) and mergers of two white dwarfs are thought to trigger type Ia supernovae.
Such events are highly dynamic and do not possess enough symmetries to be accurately described by 
(semi-) analytical methods. They require a careful, three-dimensional numerical modelling of gravity,
the hydrodynamics  and the relevant ``microphysical'' ingredients such as neutrino processes, nuclear
burning or a nuclear equation of state.

\subsection{When/why SPH?}

Many of such modelling efforts have involved the Smoothed Particle Hydrodynamics method (SPH),
originally suggested by \cite{gingold77} and by \cite{lucy77}.
There are a number of excellent numerical methods to deal with problems of ideal fluid dynamics,
but each numerical method has its particular merits and challenges, and it is usually pivotal in terms 
of work efficiency to choose the best method for the problem at hand. For this reason, we want to 
collect here the major strengths of, but also point out challenges for the SPH method.

The probably most outstanding feature of SPH is that it allows in a straight forward way
to \emph{exactly conserve} mass, energy, momentum and angular momentum \emph{by construction}, 
i.e., independent of the numerical resolution. This property is ensured by appropriate symmetries in the 
SPH particle indices in the evolution equations, see Section~\ref{chap:SPH}, together with gradient estimates (usually kernel gradients, 
but see Section~\ref{sec:SPH_with_integral_gradients}) that are anti-symmetric with respect to the 
exchange of two particle indices.\epubtkFootnote{See Section~2.3 of \cite{rosswog09b} for a very detailed 
discussion of conservation issues in SPH.} For example, as will be illustrated in Section~\ref{sec:appl_WDWD}, the mass transfer
between two stars and the resulting orbital dynamics are extremely sensitive to the accurate 
conservation of angular momentum. Eulerian methods are usually seriously challenged in 
accurately simulating the orbital dynamics in the presence of mass transfer, but this can be achieved 
when special measures are taken, see, for example, \cite{dsouza06} and \cite{motl07}.

Another benefit that comes essentially for free is the natural adaptivity of SPH. Since the particles move
with the local fluid velocity, they naturally trace the flow motion. As a corollary, simulations are 
not bound, like usually in Eulerian simulations, to a pre-defined ``simulation volume'', but instead they can 
follow whatever the astrophysical system decides to do, be it a collapse  or a rapid
expansions or both in different parts of the flow.  This automatic ``refinement on density'' is also closely related
to the fact that vacuum does not pose a problem: such regions are simply devoid of SPH particles and no
computational resources are wasted on regions without matter. In Eulerian methods, in contrast, 
vacuum is usually treated as a ``background fluid'' in which the ``fluid of interest'' moves
and special attention needs to be payed to avoid artefacts caused by the interaction of these two fluids. 
For example, the neutron star surface of a binary neutron star system close to merger moves
with an orbital velocity of $\sim 0.1c$ against the ``vacuum'' background medium. This can cause 
strong shocks and it may become challenging to disentangle, say, physical neutrino emission from 
the one that is entirely due to the artificial interaction with the background medium. There are, however,
also hydrodynamic formulations that would in principle allow to avoid such an artificial ``vacuum''
\citep{duez02}.

On the other hand, SPH's ``natural tendency to ignore vacuum'' may also become a disadvantage in cases 
where the major interest of the investigation is a tenuous medium close to a dense one, say, for gas that 
is accreted onto the surface of a compact star. Such cases are probably more efficiently handled with an adaptive 
Eulerian method that can refine on physical quantities that are different from density. Several examples of 
hybrid approaches between SPH and Eulerian methods are discussed in Section \ref{sec:WDWD_SNIa}.

SPH is Galilean invariant and thus independent of the chosen computing 
frame. The lack of this property can cause serious artefacts for Eulerian schemes if the simulation is
performed in an inappropriate reference frame, see \cite{springel10b} for a number of examples.
Particular examples related to binary mergers have been discussed in \cite{new97} and \cite{swesty00}:
simulating the orbital motion of a binary system in a space-fixed frame can lead to an entirely 
spurious inspiral and merger, while simulations in a corotating frame may yield accurate results.
For SPH, in contrast, it does not matter in which frame the calculation is performed.

Another strong asset of SPH is its exact advection of fluid properties: an attribute attached to an 
SPH particle is simply carried along as the particle moves. This makes it particularly easy to, say, 
post-process the nucleosynthesis from a particle trajectory, without any need for additional ``tracer particles''. 
In Eulerian methods high velocities with respect to the computational grid can seriously compromise the
accuracy of a simulation. For SPH, this is essentially a ``free lunch'', see for example Figure~\ref{fig:advection},
where a high-density wedge is advected with a velocity of 0.9999 c through the computational domain.

The particle nature of SPH also allows for a natural transition to n-body methods.
For example, if ejected material from a stellar encounter becomes ballistic so that hydrodynamic
forces are negligible, one may decide to simply follow the long-term evolution of  point particles 
in a gravitational potential instead of a fluid. Such a treatment can make time scales accessible 
that cannot be reached within a hydrodynamic approach
\citep{faber06b,rosswog07a,lee07,ramirezruiz09,lee10a}.

SPH can straight forwardly be combined with highly flexible and accurate
gravity solvers such as tree methods.  Such approaches are extremely powerful for problems in which a 
fragmentation with complicated geometry due to the interplay of gas dynamics and self-gravity occurs, 
say in star or planet formation. 
Many successful examples of couplings of SPH with trees exist in the literature. Maybe the first one was the use the
Barnes-Hut oct-tree algorithm \citep{barnes86} within the TREESPH code \citep{hernquist89}, closely
followed by the implementation of a mutual nearest neighbour tree due to Press for the simulation 
of white dwarf encounters \citep{benz90b}. By now a number of very fast and flexible
tree variants are routinely used in SPH (e.g., \cite{dave97,carraro98a,springel01a,wadsley04,springel05a,wetzstein09,nelson09}) 
and for a long time SPH-tree combinations were at the leading edge ahead of Eulerian approaches
that only have caught up later, see for example \cite{kravtsov99,tessier02}. More recently, also ideas 
from the fast multipole method have been borrowed \citep{dehnen00,dehnen02,gafton11,dehnen14}
that allow for a scaling with the particle number  $N$ that is close to $O(N)$ or even below, rather than
the $O(N \log N)$ scaling that is obtained for traditional tree algorithms.

But like any other numerical method, SPH also has to face some challenges. One particular example that
had received a lot of attention in recent years, was the struggle of standard SPH formulations
to properly deal with some fluid instabilities \citep{thacker00,agertz07,springel10a,read10}. As will be discussed in 
Section~\ref{sec:volume_elements}, the problem is caused
by surface tension forces that can emerge near contact discontinuities. This insight triggered
a flurry of suggestions on how to improve this aspect of SPH 
\citep{price08a,cha10,read10,hess10,valcke10,junk10,abel11,gaburov11,murante11,read12,hopkins13,saitoh13}. 
Arguably the most robust way to deal with this is the reformulation of SPH in terms of different volume
elements as discussed in Section~\ref{sec:volume_elements}, an example is shown in Section~\ref{sec:KH}.

Another notoriously difficult field is the inclusion of magnetic fields into SPH. This is closely related
to preserving the $\nabla \cdot \vec{B}=0$ constraint during the MHD evolution, but also here 
there has been substantial progress in recent years
\citep{boerve01,boerve04,price04a,price04b,price05,price06,boerve06,rosswog07a,dolag09,buerzle11a,buerzle11b,tricco12,tricco13}.
Moreover, magnetic fields may be very important in regions of low density and due to SPH's ``tendency
to ignore vacuum'', such regions are poorly sampled.
For a detailed discussion of the current state of SPH and magnetic fields we refer to the 
recent review of \cite{price12a}.

Artificial dissipation is also often considered as a major drawback. However, if dissipation is steered properly,
see Section~\ref{sec:Newtonian_shocks}, the performance should be very similar to the one of approximate 
Riemann solver approaches. 
A Riemann solver approach may from an aesthetical point of view be more appealing, though, and a number of such
approaches have been successfully implemented and tested  \citep{inutsuka02,cha03,cha10,murante11,puri14}.

Contrary to what was often claimed in the early literature, however, SPH is not necessarily a very efficient 
method. It is true that if only the bulk matter distribution is of interest, one can often obtain robust
results already  with  an astonishingly small number of SPH particles. To obtain accurate results for the 
thermodynamics of the flow, however, still usually requires large particle numbers. In large SPH simulations
it becomes a serious challenge to maintain cache-coherence since particles that were initially close 
in space and computer memory can later on become completely scattered throughout different (mostly slow) 
parts of the memory.
This can be improved by using cache-friendly variables (aggregate data that are frequently used
together into common variables, even if they have not much of a physical connection) and/or
by various sorting techniques to re-order particles in memory according to their spatial location.
This can be done --in the simplest case-- by using a uniform hash grid, but in many astrophysical
applications hierarchical structures such as trees are highly preferred for sorting the particles, see e.g., 
\cite{warren95,springel05a,nelson09,gafton11}. While such measures can easily improve the performance
by factors of a few, they come with some bookkeeping overhead which, together with, say, individual
time steps and particle sinks or sources, can make codes rather unreadable and error-prone.

Finally, we will briefly discuss in Section~\ref{sec:IC} the construction of accurate initial 
conditions where the particles are in a true (and stable) numerical equilibrium. This is yet 
another a non-trivial SPH issue.

\subsection{Roadmap through this review}
This text is organized as follows:
\begin{itemize}
\item In Section~\ref{chap:kernel_approx} we discuss those kernel approximation 
techniques that are needed for the discussed SPH discretizations. We begin with
the basics and then focus on issues that are needed to appreciate some of the recent
improvements of the SPH method. Some of these issues are by their very nature
rather technical. Readers that are familiar with basic kernel interpolation techniques
could skip this section at first reading.
\item In Section~\ref{chap:SPH} we discuss SPH discretizations of ideal fluid dynamics
for both the Newtonian and the relativistic case. Since several reviews have
appeared in the last years, we only concisely summarize the more traditional formulations
and focus on recent improvements of the method.
\item The last section, Section~\ref{sec:appl}, is dedicated to astrophysical applications.
We begin with double white dwarf encounters (Section~\ref{sec:appl_WDWD}) and then
turn to encounters between two neutron stars and between a neutron star and a stellar-mass black 
hole (Section~\ref{sec:appl_NSNS_NSBH}). In each of these cases, our emphasis is 
on the more common, gravitational wave-driven binary systems, but we  also discuss 
dynamical collisions as they may occur, for example, in a globular cluster.
\end{itemize}
Sections~\ref{chap:kernel_approx} and \ref{chap:SPH} provide the basis for an 
understanding of SPH as a numerical method and they should pave the way 
to the most recent developments. The less numerically inclined reader may, however,
just catch the most basic SPH ideas from Section~\ref{sec:Newt_vanilla} and then 
jump to his/her preferred astrophysical topic in Section~\ref{sec:appl}.  The modular 
structure of  sections \ref{chap:kernel_approx} and \ref{chap:SPH}  should allow, 
whenever needed, for a selective consultation on the more technical issues of SPH.

\newpage

\section{Kernel Approximation}
\label{chap:kernel_approx}

SPH uses kernel approximation techniques to calculate densities and pressure gradients
from a discrete set of freely moving particles, see Section~\ref{chap:SPH}.
Since kernel smoothing is one of the key ideas of the SPH technique, we will begin by collecting a number of basic
relations that are necessary to understand how the method works and to judge the accuracy of different 
approaches. The discussed approximation techniques will be applied in Section~\ref{chap:SPH}, both for the 
traditional formulations  of SPH and for the recent improvements of the method.

\subsection{Function interpolation}

A kernel-smoothed representation of a function $A$ can be obtained via
\be
\langle A \rangle (\vec{r})= \int A(\vec{r}') W(\vec{r} - \vec{r}',h) \, dr',
\label{eq:integ_approx}
\ee
where $W$ is a smoothing kernel. The width of the kernel function is determined by the quantity $h$, which, in 
an SPH context, is usually referred to as the ``smoothing length''.  Obviously, the kernel has the dimension 
of an inverse volume, it should be normalized to unity, $\int W(\vec{r}',h) \, dr'= 1$, and
it should possess the $\delta$-property, 
\be
\lim_{h \rightarrow 0} \langle A \rangle (\vec{r})= A(\vec{r}),
\ee
so that  the original function is reproduced exactly in the limit of  a vanishing kernel support.
The choice of kernel functions is discussed in Section~\ref{sec:kernel_choice} in more detail, for now we 
only assume that the kernels are radial, i.e., $W(\vec{r})= W(|\vec{r}|)$, which leads to the very useful 
property
\be
\nabla_a W(|\vec{r}_a - \vec{r}_b|,h)= - \nabla_b W(|\vec{r}_a - \vec{r}_b|,h)
\label{eq:anti_sym}
\ee
that allows for a straight forward conservation of Nature's conservation laws. This topic is laid out in much detail
in Section~2.4 of  \cite{rosswog09b} and we refer the interested reader to this text for a further discussion and technical details.
There are plausible arguments that suggest, for example, spheroidal kernels \citep{fulbright95,shapiro96}, but
such approaches do usually violate angular momentum conservation. In the following we will only discuss radial kernels.
Figure~\ref{fig:interaction_sketch} sketches how a particle $a$ interacts with its neighbour particles $b$ via a radial kernel of 
support radius $Q h_a$. Let us assume that a  function $A$ is known at the positions $\vec{r}_b$, $A_b= A(\vec{r}_b)$.
One can then approximate Eq.~(\ref{eq:integ_approx}) as
\be
\langle A \rangle (\vec{r})\simeq  \sum_b V_b \;  A_b W(\vec{r} - \vec{r}_b,h),
\label{eq:std_SPH_interpolant}
\ee
where $V_b$ is the volume associated with the fluid parcel (``particle'') at $\vec{r}_b$.
In nearly all SPH formulations that are used in astrophysics
the particle volume is estimated as $V_b= m_b/\rho_b$ with $m_b$ being the (fixed) particle mass and 
$\rho_b$ its mass density. While being a straight forward choice, this is by no means the only option. 
Recently, SPH formulations have been proposed \citep{saitoh13,hopkins13,rosswog15b} that are based 
on different volume elements and that can yield substantial improvements.  
In the following, we will therefore keep a general volume element $V_b$ in most equations and only 
occasionally we give the equations for specific volume element choices.

\epubtkImage{}{%
\begin{figure}[htb] 
  \centerline{\includegraphics[width=9cm,angle=0]{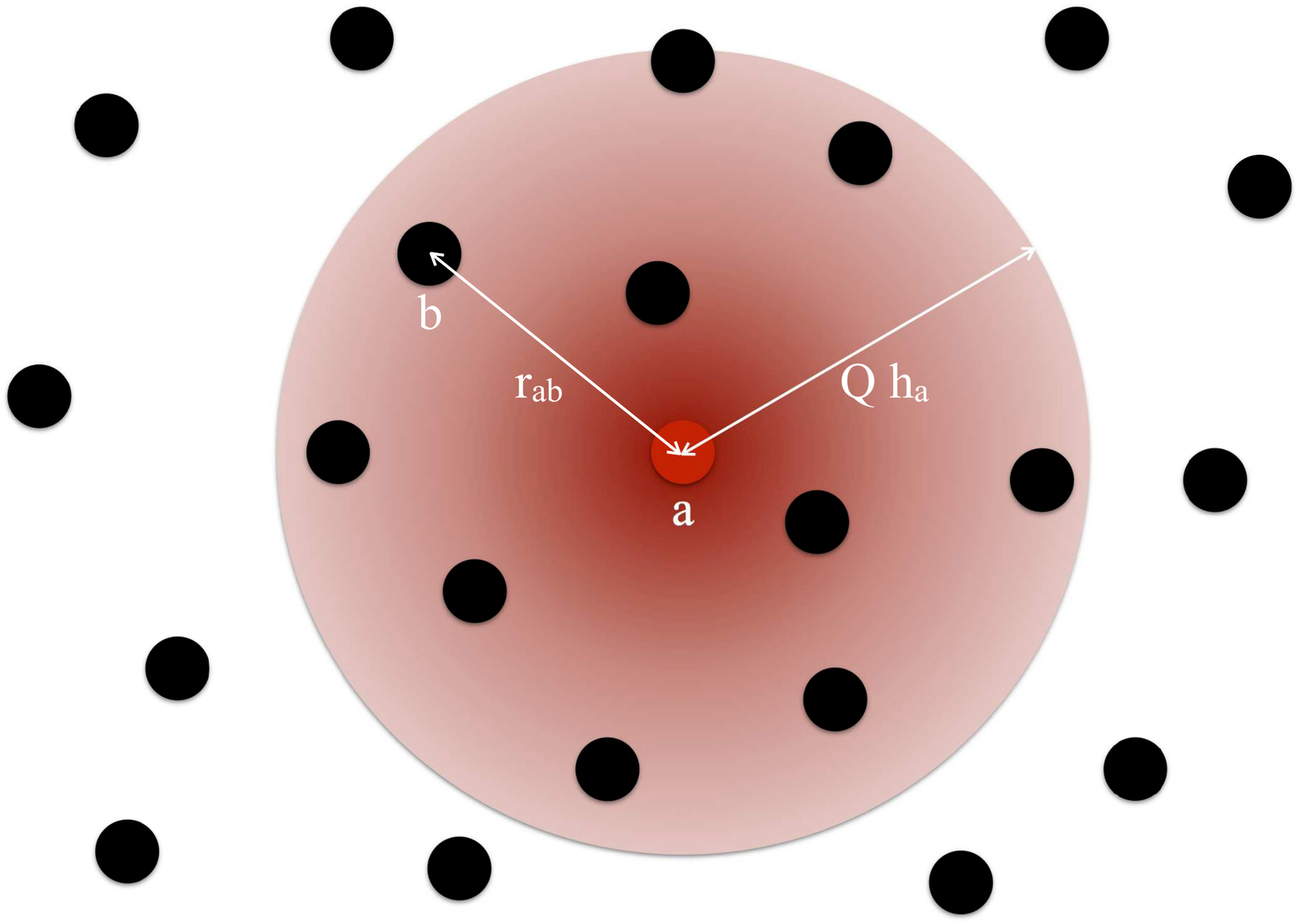}}
  \caption{Sketch of the interaction of a particle $a$, with its neighboring particles. 
                To avoid a computationally expensive interaction of each particle with all other particles, 
                kernels with a finite support (indicated by the shaded region) are usually used. 
                The support size of a particle $a$ is set as a multiple, $Q h_a$, of its smoothing length, 
                $h_a$. Often, $Q=2$ is used.}
   \label{fig:interaction_sketch}
\end{figure}}

The assessment of the kernel approximation quality in a practical SPH simulation 
is actually a rather non-trivial problem. Such an analysis is straight forward for the continuum kernel 
approximation \citep{benz90a,monaghan92}, or, in discrete form, for particles that are distributed on a regular 
lattice \citep{monaghan05},  but both of these estimates have limited validity for practical SPH simulations. 
In practice a discrete kernel approximation is used and particles are neither located on a regular grid nor 
are they randomly distributed, they are ``disordered, but in an orderly way'' \citep{monaghan05}. 
The exact particle distribution depends on the dynamics of the flow and on the kernel that is used. An 
example of a particle distribution that arises in a practical simulation (a Kelvin--Helmholtz
instability) is shown in Figure~\ref{fig:particle_dist}. 
Note in particular that the particles sample space in a regular way and no ``clumping'' occurs as would
be the case for random particle positions. Why the particles arrange in such a regular way is discussed
in Section~\ref{sec:self_regularization}.
Apart from the particle distribution, the accuracy depends, of
course, on the chosen kernel function $W$ and this is discussed further in Section~\ref{sec:kernel_choice}
 (see also Figure~\ref{fig:particle_dist_2}). 
Since the kernel function determines the distribution into which the particles settle, the kernel function 
and the particle distribution cannot be considered as independent entities. This makes the analytical 
accuracy assessment of SPH very difficult and also suggests to interpret results based on a prescribed, fix 
particle distribution with some caution. In any case, results should be further scrutinized in practical benchmark 
tests with known solutions. Initial conditions are another very important issue for the accuracy of SPH 
simulations, they are further discussed in Section~\ref{sec:IC}.

\epubtkImage{}{%
  \begin{figure}[htb]
  \centerline{\includegraphics[width=10cm,angle=-90]{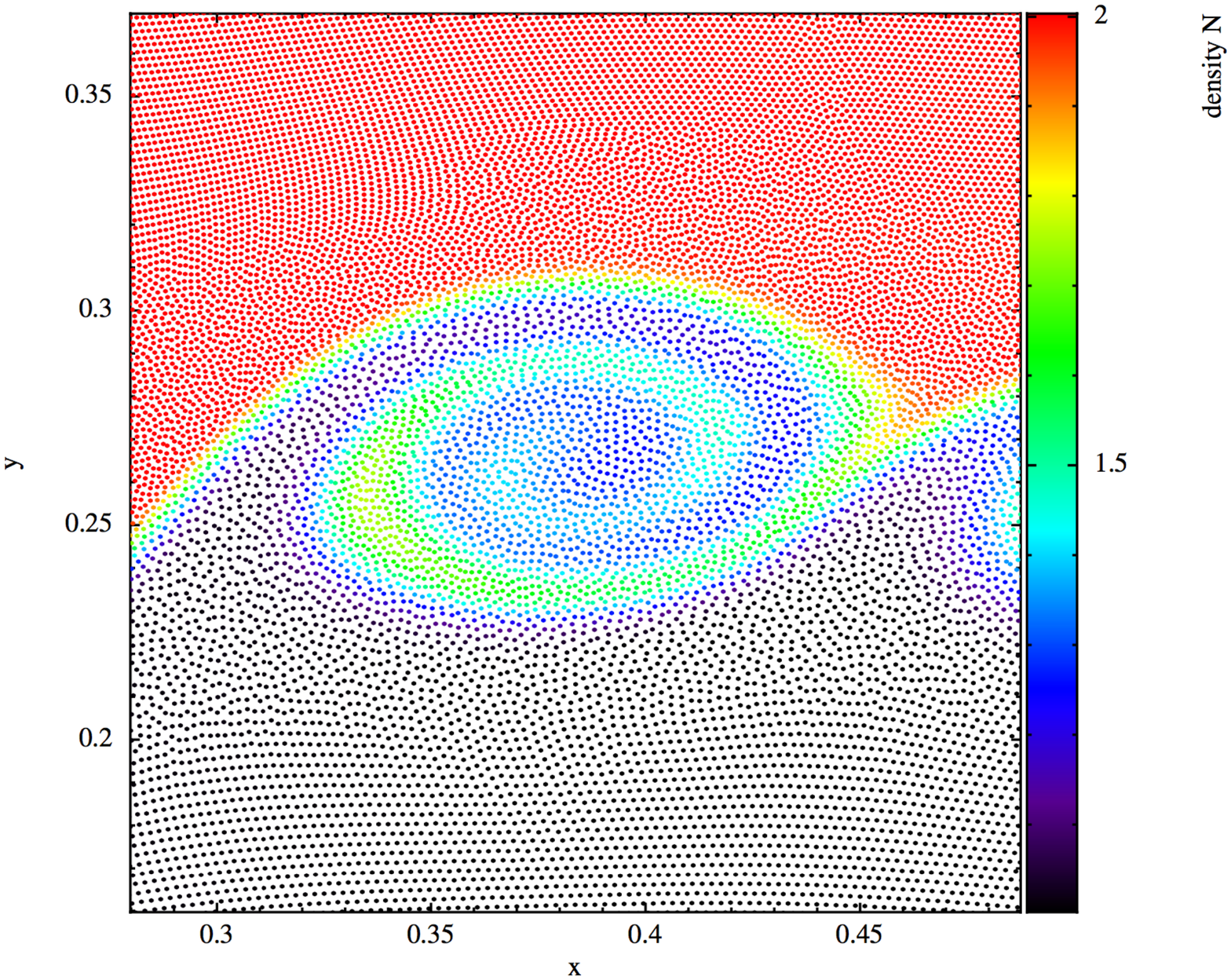}} 
 \caption{SPH particle distribution in a Kelvin--Helmholtz instability test. 
               Note that the particles are
               {\em not} distributed randomly, but instead show a high degree of regularity. This 
               is crucial for the accuracy of the SPH kernel interpolation, see the quality 
               indicators $\mathcal{Q}_1 ... \mathcal{Q}_4$ in Eqs.~(\ref{eq:quality_int}) and  
               (\ref{eq:gradient_quality}). This example uses  a high-order Wendland kernel,
               see Section~\ref{sec:kernel_choice},  which enforces  a particularly regular particle distribution.}
   \label{fig:particle_dist}
\end{figure}}

To obtain indicators for the quality of a function interpolation, it is instructive to assume that 
the function $A$ is smooth in the neighborhood of the particle of interest, $a$, and that it can 
be expanded in a Taylor-series around $\vec{r}_a$:
\be
A_b= A_a +  (\nabla A)_a \cdot (\vec{r}_{b}-\vec{r}_a) + \frac{1}{2} 
(\p_i \p_j  A)_a (\vec{r}_{b}-\vec{r}_a)^i (\vec{r}_{b}-\vec{r}_a)^j  + O(r_{ab}^3),
\label{eq:Taylor_A}
\ee
where $r_{ab}= |\vec{r}_a - \vec{r}_b|$ and the Einstein summation convention has been used.
Inserting the expansion into the discrete approximation Eq.~(\ref{eq:std_SPH_interpolant})
and requiring that $\langle A \rangle_a$ should be a close approximation of $A_a$,
one finds the ``interpolation quality indicators'' 
\be
\mathcal{Q}_1: \; \sum_b  V_b W_{ab}(h_a) \approx 1\quad {\rm and} \quad 
\mathcal{Q}_2: \; \sum_b  V_b (\vec{r}_{b}-\vec{r}_a) W_{ab}(h_a) \approx 0,
\label{eq:quality_int}
\ee
where $W_{ab}(h_a)= W(|\vec{r}_{a}-\vec{r}_b|/h_a)$. The first relation simply states that 
the particles should provide a good approximation to a partition of unity.

\subsection{Function derivatives}
\label{sec:SPH_derivs}

We restrict the discussion here to the first-order derivatives that we will need in Section~\ref{chap:SPH}, 
for higher-order derivatives that are required for some applications we refer the interested reader to the literature 
\citep{espagnol03,monaghan05,price12a}. 
There are various ways to calculate gradients for function values that are known at given particle positions.
Obviously, the accuracy of the gradient estimate is of concern, but also the symmetry in the 
particle indices since it can allow to enforce exact numerical conservation of physically 
conserved quantities. An accurate gradient estimate without built-in conservation can be less useful
in practice than a less accurate estimate that exactly obeys Nature's conservation laws, see Section~5 in
\cite{price12a} for a striking example of how a seemingly good gradient without built-in conservation can lead
to pathological particle distributions. The challenge is to combine exact conservation with an accurate 
gradient estimate.

\subsubsection{Direct gradient of the approximant}
\label{sec:direct_gradient}

The straight forward (though not most accurate) gradient estimate is the direct gradient of the interpolant
Eq.~(\ref{eq:std_SPH_interpolant})\epubtkFootnote{In the following we will drop the distinction between a function and its
approximation to alleviate the notation.}
\be
(\nabla A)^{(0)} (\vec{r})= \sum_b V_b \;  A_b \nabla W(\vec{r} - \vec{r}_b,h).
\label{eq:std_SPH_gradient}
\ee
Proceeding as above, we can insert again Eq.~(\ref{eq:Taylor_A}) into Eq.~(\ref{eq:std_SPH_gradient}),
\bea
(\nabla A)_a^{(0)} 
             &=& A_a \sum_b V_b \nabla_a W_{ab}(h_a) + \sum_b V_b (\nabla A \cdot (\vec{r}_{b}-\vec{r}_{a}) )
\nabla_a W_{ab}(h_a) + \dots,\label{eq:gradient_Taylor_expansion}
\eea
which delivers the ``gradient quality indicators'' 
\be
\mathcal{Q}_3: \; \sum_b  V_b\;  \nabla_a W_{ab}(h_a) \approx 0 \quad {\rm and} \quad
\mathcal{Q}_4: \; \sum_b  V_b \; (\vec{r}_{b}-\vec{r}_{a})^i   (\nabla_a W_{ab}(h_a))^j \approx \delta^{ij}
\label{eq:gradient_quality}
\ee
from the requirement that the estimate closely approximates the gradient of $A$. 
$\mathcal{Q}_3$ is simply the gradient of the quality indicator $\mathcal{Q}_1$ and therefore
again an expression of the partition of unity requirement. Note, however,
that even when all function values are the same, $A_b= A_0$, the gradient estimate does not necessarily 
vanish exactly. This is a direct consequence of Eq.~(\ref{eq:quality_int}) not being an exact partition
of unity. Note that therefore the reproduction of even constant functions is not enforced and this is often referred 
to as lack of zeroth order consistency. There are, however, benefits from a particle distribution noticing 
its imperfections, since, together with exact conservation, this provides an automatic ``re-meshing'' 
mechanism that drives the particles into a regular distribution \citep{price12a}. Numerical experiments 
show that the longterm evolution of a particle distribution with such a re-meshing mechanism built in 
leads to much more accurate results than seemingly more accurate gradient estimates without such a mechanism. 
The reason is that in the first case the particles arrange themselves in a regular configuration (such as in 
Figure~\ref{fig:particle_dist}) where the quality indicators are fulfilled to a high degree, while in the latter 
case pathological particle distributions can develop that sample the fluid only very poorly.

%
\subsubsection{Constant-exact gradients}

An immediate way to improve the gradient estimate is to simply subtract the gradient 
of the residual in the approximate partition of unity, see Eq.~(\ref{eq:quality_int}), or,
equivalently, the first error term in Eq.~(\ref{eq:gradient_Taylor_expansion}). The resulting
gradient estimate 
\be
(\nabla A)_a^{(1)}=  \sum_b V_b (A_b - A_a)  \nabla_a W_{ab}(h_a),
\label{eq:const_exact_gradient}
\ee
now manifestly vanishes for a constant function $A$, i.e., if all $A_k$ are the same, independent
of the particle distribution.

\subsubsection{Linear-exact gradients}

Exact gradients of linear functions can be constructed in the following way \citep{price04c}.
Start with the RHS of Eq.~(\ref{eq:std_SPH_gradient}) at $\vec{r}_a$ and again 
insert the Taylor expansion of $A_b$ around $\vec{r}_a$
\be
\sum_b V_b \;  A_b \nabla_a W_{ab}= \sum_b V_b \;  \left\{ A_a + (\nabla A)_a \cdot (\vec{r}_{b}-\vec{r}_{a}), 
+ \dots \right\} \nabla_a W_{ab}
\ee
which can be re-arranged into
\be
\sum_b V_b \;  (A_b - A_a) (\nabla_a W_{ab})^i= (\nabla A)_a^k M^{ki}.
\ee
Here the sum over the common index $k$ is implied and the matrix is given by
\be
M^{ki}= \sum_b V_b (\vec{r}_b - \vec{r}_a)^k (\nabla_a W_{ab})^i.
\label{eq:MIK}
\ee
Solving for the gradient component then yields
\be
(\nabla A)_a^k= (M^{ki})^{-1} \sum_b V_b (A_b-A_a) (\nabla_a W_{ab})^i.
\label{eq:lin_exact_gradient}
\ee
Note that the sum is the same as in the constant-exact gradient estimate Eq.~(\ref{eq:const_exact_gradient}),
corrected by the matrix $M^{-1}$ which depends on the properties of the particle distribution.
It is straight forward to double-check that this exactly reproduces the gradient of
a linear function. Assume that $A(\vec{r})= A_0 + (\nabla A)_0 \cdot (\vec{r}-\vec{r}_0)$ so that
\be
A_b - A_a= (\nabla A)_0 \cdot (\vec{r_b}-\vec{r}_a).
\ee
If this is inserted into Eq.~(\ref{eq:lin_exact_gradient}) one finds, as expected,
\be
(\nabla A)_a^k= (M^{ki})^{-1}  \sum_b (\nabla A)_0^l (\vec{r_b}-\vec{r}_a)^l 
(\nabla_a W_{ab})^i= (\nabla A)_0^l \; (M^{ki})^{-1} M^{li}= (\nabla A)_0^k.
\ee
Obviously, the linear-exact gradient comes at the price of inverting a $D \times D$ matrix in $D$ dimensions,
however, since the inversion of this small matrix can be done analytically, this does not represent a major 
computational burden.

\subsubsection{Integral-based gradients}
\label{sec:integral_gradients}

Integral-based higher-order derivatives \citep{brookshaw85,monaghan05} have for a long time been appreciated for
their insensitivity to particle noise. Surprisingly, integral-based estimates for first-order derivatives have only recently 
been explored \citep{garcia_senz12,cabezon12a}.
Start from the vector
\be
\tilde{\vec{I}}_A (\vec{r})= \int \left[ A(\vec{r}) - A(\vec{r}')\right] \; (\vec{r} - \vec{r}') \; 
W(|\vec{r} - \vec{r}'|, h) \, dV'
\ee
and, similar to above, insert a first-order Taylor expansion of $A(\vec{r}')$ around $\vec{r}$ (sum over $k$)
\be
\tilde{I}_A^i (\vec{r})= \int \left[ (\nabla A)^k_{\vert \vec{r}} \; (\vec{r} - \vec{r}')^k 
+ O\left((\vec{r} - \vec{r}')^2 \right) \right] \; (\vec{r} - \vec{r}')^i \; 
W(|\vec{r} - \vec{r}'|, h) \, dV',
\ee
so that the gradient component representation (exact for linear functions) is given by
\be
(\nabla A)^k_{\vert \vec{r}}= \tilde{C}^{ki}(\vec{r}) \tilde{I}_A^i(\vec{r}),
\ee
where the matrix $\tilde{C}^{ki}$ is the inverse of
\be
\tilde{T}^{ki}(\vec{r})=  \tilde{T}^{ik}(\vec{r})= \int (\vec{r} - \vec{r}')^k (\vec{r} - \vec{r}')^i W(|\vec{r} - \vec{r}'|, h) \, dV'.
\ee
The matrix $\tilde{T}^{ki}$ only depends on the positions while $\tilde{\vec{I}}_A$ contains 
the function to be differentiated. If we replace the integral by summations (and drop the tilde), the integral-based
gradient estimate reads (sum over $d$)
\be
(\nabla A)^k_{\vert \vec{r}}=  C^{kd}(\vec{r}) \; I^d_A(\vec{r}), 
\label{eq:full_IA_gradient}
\ee
where $(C^{kl})= (T^{kl})^{-1}$ and
\bea
T^{kl}(\vec{r}) &=& \sum_b V_b W(|\vec{r} - \vec{r}_b|, h) \; (\vec{r} - \vec{r}_b)^k (\vec{r} - \vec{r}_b)^l \\
I^l_A(\vec{r})  &=& \sum_b V_b W(|\vec{r} - \vec{r}_b|, h) \; \left[ A(\vec{r}) - A(\vec{r}_b)\right] \; (\vec{r} - \vec{r}_b)^l. 
\label{eq:matrices}
\eea
It is worth mentioning that for a radial kernel its gradient can be written as
\be
\nabla_a W_{ab}(h_a)
= (\vec{r}_b - \vec{r}_a) Y_{ab}(h_a),
\label{eq:kernel_gradient}
\ee
where  $Y$ is also a valid, positively definite and compactly supported kernel function. Therefore,
if Eq.~(\ref{eq:kernel_gradient}) is inserted in Eqs.~(\ref{eq:MIK}) and (\ref{eq:lin_exact_gradient}), 
one recovers Eq.~(\ref{eq:full_IA_gradient}), i.e., the linear-exact and the integral-based gradient 
are actually equivalent for radial kernels.

For a good interpolation, where the quality indicator $\mathcal{Q}_2$ in Eq.~(\ref{eq:quality_int}) vanishes 
to a good approximation, one can drop the term containing $A(\vec{r})$
\be
I^l_A(\vec{r})  \simeq  - \sum_b V_b W(|\vec{r} - \vec{r}_b|, h) \; A_b \; (\vec{r} - \vec{r}_b)^l \, ,
\ee
so that the gradient estimate in integral approximation (IA) explicitly reads (sum over $d$)
\be
(\nabla A)^k_{\rm IA}= \sum_b V_b A_b  C^{kd} (\vec{r}_b - \vec{r})^d \; 
W(|\vec{r} - \vec{r}_b|, h) \equiv \sum_b V_b A_b G^k_b(\vec{r}).
\label{eq:IA_gradient}
\ee
Comparison of Eq.~(\ref{eq:IA_gradient}) with Eq.~(\ref{eq:std_SPH_gradient}) suggests that $\vec{G}_b(\vec{r})$ takes over
the role that is usually played by $\nabla W(\vec{r} - \vec{r}_b,h)$:
\be
\nabla W(\vec{r} - \vec{r}_b,h) \rightarrow \vec{G}_b(\vec{r}).
\label{eq:nabla_W_to_G}
\ee

While being slightly less accurate than Eqs.(\ref{eq:full_IA_gradient})\,--\,(\ref{eq:matrices}), see Section~\ref{sec:accuracy},
the approximation Eq.~(\ref{eq:IA_gradient}) has the major advantage that $\vec{G}$ is anti-symmetric
with respect to the exchange of $\vec{r}$ and $\vec{r}_b$, just as the direct gradient of the
radial kernel, see Eq.~(\ref{eq:anti_sym}). Therefore it allows in a straight-forward way to enforce exact 
momentum conservation,\epubtkFootnote{For an explicit discussion of how conservation works in SPH, see
Section~2.4 in \cite{rosswog09b}.} though with a substantially more accurate gradient estimate. This type
of gradient has in a large number of benchmark tests turned out to be superior to the traditional SPH
approach.

\subsubsection{Accuracy assessment of different gradient prescriptions}
\label{sec:accuracy}

We perform a numerical experiment to measure the accuracy of different gradient prescriptions for
a regular particle distribution. Since, as outlined above, kernel function and particle distribution
are not independent entities, such tests at fixed particle distribution are a useful accuracy indicator,
but they should be backed up by tests in which the particles can evolve dynamically. We place SPH particles
in a 2D hexagonal lattice configuration in $[-1,1] \times [-1,1]$. 
Each particle is assigned a pressure value according $P(x,y)= 2 + x$ and we use
the different prescriptions discussed in Sections~\ref{sec:direct_gradient} \,--\,\ref{sec:integral_gradients} to 
numerically measure the pressure gradient. The relative average errors, $\epsilon= N^{-1}\sum_{b=1}^{N} \epsilon_b$ with
$\epsilon_b= |(\p_x P)_b - 1|$, as a function of the kernel support are shown in Figure~\ref{fig:gradient_prescriptions}. 
The quantity $\eta$ determines the smoothing length via $h_b= \eta \left(m_b/\rho_b\right)^{1/D}$, while $D$ denotes
the number of spatial dimensions. For this numerical experiment the standard cubic spline kernel M$_4$, see 
Section~\ref{sec:kernel_choice}, has been used. We apply the ``standard'' SPH-gradient,  
Eq.~(\ref{eq:std_SPH_gradient}), the approximate integral-based gradient (``IA gradient''), 
Eq.~(\ref{eq:IA_gradient}), the full integral-based gradient (``fIA gradient''), 
Eq.(\ref{eq:full_IA_gradient}), and the linearly exact-gradient (``LE gradient''), 
Eq.~(\ref{eq:lin_exact_gradient}). Note that for this regular particle distribution,
the constant-exact gradient, Eq.~(\ref{eq:const_exact_gradient}), is practically indistinguishable 
from the standard prescription, since it differs only by a term that is
proportional to the first ``gradient quality indicator'', Eq.~(\ref{eq:gradient_quality}), 
which vanishes here to a very good approximation. Clearly, the more sophisticated gradient prescriptions 
yield vast improvements of the gradient accuracy. Both the LE and fIA gradients reproduce the exact
value to within machine precision, the IA gradient which has the desired anti-symmetry property, improves
the gradient accuracy of the standard SPH estimate by approximately 10 orders of magnitude.

As expected from the term that was neglected in Eq.~(\ref{eq:matrices}), 
the accuracy of the ``IA gradient'' deteriorates substantially (to an accuracy level 
comparable to the standard prescription) for a less regular particle distribution, see \cite{rosswog15b}.
Therefore, the usefulness of the IA gradient depends on how regularly the SPH particles
are distributed in a  practical simulation. The original work by \cite{garcia_senz12} and our own extensive
numerical experiments \citep{rosswog15b}, however, show that the IA gradient is also in practical numerical tests
highly superior to the standard kernel gradient prescription.

\epubtkImage{}{%
  \begin{figure}[htb]
    \centerline{\includegraphics[angle=0,width=0.8\textwidth]{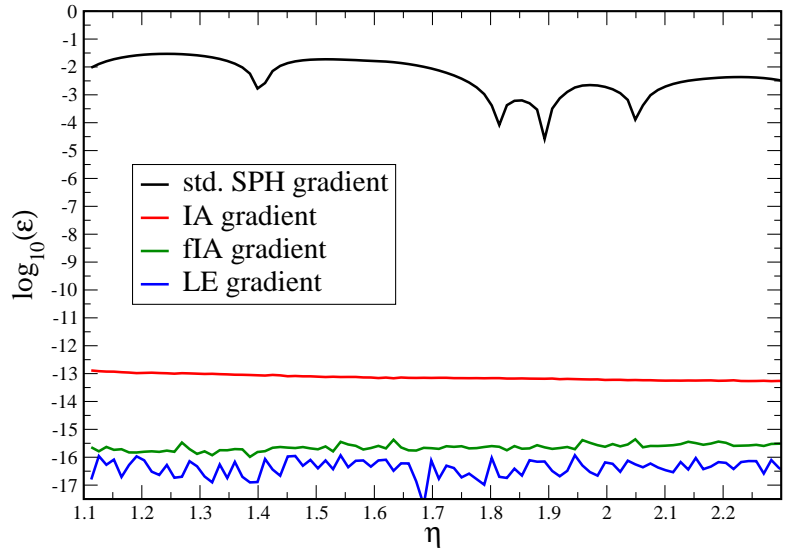}}
    \caption{Accuracy of different gradient prescriptions. Shown are the average relative errors in the numerical determination of a pressure gradient for different gradient prescriptions, see Eq.~(\ref{eq:std_SPH_gradient}) for ``std. SPH gradient'', Eq.~(\ref{eq:IA_gradient}) for ``IA gradient'', Eq.~(\ref{eq:full_IA_gradient}) for ``fIA gradient'' and Eq.~(\ref{eq:lin_exact_gradient}) for ``LE gradient''. The more sophisticated gradient prescriptions offer vast improvements over the standard SPH gradient. However, of those latter ones only the IA gradient has the same symmetry properties of the standard SPH prescription which allows for a straight forward enforcing of Nature's conservation laws. Adapted from \cite{rosswog15b}.}
    \label{fig:gradient_prescriptions}
\end{figure}}

\subsection{Which kernel function to choose?}
\label{sec:kernel_choice}

In the following, we briefly explore the properties of a selection of kernel functions.
We write normalized SPH kernels in the form\epubtkFootnote{We use the convention that $W$ 
refers to the full normalized kernel while $w$ is the un-normalized shape of the kernel.}
\be
W(|\vec{r}-\vec{r}'|,h)= \frac{\sigma}{h^D} w(q), \label{eq:normalized_SPH_kernel}
\ee
where $h$ is the smoothing length that determines the support of $W$, $q= |\vec{r}-\vec{r}'|/h$ 
and $D$ is the number of spatial dimensions. The normalizations are obtained from
\be
\sigma^{-1}= \left\{\begin{array}{ll}  
                       2 \int_0^{Q} w(q) dq  & {\rm in \; 1D}\\\\
                       2 \pi \; \int_0^{Q} w(q)  \; q \, dq & {\rm in \; 2D}\\\\
                       4 \pi \; \int_0^{Q} w(q)  \; q^2 \, dq & {\rm in \; 3D},
                        \end{array}\right. \\
\label{eq:normalization}
\ee
where $Q$ is the kernel support (= 2 for most of the following kernels).
In the following we will give the kernels in the form that is usually found in the literature. 
For a fair comparison in terms of computational effort/neighbor number we stretch the kernels 
in all plots and experiments to a support of 2$h$. So if a kernel has a normalization $\sigma_{lh}$
for a support of $l h$,  it has normalization $\sigma_{kh}= (l/k)^D \sigma_{lh}$ if it is stretched
to a support of $k h$.

\epubtkImage{}{%
  \begin{figure}[htbp]
    \centerline{\includegraphics[width=0.8\textwidth,angle=0]{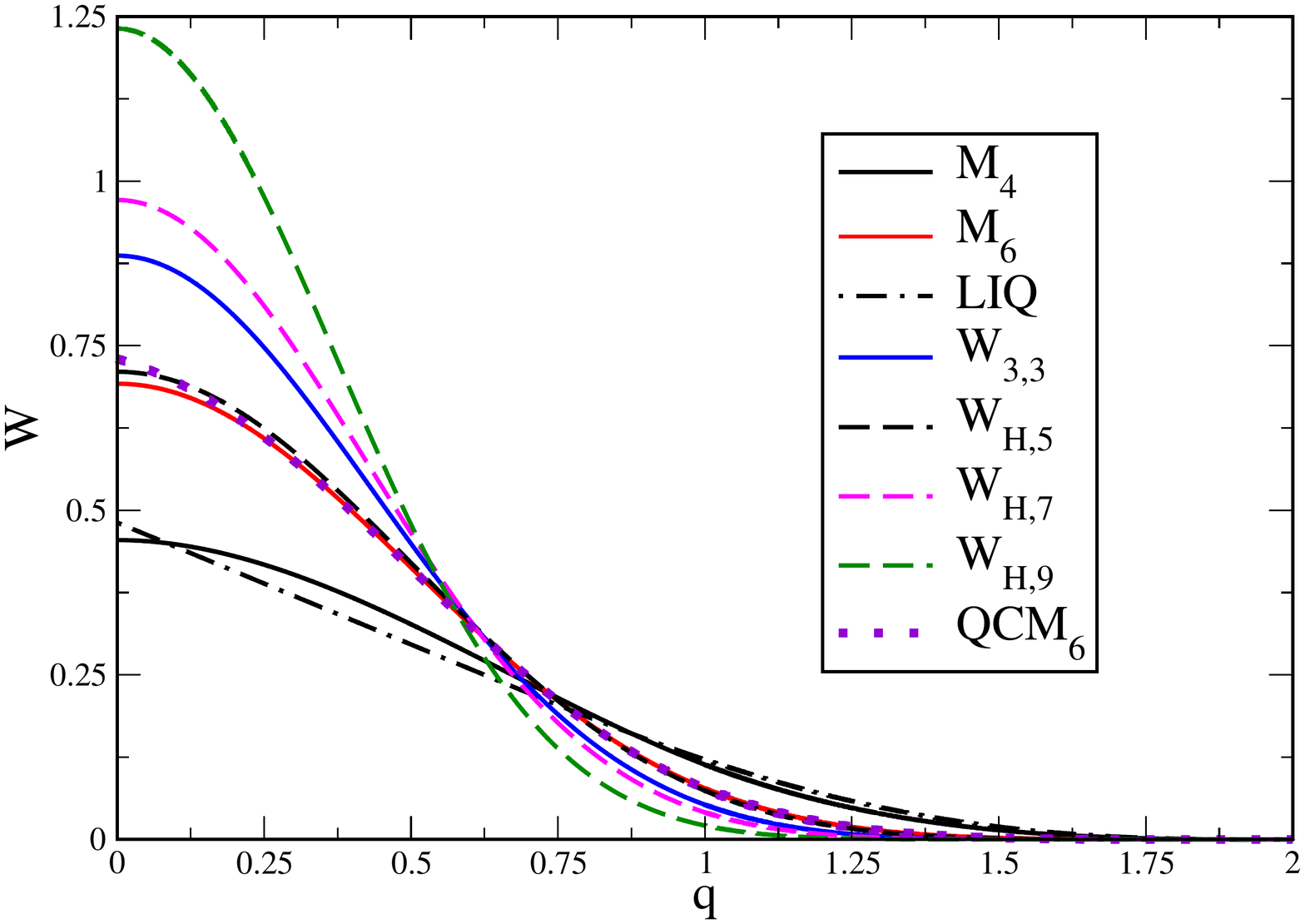}}
    \centerline{\includegraphics[width=0.8\textwidth,angle=0]{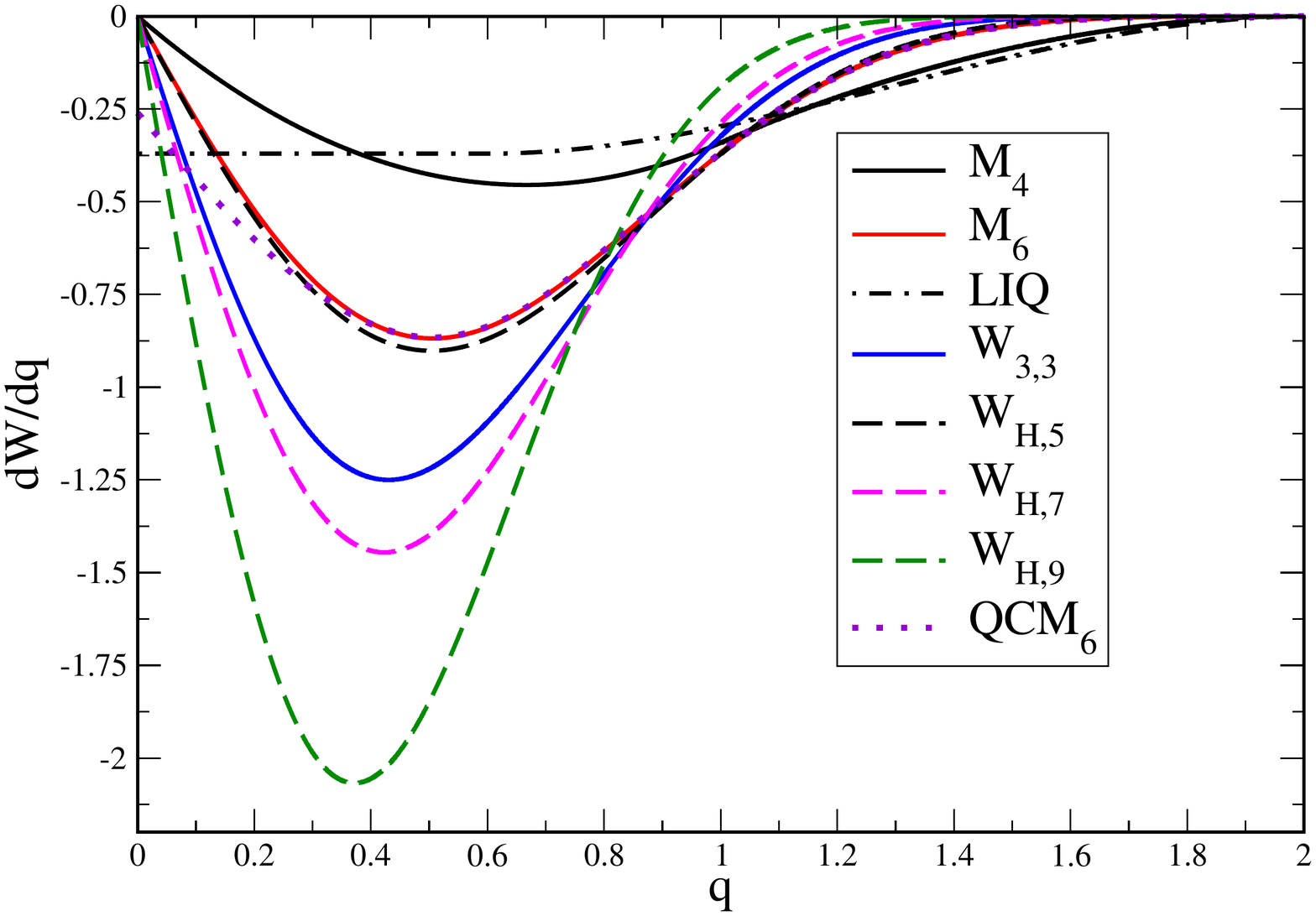}}
    \caption{Comparison of a selection of kernels. The kernel value $W$ ($h$=1) is displayed
          in the upper and  its derivatives in the lower panel. Note that the \M6 and the QCM$_6$
          kernels have, for ease of comparison, been scaled to a support of $2h$.}
    \label{fig:kernels}
\end{figure}}

\subsubsection{Kernels with vanishing central derivatives}

``Bell-shaped'' kernels with their vanishing derivatives at the origin are rather insensitive 
to the exact positions of nearby particles and therefore they are good 
density estimators \citep{monaghan92}. For this 
reason they have been widely used in SPH simulations. The kernels that we discuss here
and their derivatives are plotted in Figure~\ref{fig:kernels}. More recently, kernels with non-vanishing
central derivatives have been (re-)suggested, some of them are discussed in Section~\ref{sec:peaked_kernels}.

\subsubsection*{B-spline functions: M$_4$ and M$_6$ kernels}

The most commonly used SPH kernels are the so-called B-spline 
functions \citep{schoenberg46}, $M_n$, which are generated as Fourier transforms:
\be
M_n(x,h)= \frac{1}{2 \pi} \int_{-\infty}^\infty \left[ \frac{\sin(kh/2)}{kh/2}\right]^n \cos(kx) dk.
\ee
The smoothness of the $M_n$ functions increases with $n$ and they are
continuous up to the $(n-2)$-th derivative. Since SPH requires at the very least the continuity in the 
first and second derivative, the cubic spline kernel M$_4$ 
\be
w_4(q)=  \left\{\begin{array}{ll}  \frac{1}{4}(2-q)^3 - (1-q)^3
& 0 \le q < 1\\
         \frac{1}{4}(2-q)^3 & 1 \le q < 2\\
         0      & {\rm else}
         \end{array}\right.
\ee
is the lowest-order SPH kernel that is a viable option, it is often considered the ``standard choice'' in SPH.
As we will show below,  much more accurate choices are available at a moderate extra cost.
The normalization factor $\sigma$ of  M$_4$ has values of $[2/3,10/(7 \pi),1/\pi]$ in 1, 2 and 3 dimensions.

The quintic spline kernel M$_6$ has also occasionally been used in SPH simulations. It reads
(truncated at support $Q= 3$) 
\be
w_6(q)= \left\{\begin{array}{ll}  
         (3-q)^5 - 6(2-q)^5 + 15(1-q)^5 & 0 \le q < 1\\
         (3-q)^5 - 6(2-q)^5                      & 1 \le q < 2\\
         (3-q)^5                                        & 2 \le q < 3\\
                           0      & {\rm else}
         \end{array}\right. 
\ee
with normalizations of $[1/120, 7/(478 \pi), 1/(120 \pi)]$ in 1, 2 and 3 dimensions. Although
this is the commonly used form, we re-scale the \M6 kernel in all plots to a support of $Q= 2$ to 
enable a fair and easy comparison with other kernels in terms of computational effort (i.e., 
neighbor number).

\subsubsection*{A parametrized family of Kernels}

More recently, a one parameter family of kernels has been suggested \citep{cabezon08}
\be
W_{{\rm H},n}= \frac{\sigma_{{{\rm H},n}}}{h^D} \left\{\begin{array}{ll}  
         1 & q = 0\\
         \left( \frac{\sin [\frac{\pi}{2} q]}{\frac{\pi}{2} q}\right)^n    & 0 < q \le 2\\
         0      & {\rm else,}
         \end{array}\right. 
\ee
where $n$ determines the smoothness and the shape of the kernel, see Figure~\ref{fig:kernels}.
The normalization of this kernel family, $\sigma_{{\rm H},n}$,  for integer $n$ from 3 to 9 is given in 
Table~\ref{tab:kernel_params}. In \cite{cabezon08}, their Table~2, a fifth order polynomial is 
given that provides the normalization for continuous $n$ between 3 and 7.
The $W_{{\rm H},3}$ kernel is very similar to the  M$_4$ (not shown),
while $W_{{\rm H},5}$ is a close approximation of M$_6$, see Figure~\ref{fig:kernels},
 provided they have the same support.

\begin{table}
\caption{Parameters of W$_{H,n}$ and QCM$_6$ kernels.}
\label{tab:kernel_params}
\centering
{\small
\begin{tabular}{ c | c c c c c c c }
\multicolumn{8}{c}{\textbf{Normalization $\sigma_{H,n}$ of W$_{H,n}$ kernels}}\\\noalign{\vskip 2mm}
               & n= 3    & n= 4             & n= 5            & n= 6            & n= 7             & n= 8             & n= 9 \\
\hline \\
1D  & 0.66020338 & 0.75221501 & 0.83435371 & 0.90920480 & 0.97840221 & 1.04305235  & 1.10394401\\
2D  & 0.45073324 & 0.58031218 & 0.71037946 & 0.84070999 & 0.97119717 & 1.10178466  & 1.23244006\\
3D  & 0.31787809 & 0.45891752 & 0.61701265 & 0.79044959 & 0.97794935 & 1.17851074  & 1.39132215\\
\end{tabular}}

\vspace*{0.5cm}

{\small
\begin{tabular}{c|r}
\multicolumn{2}{c}{\textbf{Parameters of the QCM$_6$ kernel}}\\\noalign{\vskip 2mm}
\hline \\
$q_c$ &   $0.75929848$     \\
A        &  $11.01753798$    \\
B         & $-38.11192354$     \\
C        & $-16.61958320$   \\
D        &  $69.78576728$    \\
$\sigma_{1D}$  &  $8.24554795  \times10^{-3}$ \\
$\sigma_{2D}$  &  $4.64964683  \times10^{-3}$ \\
$\sigma_{3D}$  &  $2.65083908  \times10^{-3}$ \\
\end{tabular}}
\end{table}

\subsubsection*{Wendland kernels}

An interesting class of kernels with compact support and positive Fourier transforms
are the so-called Wendland functions \citep{wendland95}. In several areas of applied mathematics
Wendland functions have been well appreciated for their good interpolation properties, but they 
have not received much attention as SPH kernels and have only recently been explored in more
detail \citep{dehnen12,hu14a,rosswog15b}.  \cite{dehnen12} have pointed out in particular that these kernels
are not prone to the pairing instability, see Section~\ref{sec:pairing}, despite having a vanishing central derivative.
These kernels have been explored in a large number of benchmark tests \citep{rosswog15b} and 
they are high appreciated for their ``cold fluid properties'':  the particle distribution remains highly
ordered even in dynamical simulations  (e.g., Figure~\ref{fig:particle_dist_2}, upper right panel) 
and only allows for very little noise.

Here we only experiment with one particular example, the $C^6$ smooth
\be
W_{3,3}= \frac{\sigma_W}{h^3} \left(1 - q \right)_{+}^8 \left(  32 q^3 + 25 q^2 +  8 q + 1 \right)
\label{eq:wend33},
\ee
see e.g., \cite{schaback06}, where the symbol $(.)_+$ denotes the cutoff function $\max (.,0)$. 
The normalization $\sigma_W$ is 78/(7$\pi$) and 1365/(64 $\pi$) in  2 and 3 dimensions.

\subsubsection{Kernels with  non-vanishing central derivatives}
\label{sec:peaked_kernels}

A number of kernels have been suggested in the literature whose derivatives remain finite in  
the centre so that the repulsive forces between particles never vanish. The major
motivation behind such kernels is to achieve a very regular particle distribution
and in particular to avoid the pairing instability, see Section~\ref{sec:pairing}. However, as recently pointed
out by \cite{dehnen12}, the pairing instability is not necessarily related to a vanishing
central derivative, instead, non-negative Fourier transforms have been found as a necessary condition
for stability against pairing. Also kernels with vanishing central derivatives can possess this properties. 
We explore here only two peaked kernel functions, one, the Linear Quartic core (LIQ) 
kernel, that has been suggested as a cure to improve SPH's performance in Kelvin--Helmholtz
instabilities and another one, the Quartic Core M$_6$ (QCM$_6$) kernel, mainly for pedagogical 
reasons to illustrate how an even very subtle change in the central part of the kernel can 
seriously deteriorate the approximation quality. For a more extensive account on kernels with 
non-vanishing central derivatives we refer to the literature \citep{thomas92,fulk96,valcke10,read10}.

\subsubsection*{Linear quartic kernel}

The centrally peaked ``Linear Quartic'' (LIQ) kernel \citep{valcke10} reads
\be
W_{\rm LIQ}(q)= \frac{\sigma_{\rm LIQ}}{h^D}   \left\{
  \begin{array}{ l l l}
     F - q   \hspace*{3.7cm} {\rm for \; } q \le x_s\\
     A q^4 + B q^3 + C q^2 + D q + E  \;\; \; {\rm for \; }  x_s < q \le 1\\
     0   \hspace*{4.4cm} {\rm else } 
       \end{array} \right.\ee
with $x_s=0.3, A=-500/343, B=1300/343, C=-900/343, D=-100/343, E=200/343$ and $F= 13/20$. The 
normalization constant $\sigma_{\rm LIQ}$ is 1000/447, 3750/403 $\pi$ and 30000/2419$\pi$ in one, two and three dimensions.

\subsubsection*{Quartic core M$_6$ kernel}

We also explore a very subtle modification of the well-known and appreciated  M$_6$ kernel
so that it exhibits a small, but non-zero derivative in the centre.
This ``Quartic core $M_6$ kernel'' (QCM$_6$) 
is constructed by replacing the second derivative of the $M_6$ kernel for $q<q_c$ by a 
parabola whose parameters have been chosen so that the kernel fits smoothly and differentiably the 
$M_6$ kernel at the transition radius defined by  $d^2 w_6/dq^2(q_c)= 0$ (numerical value
$q_c= 0.75929848$). The QCM$_6$-kernel then reads:
\be
W_{\rm QCM_6}(q)= \frac{\sigma_{\rm QCM_6}}{h^D}  \left\{\begin{array}{ll}  
         A q^4 + B q^2 + C q + D& 0 \le q < q_c\\
         (3-q)^5 - 6(2-q)^5 + 15(1-q)^5& q_c \le q < 1\\
         (3-q)^5 - 6(2-q)^5            & 1 \le q < 2\\
         (3-q)^5                       & 2 \le q < 3\\
                           0      & {\rm else.}
         \end{array}\right. 
\ee
The coefficients $A, B, C$ and $D$ are determined by the conditions $w_{\rm QCM_6}(q_c)= w_6(q_c)$,
$w'_{\rm QCM_6}(q_c)= w'_6(q_c)$, $w''_{\rm QCM_6}(q_c)= w''_6(q_c)$ and $w'''_{\rm QCM_6}(q_c)= w'''_6(q_c)$, 
where the primes indicate the derivatives with respect to $q$. The resulting numerical coefficients are 
given in Table~\ref{tab:kernel_params}. Note that QCM$_6$ is continuous everywhere up to the third derivative.
As can be seen in Figure~\ref{fig:kernels}, the QCM$_6$ kernel (violet dots) deviates only subtly from
M$_6$ (solid red line), but even this very subtle modification does already seriously compromise the 
accuracy of the kernel. In a recent study \citep{rosswog15b}, it was found, however, that this kernel
has the property of producing only very little noise, particles placed initially on a quadratic or hexagonal 
lattice (in pressure equilibrium) remain on this lattice configuration.

The different kernels and their derivatives are summarized in Figure~\ref{fig:kernels}. As mentioned above,
the M$_6$, QCM$_6$ and W$_{3,3}$ kernels have been rescaled to a support of $Q=2$ to allow for an easy comparison.
Note how the kernels become more centrally peaked with increasing order, for example, the W$_{\rm H,9}$ kernel
only deviates noticeably from zero inside of $q=1$, so that it is very insensitive to particles entering or
leaving its support near $q=2$.

\subsubsection{Accuracy assessment of different kernels}

\subsubsection*{Density estimates}

We assess the density estimation accuracy of the different kernels in a numerical experiment.
The particles are placed in a 2D hexagonal lattice configuration in $[-1,1] \times [-1,1]$. This
configuration corresponds to the closest packing of spheres with radius $r_{\rm s}$ where each particle 
possesses an effective area of $A_{\rm eff}= 2 \sqrt{3} r_{\rm s}^2$. Each particle is now assigned the same mass 
$m_b= \rho_0 A_{\rm eff}$ to obtain the uniform density $\rho_0$. Subsequently the densities at 
the particle locations, $\rho_b$, are calculated via the standard SPH expression for the density,
Eq.~(\ref{eq:dens_sum}), and the average error $\epsilon= N^{-1}\sum_{b=1}^{N} \epsilon_b$ with
$\epsilon_b= |\rho_b-\rho_0|/\rho_0$ is determined. Figure~\ref{fig:density_errors}, upper panel,
shows the error as a function of $\eta$ which parameterizes the kernel support size via 
$h_b= \eta \left(m_b/\rho_b\right)^{1/D}$, $D$ being again the number of spatial dimensions.

\epubtkImage{}{%
  \begin{figure}[htbp]
    \centerline{\includegraphics[width=0.8\textwidth,angle=0]{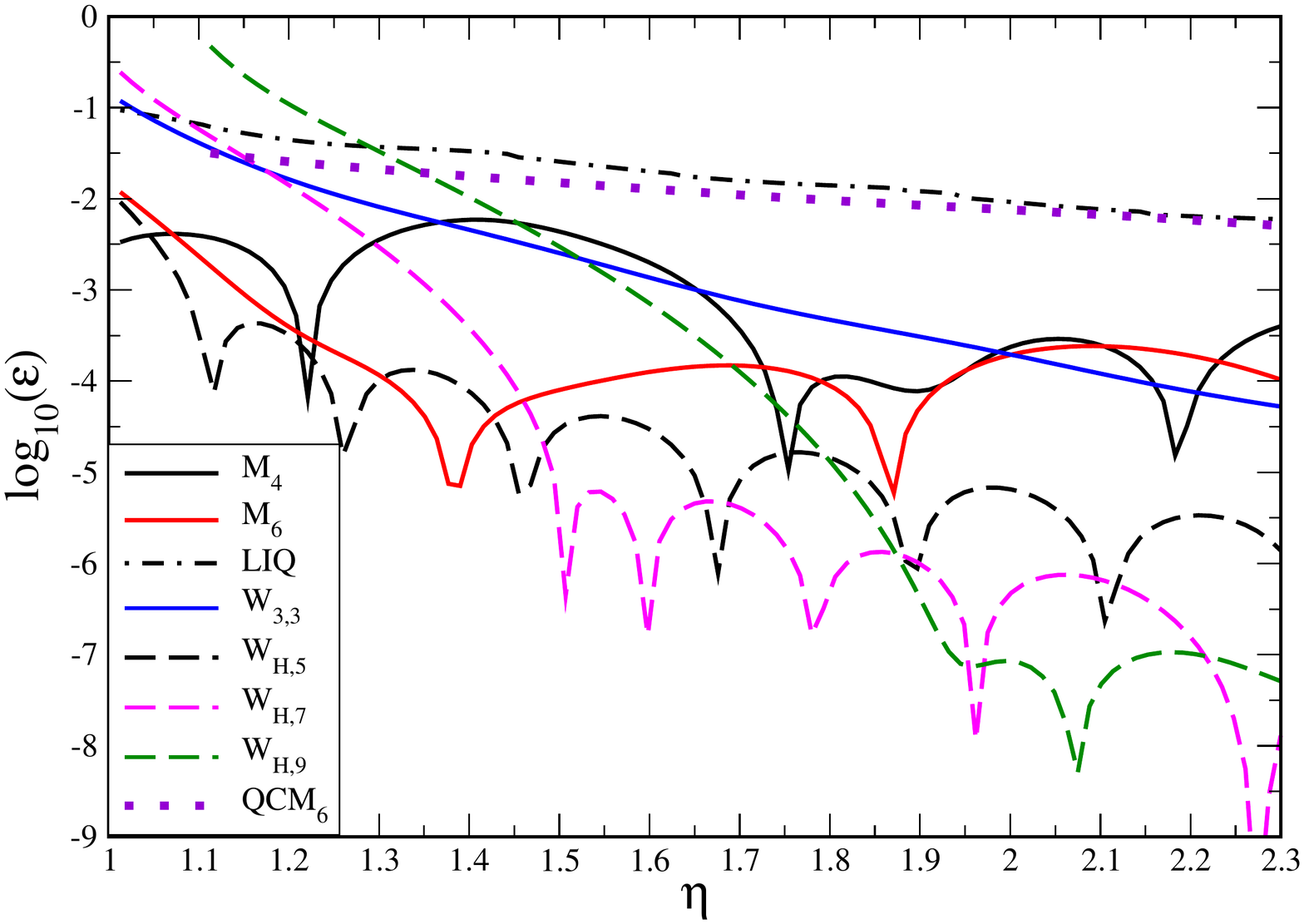}}
    \centerline{\includegraphics[width=0.8\textwidth,angle=0]{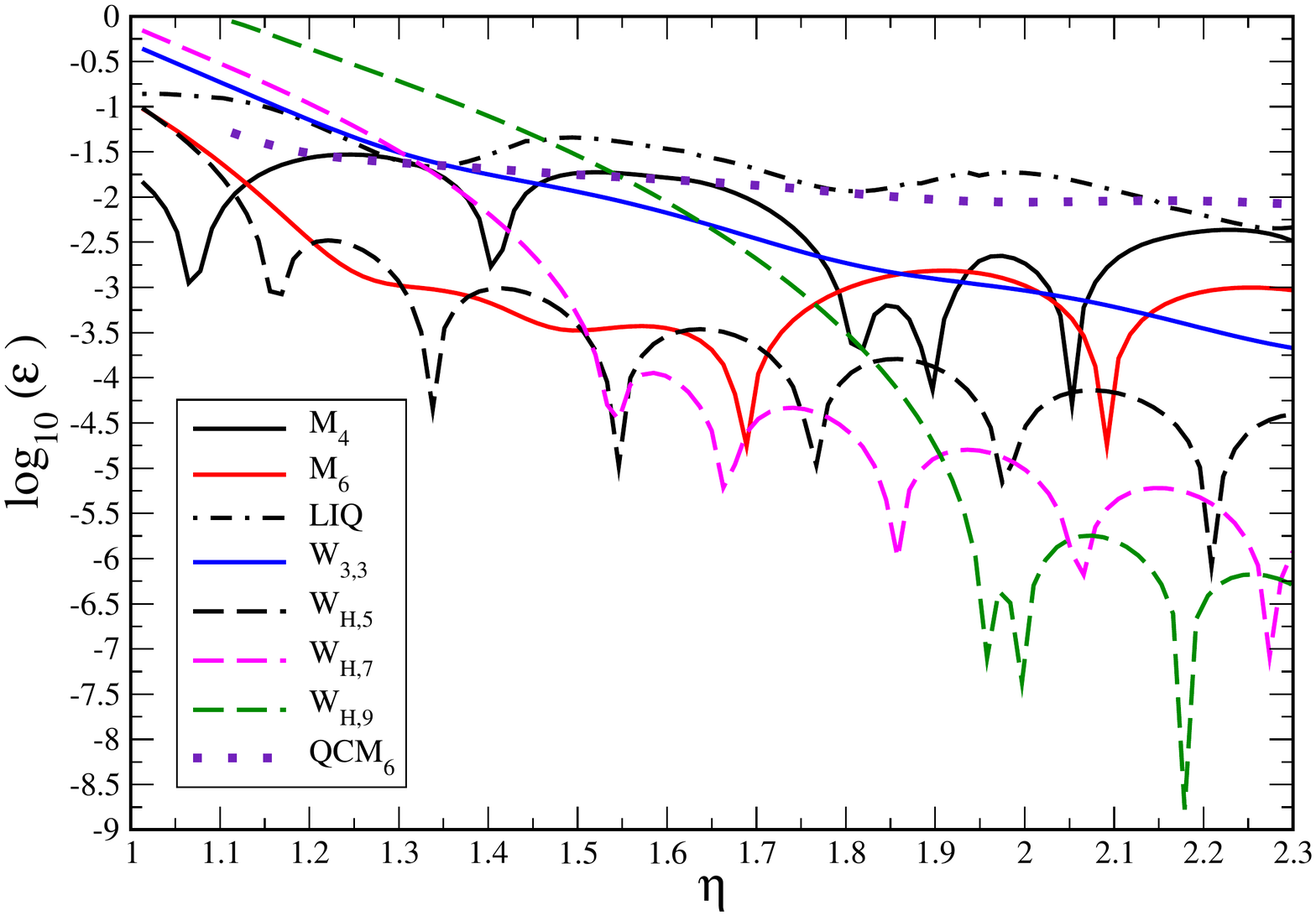}}
    \caption{Accuracy of different kernels. The upper/lower panel shows the accuracy of density/gradient
             accuracy for different kernels. The quantity $\eta$
             determines the smoothing length, $h \propto \eta$.  Particles are distributed on a 2D
             hexagonal lattice and they are assigned a constant mass for the density test and a linearly
             increasing pressure distribution in the gradient test. Note that the ``standard'' SPH kernel
             (``M$_4$'') does not perform particular well in either task. Adapted from \cite{rosswog15b}.}
    \label{fig:density_errors}
\end{figure}}

Interestingly, the ``standard'' cubic spline kernel (``CS'', solid black line in the figure)
actually does not perform well. At typical values of $\eta$ near 1.3 
the relative density error is a few times $10^{-3}$. Just replacing it by the quintic spline
kernel (``\M6'', solid red line in the figure) improves the density estimate for similar values of $\eta$
already by two orders of magnitude.\epubtkFootnote{Keep in mind that \M6 has been rescaled everywhere to a support
of $2h$ to ensure a fair comparison.}  
The Wendland kernel (``$W_{3,3}$'', dashed blue line in the figure) continuously decreases the error with increasing
$\eta$ and therefore does not show the pairing instability at any value of $\eta$ \citep{dehnen12}. It maintains a 
very regular particle distribution with very little noise \citep{rosswog15b}, 
see also the Gresho--Chan vortex test that is shown in Section~\ref{sec:gresho}.
In these fixed particle distribution tests the \whn-kernels perform particularly well. At large smoothing length
they deliver exquisite density estimates. For example, the W$_{\rm H,9}$ kernel is at $\eta > 1.9$ more than 
two orders of magnitude more accurate than \M6.

The centrally peaked kernels turn out to be rather poor density estimators. The shown LIQ 
kernel (filled black circles) needs a large $\eta$ beyond 2 to achieve a 
density estimate better than 1\%. Even the subtle modification of the central core in the QCM$_6$ 
kernel, substantially deteriorates the density estimate (violet dots) in comparison to the original M$_6$ kernel. 
One needs a value of $\eta>1.8$ for a density accuracy that is better than 1\%, at this large $\eta$ the 
\wh7 kernel is approximately four orders of magnitude more accurate. After extensive numerical experiments 
a recent study \citep{rosswog15b} still gave overall preference to the Wendland kernel Eq.~(\ref{eq:wend33}).
It gave less accurate results on ``frozen'' particle distributions than, say, the $W_{\rm H,9}$-kernel, but
in dynamical experiments where particles were allowed to move, the Wendland kernel maintained
a substantially lower noise level.

\subsubsection*{Gradient estimates}

We perform a similar experiment to measure the gradient accuracy. The particles are set up as before
and each particle is assigned a pressure $P(x,y)= 2 + x$.
We use the straight forward SPH estimate
\be
(\nabla P)_a= \sum_b \frac{m_b}{\rho_b} P_b \nabla_a W_{ab}(h_a)
\label{eq:press_grad}
\ee
to calculate the pressure gradient. 
The average error, $\epsilon= N^{-1}\sum_{b=1}^{N} \epsilon_b$ with $\epsilon_b \equiv |\nabla P - (\nabla P)_b|/
|\nabla P|$ is shown in Figure~\ref{fig:density_errors}, right panel, for different 
kernels as a function of $\eta$.

Again, the ``standard'' cubic spline kernel (solid black) does not perform particularly well, only
for $\eta>1.8$ reaches the gradient estimate an accuracy better than 1\%. In a dynamical situation, however,
the particle distribution would at such a large kernel support already fall prey to the pairing instability.
For moderately small supports, say $\eta<1.6$,
the \M6 kernel is substantially more accurate. Again, the accuracy of the Wendland kernel increases monotonically
with increasing $\eta$, and the three \whn-kernels perform best in this static test.  As in the case of the 
density estimates, the peaked kernels perform rather poorly and only achieve a 1\% accuracy for extremely large
values of $\eta$.

\subsubsection{Kernel choice and pairing instability}
\label{sec:pairing}

Figure~\ref{fig:density_errors} suggests to increase the kernel support to achieve greater
accuracy in density and gradient estimates (though at the price of reduced spatial resolution).
For many bell-shaped kernels, however, the support cannot be increased arbitrarily since
for large smoothing lengths the so-called ``pairing instability'' sets in where particles
start to form pairs. In the most extreme case, two particles can merge into effectively one.
Nevertheless, this is a rather benign instability. An example  is shown in Figure~\ref{fig:particle_dist_2},
where for the interaction of a blast wave with a low-density bubble\epubtkFootnote{This example is discussed 
in more detail in \cite{rosswog15b}.} we have once chosen (left panels) a kernel-smoothing length 
combination (M$_4$ kernel with $\eta= 2.2$) that leads to strong particle pairing and 
once (right panels) a combination (Wendland kernel with $\eta=2.2$) that produces a very regular
particle distribution. Note that despite the strong pairing in the left column, the continuum properties 
(the lower row shows the density as an example) are still reasonably well reproduced. The pairing 
simply leads to a loss of resolution, though at the original computational cost.

This instability has traditionally been explained by means of the inflection point in the kernel
derivatives of bell-shaped kernels, see Figure~\ref{fig:kernels}, that leads to decreasing and finally
vanishing repelling forces once the inflection point has been crossed. \cite{dehnen12} in contrast
argue that the stability of the Wendland kernels with respect to pairing is due to the density error being 
monotonically declining, see Figure~\ref{fig:density_errors}, and non-negative kernel Fourier transforms. 
They investigated in particular a number of bell-shaped Wendland kernels with 
strictly positive Fourier transforms and they did not find any sign of the instability despite the
vanishing central derivatives. This is, of course, a desirable feature for convergence studies since the neighbor
number can be increased without limit. 

\epubtkImage{}{%
  \begin{figure}[htbp]
  \centerline{\includegraphics[angle=90,width=1.2\textwidth]{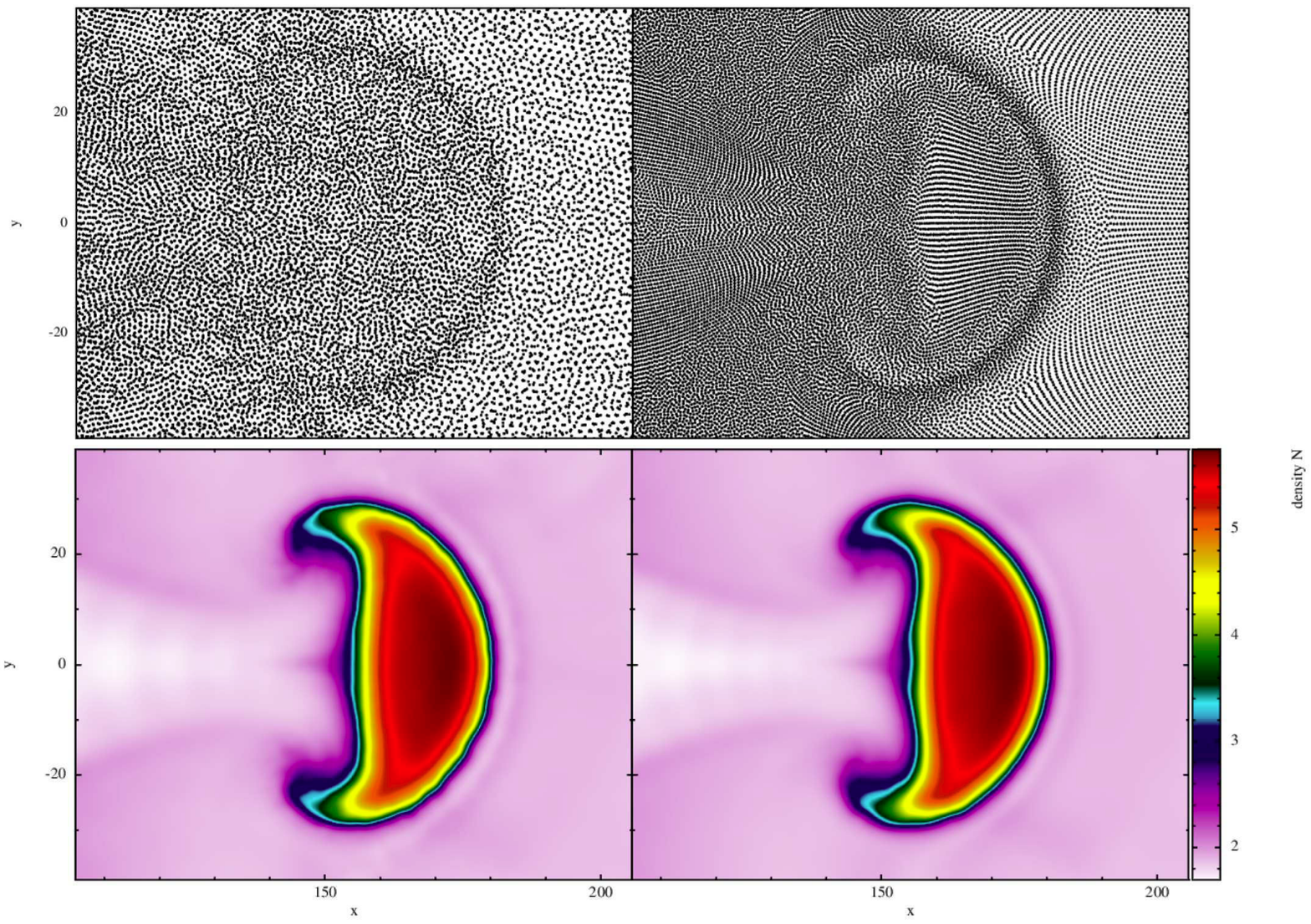}}
  \caption{Numerical experiment to illustrate the impact of the kernel and the smoothing length on the particle
               distribution. Shown is the interaction of a blast wave with a low-density bubble of equal pressure.
               Both experiments were conducted with a large smoothing length ($\eta=2.2$ in Eq.~(\ref{eq:eta})), 
               in the left column the ``standard'' cubic spline kernel is used, in the right column a high-order 
               Wendland kernel is used, see Eq.~(\ref{eq:wend33}). At these very large smoothing lengths, the cubic
               spline kernel becomes pairing unstable, i.e., the particles form a clumpy distribution. The Wendland kernel in contrast
               produces a very regular particle distribution. Note that despite the somewhat pathological
               particle distribution in the cubic spline kernel case, the density (lower row) is still reasonable well
               reproduced.}
   \label{fig:particle_dist_2}
\end{figure}}

\subsection{Summary: kernel approximation}

The improvement of the involved kernel approximation techniques is one obvious 
way how to further enhance the accuracy of SPH.
One should strive, however, to preserve SPH's most important asset, its exact
numerical conservation. The simplest possible, yet effective,  improvement is 
to just replace the kernel function, see Section~\ref{sec:kernel_choice}. We have briefly
discussed a number of kernels and assessed their accuracy in estimating a uniform
density and the gradient of a linear function for the case where the SPH particles
are placed on a hexagonal lattice. The most widely used kernel function, the cubic 
spline M$_4$,  does actually not perform particularly well, neither for the density 
nor the gradient estimate. At moderate extra cost, however, one can use substantially more accurate kernels, for
example the quintic spline kernel, M$_6$, or the higher-order members of the W$_{H,n}$ family. Another,
very promising kernel family are the Wendland functions. They are not prone to the pairing 
instability and therefore show much better convergence properties than kernels that start 
forming pairs beyond a critical support size. Moreover, the Wendland kernel that we explored
in detail  \citep{rosswog15b} is very reluctant to allow for particle  
motion on a sub-resolution scale and it maintains a very regular particle distribution, 
even in highly dynamical tests. The explored peaked kernels, in contrast, performed 
rather poorly in both estimating  densities and gradients.


\section{SPH Formulations of Ideal Fluid Dynamics}
\label{chap:SPH}

Smoothed Particle Hydrodynamics (SPH) is a completely mesh-free, Lagrangian method
that was originally suggested in an astrophysical context \citep{gingold77,lucy77}, but by now it has
also found many applications in the engineering world, see \cite{monaghan12a} for a starting point.
Since a number of detailed reviews exists, from the ``classics'' \citep{benz90a,monaghan92} to 
more recent ones \citep{monaghan05,rosswog09b,springel10a,price12a}, we want to 
avoid becoming too repetitive about SPH basics and therefore put the emphasis here on recent developments.
Many of them have very good potential, but have not yet fully made their way into practical 
simulations. Our emphasis here is also meant as a motivation for computational astrophysicists 
to keep their simulation tools up-to-date in terms of methodology. A very explicit account 
on the derivation of various SPH aspects has been provided in \cite{rosswog09b}, therefore 
we will sometimes refer the interested reader to this text for more technical details.

The basic idea of SPH is to represent a fluid by freely moving interpolation points -- the
particles -- whose evolution is governed Nature's  conservation laws. 
These particles move with the local fluid velocity and their densities and gradients 
are determined by the kernel approximation techniques discussed in Section~\ref{chap:kernel_approx}, 
see Figure~\ref{fig:sph_flow}. The corresponding evolution equations can be formulated in such a way that mass, energy, 
momentum and angular momentum are conserved by construction, i.e., they are fulfilled independent 
of the numerical resolution.

In the following we use the convention that the considered particle is labeled with ``a'',
its neighbors with ``b'' and a general particle with ``k'', see also the sketch in Figure~\ref{fig:interaction_sketch}. 
Moreover, the difference between two vectors is denoted as $\vec{A}_{ab}= \vec{A}_a - \vec{A}_b$ 
and the symbol $B_{ab}$ refers to the arithmetic average $B_{ab}= (B_a+B_b)/2$ of two scalar functions.

\epubtkImage{}{%
  \begin{figure}[htb]
    \centerline{\includegraphics[angle=90,width=0.6\textwidth]{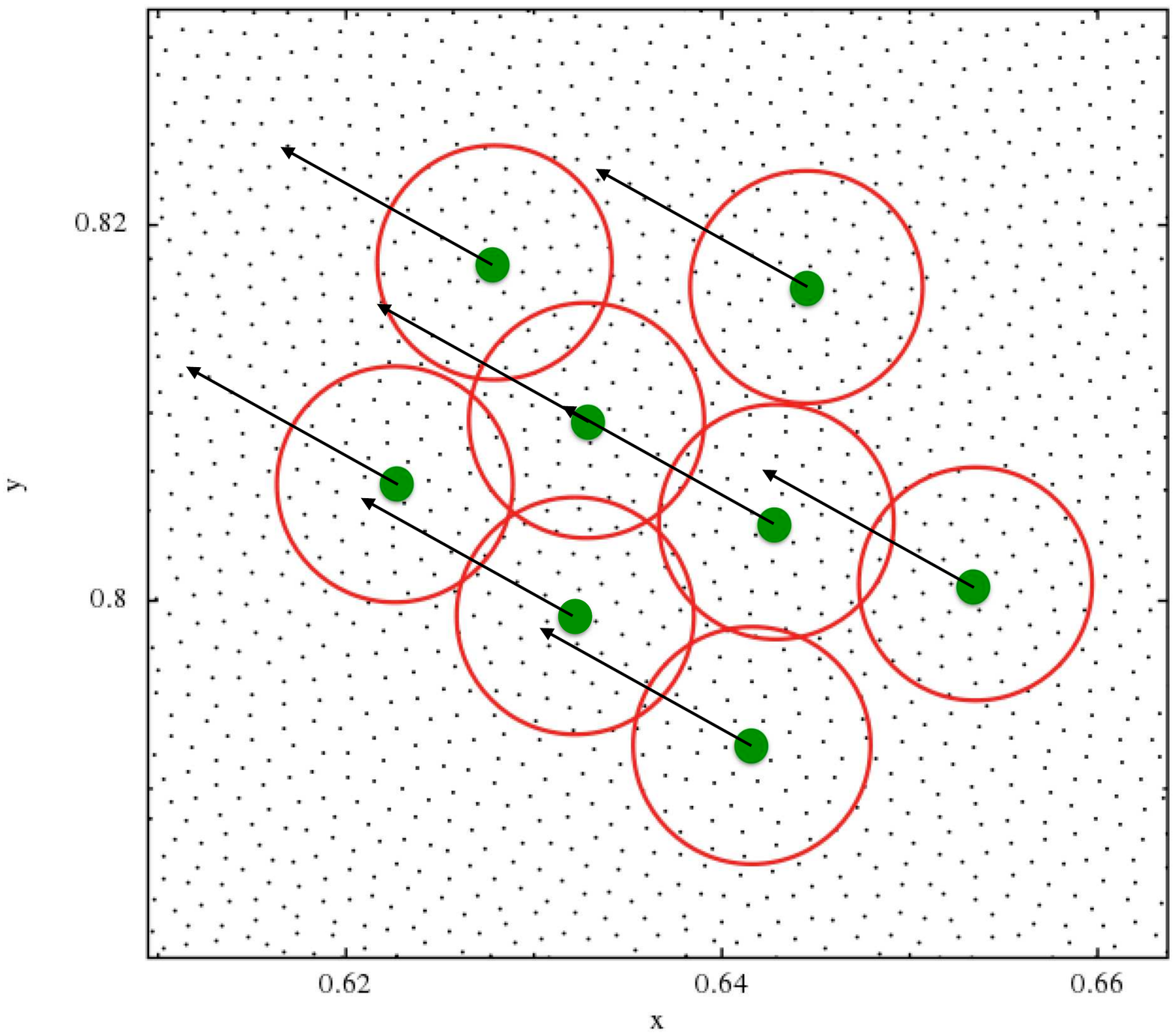}}
    \caption{Sampling of a fluid flow by SPH particles: the continuum is represented by a finite
                  number of sampling points (``particles'') that move with the local fluid velocity.
                  The particles share the total mass and move in a way that they conserve energy,
                  momentum and angular momentum. For some particles (green, filled circles) 
                  also their support (``sphere of influence'') is indicated by red circles.}
    \label{fig:sph_flow}
\end{figure}} 

\subsection{Choice of the SPH volume element}
\label{sec:volume_elements}

Up to now the choice of the volume element  in the kernel approximation, see Section~\ref{chap:kernel_approx},
has been kept open. The traditional choice is simply $V_b= m_b/\rho_b$, where $\rho$
is calculated as a kernel-weighted sum over nearby masses, see Eq.~(\ref{eq:dens_sum}) 
and $m$ is the  SPH particle mass. As will be discussed in Sections~\ref{sec:SR_SPH} and  
\ref{sec:GR_SPH}, this translates in a straight forward manner to relativistic SPH.
In the latter case the particle mass $m$ is replaced by the SPH particle's baryon number $\nu$ 
and  the mass density $\rho$ is replaced by the baryon number density as calculated in the
``computing frame'', see below.\\
 It has recently been realized \citep{saitoh13} that this ``standard''  choice of volume 
element is responsible for the spurious surface tension forces that can occur in many SPH 
implementations near contact discontinuities. At such discontinuities 
the pressure is continuous, but density $\rho$ and internal energy $u$ exhibit a jump.
For a polytropic equation of state, $P= (\Gamma-1) u \rho$, the product of density and 
internal energy must be the same on both sides to ensure a single value of 
$P$ at the discontinuity, 
i.e., $\rho_1 u_1= \rho_2 u_2$, where the subscripts label the two sides of the discontinuity. 
This means in particular, that the jumps in $u$ and $\rho$ need to be consistent with each other, 
otherwise the mismatch can cause spurious forces that have an effect like a surface tension and
can suppress weakly triggered fluid instabilities, see for example  
\cite{thacker00,agertz07,springel10a,read10}. In the ``standard'' SPH formulation such a mismatch can
occur because the density estimate is smooth, but the internal energy enters the SPH equations 
as an un-smoothed quantity. One may question, however, whether an unresolvably sharp transition
in $u$ is a viable numerical initial condition in the first place.

The problem can be alleviated if also the internal energy is smoothed by applying some 
artificial thermal conductivity and this has been shown to work well for Kelvin--Helmholtz 
instabilities \citep{price08a}. But, it is actually a non-trivial problem 
to design appropriate  triggers that supply conductivity exclusively where needed and not 
elsewhere. Artificial conductivity applied where it is undesired  can have catastrophic 
consequences, for example by removing physically required energy/pressure gradients for a star in
hydrostatic equilibrium.

An alternative cure comes from using different volume elements in
the SPH discretization process. The first step in this direction was probably taken by \cite{ritchie01}
who realized that by using the internal energy as a weight in the SPH density estimate a much sharper
density transition could be achieved than by the standard SPH density sum where each particle is,
apart from the kernel, only weighted by its mass.
In \cite{saitoh13} it was pointed out that SPH formulations that do not 
include density explicitly in the equations of motion do avoid the pressure becoming multi-valued 
at contact discontinuities. Since the density usually enters the equation of motion via the 
choice of the volume element, a different choice can possibly avoid the problem 
altogether. This observation is consistent with the findings of \cite{hess10} who used a 
particle hydrodynamics method, but calculated the volumes via a
Voronoi tessellation rather than via smooth density sum. In their approach no spurious surface 
tension effects have been observed. Closer to the original SPH spirit is the class of 
kernel-based particle volume estimates that have recently been suggested by \cite{hopkins13} 
as a generalization of the approach from \cite{saitoh13}. Recently, such volume elements
have been generalized for the use in special-relativistic studies \citep{rosswog15b}.

Assume that we have an (exact or approximate) partition of unity
so that 
\be
\sum_b \Phi_b(\vec{r})= 1
\label{eq:PU}
\ee
and a function $U$ can be approximated as
\be
\tilde{U}(\vec{r}) \approx \sum_b U_b \Phi_b(\vec{r}),
\label{eq:PU_approx}
\ee
where $U_b$ are the known function values at positions $\vec{r}_b$. 
This can be used to split up space (total volume $V$) into volume
elements $V_b$
\be
V = \int dV=  \int \left(\sum_b \Phi_b(\vec{r})\right) dV= \sum_b V_b,
\ee
where Eq.~(\ref{eq:PU}) has been inserted and the particle volume
\be
V_b= \int \Phi_b(\vec{r}) \, dV \label{eq:def_volume}
\ee
has been introduced. One may use, for example, a Shepard-type partition of unity \citep{shepard68}
\be
\Phi_b(\vec{r})= \frac{W_b(\vec{r})}{\sum_k W_k(\vec{r})},
\ee
with $W_b(\vec{r})= W(|\vec{r}-\vec{r}_b|)$ being a smoothing kernel so that upon using Eq.~(\ref{eq:def_volume}) 
the particle volume becomes
\be
V_b= \int \frac{W_b(\vec{r})}{\sum_k W_k(\vec{r})} \, dV.
\ee
Making use of the approximate $\delta$-property of the kernel, this yields
\be
V_b= \frac{1}{\sum_k W_{kb}},
\ee
where $W_{kb}= W(|\vec{r}_k-\vec{r}_b|)$. This, of course, has the simple physical interpretation of locally
sampling the particle number density (keep in mind that the kernel has the dimension of an inverse volume)
and taking its inverse as the particle volume.

While this is a straight forward and plausible approach, one can in principle
generalize this volume estimate by weighting the kernel with any scalar property $X$  \citep{hopkins13}
of the particles, so that the volume becomes
\be
V_b^{(X)}= \frac{X_b}{\sum_k X_k W_{kb}} \equiv \frac{X_b}{\kappa_{X, b}} 
\label{eq:gen_vol_element}
\ee
and one can choose to calculate the density via
\be
\rho_b^{(X)}= \frac{m_b}{V_b^{(X)}}= \frac{m_b}{X_b} \kappa_{X, b},
\label{eq:density}
\ee
respectively. If the smoothing length is adjusted to a multiple of the typical particle separation,
\be
h_b= \eta (V_b^{(X)})^{1/D},
\label{eq:eta}
\ee
 $D$ being the number of spatial dimensions, the derivatives of the 
quantity $\kappa_{X,b}$ become\epubtkFootnote{The derivation follows closely the steps explained in detail 
in Section~3 of \cite{rosswog09b}.}
\be
\nabla_a \kappa_{X,b}= \frac{1}{\Omega_b} \sum_k X_k \nabla_a W_{bk}(h_b)
\quad {\rm and} \quad 
\frac{d \kappa_{X,b}}{dt}= \frac{1}{\Omega_b} \sum_k X_k \vec{v}_{bk} \cdot \nabla_b W_{bk}(h_b), 
\label{eq:kappa_derivs}
\ee
with the generalized ``grad-h terms'' being
\be
\Omega_b= 1 - \frac{m_b}{X_b} \frac{\p h_b}{\p \rho_b} \sum_k X_k \frac{\p W_{kb}(h_b)}{\p h_b}.
\ee
For the SPH discretization process one needs the derivatives of the volume elements
\be
\nabla_a V_b
= -\frac{V_b^2}{X_b \Omega_b} 
\sum_k X_k \nabla_a W_{bk}(h_b)
\label{eq:nabla_V}
\ee
and 
\be
\frac{d V_b}{dt} 
= - \frac{V_b^2}{X_b \Omega_b} \sum_k X_k \vec{v}_{bk} \cdot \nabla_b W_{bk}(h_b).
\label{eq:ddt_V}
\ee

\subsection{Newtonian SPH}
\label{sec:Newt_SPH}

In its most basic form, the task is simply to solve the Lagrangian conservation equations for mass, energy and 
momentum of an ideal fluid \citep{landau59}:
\bea
\frac{d\rho}{dt}&=& -\rho \nabla \cdot \vec{v},\\ 
\frac{du}{dt}&=& \left(\frac{P}{\rho^2}\right) \frac{d\rho}{dt}, \\
\frac{d\vec{v}}{dt}&=& -\frac{\nabla P}{\rho},\label{eq:Newt_Euler}
\eea
where $\rho$ is the mass density, $d/dt= \p_t + \vec{v} \cdot \nabla$ the Lagrangian time derivative, $\vec{v}$
the fluid velocity, $u$ the specific thermal energy and $P$ the pressure. Like in other numerical methods, 
many different discrete approximations to the continuum equations can be used and they may differ substantially
in their accuracy. In the following, we will summarize commonly used SPH discretizations that have been used
for simulations of compact objects. They differ in their derivation strategy, the resulting symmetries in the 
particle indices, the volume elements, the manner how gradients are calculated and in the way they deal with shocks.

\subsubsection{``Vanilla ice'' SPH}
\label{sec:Newt_vanilla}

We will begin with the simplest, but fully conservative, SPH formulation (``vanilla ice'') that is still used
in many astrophysical simulations. A detailed step-by-step derivation with particular attention to
conservation issues can be found in Section~2.3 of \cite{rosswog09b}. By using the volume
element $V_b= m_b/\rho_b$ and the derivative prescription Eq.~(\ref{eq:const_exact_gradient})
one finds the discrete form of the Lagrangian continuity equation 
\be
\frac{d\rho}{dt}= -\rho \nabla \cdot \vec{v} \quad \Rightarrow 
\quad \frac{d\rho_a}{dt}= \sum_b m_b \vec{v}_{ab} \cdot \nabla_a W_{ab}
\label{eq:Newt_drho_dt},
\ee
where $\vec{v}_{ab}= \vec{v}_a - \vec{v}_b$ and $W_{ab}= W(|\vec{r}_a-\vec{r}_b|,h)$. 
As an alternative to this ``density-by-integration'' approach one can also estimate the
density at a particle position $\vec{r}_a$ as a weighted sum over contributing particles
(``density-by-summation'')
\be
\rho_a= \sum_b m_b W_{ab}(h_a).
\label{eq:dens_sum}
\ee
In practice, there is little difference between the two.
Note that this density estimate corresponds to the choice $X= m$ in 
Eq.~(\ref{eq:gen_vol_element}).
Usually the particle masses are kept constant so that exact mass conservation is enforced automatically.
Most often, the specific energy $u$ is  evolved in time, its evolution equation follows directly from 
Eq.~(\ref{eq:Newt_drho_dt}) and the adiabatic first law of thermodynamics, 
\be
\p u_b/\p \rho_b = P_b/\rho_b^2
\label{eq:first_law}
\ee
as
\be
\frac{du}{dt}= \left(\frac{P}{\rho^2}\right) \frac{d\rho}{dt} \quad
\Rightarrow \quad
\frac{du_a}{dt}= \frac{P_a}{\rho_a^2}\sum_b m_b \vec{v}_{ab} \cdot \nabla_a W_{ab}.
\label{eq:Newt_du_dt}
\ee
A discrete form of the momentum equation that enforces exact conservation can be found by using
the identity
\be
\frac{\nabla P}{\rho} =  \frac{P}{\rho^2} \nabla \rho+ \nabla \left(\frac{P}{\rho}\right)
\label{eq:basic:nabla_P_rho}
\ee
in the Lagrangian momentum equation
\bea
\frac{d\vec{v}}{dt}= -\frac{\nabla P}{\rho} \quad \Rightarrow \quad
\frac{d\vec{v}_a}{dt}= - \sum_b m_b \left(\frac{P_a}{\rho_a^2} + \frac{P_b}{\rho_b^2} \right)
\nabla_a W_{ab}. 
\label{eq:Newt_momentum_equation}
\eea
If the kernel $W_{ab}$ in Eqs.~(\ref{eq:Newt_du_dt}) and (\ref{eq:Newt_momentum_equation}) is
evaluated with a symmetric combination of smoothing lengths, say with $h_{ab}= (h_a+h_b)/2$, 
then Eqs.~(\ref{eq:dens_sum}), (\ref{eq:Newt_du_dt}) and 
(\ref{eq:Newt_momentum_equation}) form, together with an equation of state, a closed 
set of equations that enforces the exact conservation of mass, energy, linear and angular momentum
by construction. For practical simulations that may possibly involve shocks, Eqs.~(\ref{eq:Newt_du_dt}) and 
(\ref{eq:Newt_momentum_equation}) need to be augmented by extra measures (e.g., by artificial 
viscosity) to ensure that entropy is produced in a shock and  that kinetic energy is properly transformed 
into thermal energy, see Section~\ref{sec:Newtonian_shocks}.

\subsubsection{SPH from a variational principle}
\label{sec:SPH_from_variational_principle}

More elegantly, a set of SPH equations can be obtained by starting from the (discretized)
Lagrangian of an ideal fluid \citep{gingold82,speith98,monaghan01,springel02,rosswog09b,price12a}. 
This ensures that  conservation of mass, energy, momentum and angular momentum is, by 
construction, built into the discretized form of the resulting fluid equations. Below, we use the 
generalized volume element Eq.~(\ref{eq:gen_vol_element}) 
without specifying the choice of the weight $X$ so that a whole class of SPH equations is produced 
\citep{hopkins13}. In this derivation the volume of an SPH particle takes over the fundamental role
that is usually played by the density sum. Therefore we will generally express 
the SPH equations in terms of volumes rather than densities, we only make 
use of the latter for comparison with known equation sets.

The Lagrangian of an ideal fluid can be written as \citep{eckart60}
\be
L= \int \rho(x) \left( \frac{v(x)^2}{2} - u(x)\right) \, dx,
\ee
and on using Eq.~(\ref{eq:PU_approx}) it can be discretized into
\be
L \simeq  \int \left\{ \sum_b \rho_b \left( \frac{v_b^2}{2} - u_b\right) \Phi_b(x) \right\}dx = 
\sum_b \rho_b V_b \left( \frac{v_b^2}{2} - u_b \right),
\ee
where we have used the definition of the particle volume Eq.~(\ref{eq:def_volume}).
For the choice $m_b= \rho_b V_b$ the standard SPH-Lagrangian
\be
L= \sum_b m_b \left( \frac{v_b^2}{2} - u_b \right)
\ee
is recovered. The Euler--Lagrange equations
\be
\frac{d}{dt}\frac{\p L}{\p \vec{v}_a} - \frac{\p L}{\p \vec{r}_a} = 0
\ee
then yield, for a fixed particle mass $m_a$,
\bea
\frac{d \vec{v}_a}{dt}&=& \frac{1}{m_a}\frac{\p L}{\p \vec{r}_a}= - \frac{1}{m_a}
                          \sum_b m_b \frac{P_b}{\rho_b^2} \frac{\p (m_b/V_b)}{\p \vec{r}_a} 
                       = \frac{1}{m_a} \sum_b P_b \frac{\p V_b}{\p\vec{r}_a},
\label{eq:momentum_from_Lagrangian}
\eea
where Eq.~(\ref{eq:first_law}) was used. 
The energy equation follows directly from the first law of thermodynamics as
\be
\frac{du_a}{dt}=  -\frac{P_a}{m_a} \frac{dV_a}{dt}.
\ee
With Eqs.~(\ref{eq:nabla_V}), (\ref{eq:ddt_V}),
$\nabla_b W_{ab}= -\nabla_a W_{ab}$ and $\nabla_a W_{bk}= \nabla_b W_{bk}(\delta_{ba} - \delta_{ka})$
the SPH equations for a general volume element of the form given in Eq.~(\ref{eq:gen_vol_element})
become
\bea
V_b                   &=& \frac{X_b}{\kappa_{X,b}}\label{eq:gen_vol_N}\\
m_a \frac{d\vec{v}_a}{dt} &=& - \sum_{b} X_a X_b \left\{  \frac{P_a}{\kappa_a^2 \Omega_a} \nabla_a W_{ab}(h_a) 
+ \frac{P_b}{\kappa_b^2 \Omega_b} \nabla_a W_{ab}(h_b)  \right\}\label{eq:gen_mom_N} \\
m_a \frac{du_a}{dt}       &=& \frac{P_a X_a}{\kappa_a^2\Omega_a} \sum_b X_b \vec{v}_{ab} \cdot \nabla_a W_{ab}(h_a).\label{eq:gen_en_N}
\eea
For the choice $X=m$ this reduces to the commonly used equation set \citep{monaghan01,price04c,rosswog07c}.
Explicit forms of the equations, also for other choices of $X$, are given in Tab.~\ref{tab:explicit_forms_Newt_SPH}.

\begin{table}
\caption{Explicit forms of the equations for special choices of the weight $X$ for Newtonian and special-relativistic SPH.}
\label{tab:explicit_forms_Newt_SPH}
\centering
{\small
\centerline{\bf \underline{Newtonian SPH}}

\vspace*{0.4cm}

\begin{tabular}{ l l}
\multicolumn{2}{c}{\textbf{Weight $X= m$}}\\
\midrule
\\
 density      & $\rho_a \; \,= \sum_b m_b W_{ab}(h_a)$ \\
momentum      & $\frac{d\vec{v}_a}{dt}= - \sum_b m_b \left\{ \frac{P_a}{\Omega_a \rho_a^2} \nabla_a W_{ab}(h_a) 
                + \frac{P_b}{\Omega_b \rho_b^2} \nabla_a W_{ab}(h_b) \right\}$   \\
energy        & $\frac{du_a}{dt}=  \frac{P_a}{\Omega_a \rho_a^2} \sum_b m_b \vec{v}_{ab} \cdot \nabla_a W_{ab}(h_a)$
\end{tabular}

\vspace*{0.5cm}

\begin{tabular}{ l l}
\multicolumn{2}{c}{\textbf{Weight $X=1$}}\\
\midrule
\\
 density      & $\rho_a \; \,= m_a \sum_b W_{ab}(h_a)$ \\
momentum      & $\frac{d\vec{v}_a}{dt} = - 
                                   \sum_b m_b \left\{ \frac{m_a}{m_b} \frac{P_a}{\Omega_a \rho_a^2} \nabla_a W_{ab}(h_a) 
                                   + \frac{m_b}{m_a} \frac{P_b}{\Omega_b \rho_b^2}   \nabla_a W_{ab}(h_b) \right\}$   \\
energy        & $\frac{du_a}{dt} = \frac{P_a m_a}{\Omega_a \rho_a^2} \sum_b \vec{v}_{ab} 
                \cdot \nabla_a W_{ab}(h_a)$
\end{tabular}

\vspace*{0.5cm}

\begin{tabular}{ l l}
\multicolumn{2}{c}{\textbf{Weight $X= P^k$}}\\
\midrule
\\
 density      & $\rho_a \; \,= m_a \sum_b \left( \frac{P_b}{P_a} \right)^k W_{ab}(h_a)$ \\
momentum      & $\frac{d\vec{v}_a}{dt}= - \sum_b m_b \left\{ \frac{m_a}{m_b}\frac{P_a^{1-k} P_b^k }{\Omega_a^2 \rho_a^2} \nabla_a W_{ab}(h_a) 
                + \frac{m_b}{m_a} \frac{P_a^k P_b^{1-k}}{\Omega_b \rho_b^2} \nabla_a W_{ab}(h_b) \right\}$   \\
energy        & $\frac{du_a}{dt}=  \frac{P_a m_a}{\Omega_a \rho_a^2} \sum_b \left( \frac{P_b}{P_a} \right)^k \vec{v}_{ab} \cdot \nabla_a W_{ab}(h_a)$
\end{tabular}

\vspace*{0.8cm}

\centerline{\bf \underline{Special-relativistic SPH}}

\vspace*{0.4cm}

\begin{tabular}{ l l}
\multicolumn{2}{c}{\textbf{Weight $X= \nu$}}\\
\midrule
\\
CF bar. num. density       & $N_a \; \,= \sum_b \nu_b W_{ab}(h_a)$ \\
spec. canon. momentum & $\frac{d\vec{S}_a}{dt}= - \sum_b \nu_b 
                                          \left\{ \frac{P_a}{\Omega_a N_a^2} \nabla_a W_{ab}(h_a) 
                                        + \frac{P_b}{\Omega_b N_b^2} \nabla_a W_{ab}(h_b) \right\}$   \\
spec. canon. energy        &  $\frac{d \epsilon_a}{dt}= - \sum_b \nu_b
                                           \left\{ \frac{P_a }{\tilde{\Omega}_a N_a^2}  \vec{v}_b \cdot 
                                           \nabla_a W_{ab}(h_a)
                                        + \frac{P_b }{\tilde{\Omega}_b}   \vec{v}_a \cdot 
                                            \nabla_a W_{ab}(h_b)\right\}$ 
\end{tabular}

\vspace*{0.5cm}

\begin{tabular}{ l l}
\multicolumn{2}{c}{\textbf{Weight $X=1$}}\\
\midrule
\\
CF bar. num. density            & $N_a \; \,= \nu_a \sum_b W_{ab}(h_a)$ \\
spec. canon. momentum      & $\frac{d\vec{S}_a}{dt} = - 
                                   \sum_b \nu_b \left\{ \frac{\nu_a}{\nu_b} \frac{P_a}{\Omega_a N_a^2} \nabla_a W_{ab}(h_a) 
                                   + \frac{\nu_b}{\nu_a} \frac{P_b}{\Omega_b N_b^2}   \nabla_a W_{ab}(h_b) \right\}$   \\
spec. canon. energy             & $\frac{d \epsilon_a}{dt}= - \sum_b \nu_b
                                           \left\{ \frac{\nu_a}{\nu_b} \frac{P_a}{\tilde{\Omega}_a N_a^2}  \vec{v}_b \cdot 
                                           \nabla_a W_{ab}(h_a)
                                        +\frac{\nu_b}{\nu_a}  \frac{P_b}{\tilde{\Omega}_b}   \vec{v}_a \cdot 
                                            \nabla_a W_{ab}(h_b)\right\}$ 
\end{tabular}

\vspace*{0.5cm}

\begin{tabular}{ l l}
\multicolumn{2}{c}{\textbf{Weight $X= P^k$}}\\
\midrule
\\
CF bar. num. density       & $N_a \; \,= \nu_a \sum_b \left( \frac{P_b}{P_a} \right)^k W_{ab}(h_a)$ \\
spec. canon. momentum      & $\frac{d\vec{S}_a}{dt}= - \sum_b \nu_b \left\{ \frac{\nu_a}{\nu_b}\frac{P_a^{1-k}P_b^k}{\Omega_a^2 N_a^2} \nabla_a W_{ab}(h_a) 
                + \frac{\nu_b}{\nu_a} \frac{P_a^k P_b^{1-k}}{\Omega_b N_b^2} \nabla_a W_{ab}(h_b) \right\}$   \\
spec. canon.  energy        &  $\frac{d \epsilon_a}{dt}= - \sum_b \nu_b
                                           \left\{ \frac{\nu_a}{\nu_b} \frac{P_a^{1-k} P_b^{k}}{\tilde{\Omega}_a N_a^2}  \vec{v}_b \cdot 
                                           \nabla_a W_{ab}(h_a)
                                        +\frac{\nu_b}{\nu_a}  \frac{P_a^k P_b^{1-k}}{\tilde{\Omega}_b}  \vec{v}_a \cdot 
                                            \nabla_a W_{ab}(h_b)\right\}$ 
\end{tabular}}
\end{table}

\subsubsection*{The GADGET equations}

The arguably most wide-spread SPH simulation code, \textsc{Gadget} \citep{springel01a,springel05a},
uses an entropy formulation of SPH. It uses the entropic function $A(s)$ occurring
in a polytropic equation of state, $P= A(s) \rho^\Gamma$. If no entropy is created,
the quantity $A$ is simply advected by the fluid element and only the momentum equation needs to be 
solved explicitly. In \textsc{Gadget} the smoothing lengths\epubtkFootnote{Note that we write the equations here as in the
original paper, i.e., with a smoothing kernel that vanishes at $1h$ rather than at $2 h$.} are adapted so that 
\be
\frac{4 \pi}{3} h_a^3 \rho_a = N_{\rm SPH} \bar{m}
\label{eq:gadget_h}
\ee
is fulfilled, where $N_{\rm SPH}$ is the typical neighbor number and $\bar{m}$ is an 
average particle mass. The Euler--Lagrange equations then yield 
\be
\frac{d\vec{v}_a}{dt}= - \sum_b m_b \left\{ \frac{f_a P_a}{\rho_a^2} \nabla_a W_{ab}(h_a) 
                + \frac{f_b P_b}{\rho_b^2} \nabla_a W_{ab}(h_b) \right\},
\label{eq:momentum_gadget}
\ee
where the $f_k$ are the grad-h terms that read
\be
f_k= \left(1 + \frac{h_k}{3 \rho_k} \frac{\p \rho_k}{\p h_k}\right)^{-1}.
\ee
Obviously, in this formulation artificial dissipation terms need to be applied also
to $A$ to ensure that entropy is produced at the right amounts in shocks.

\subsubsection{Self-regularization in SPH}
\label{sec:self_regularization}

SPH  possesses a built-in ``self-regularization'' mechanism, i.e. SPH particles feel,
in addition to pressure gradients, a force that aims at driving them towards
 an optimal particle distribution. This corresponds to (usually ad hoc introduced)
``re-meshing'' steps that are used in Lagrangian mesh methods. The ability to automatically re-mesh is closely related the lack
of zeroth order consistency of SPH that was briefly described in Section~\ref{sec:direct_gradient}: the particles  ``realize''
that their distribution is imperfect and they have to adjust accordingly. Particle methods without such a re-meshing mechanism can
quickly evolve into rather pathological particle configurations that yield, in long-term, very poor results, see \cite{price12a}
for an explicit numerical example.

To understand this mechanism better, it is instructive to expand $P_b$ in Eq.~(\ref{eq:momentum_from_Lagrangian}) 
around $P_a$ (sum over $k$)
\bea
m_a \left(\frac{d \vec{v}_a}{dt}\right)^i&=& \sum_b \left\{ P_a + (\nabla P)_a^k  (\vec{r}_b - \vec{r}_a )^k + \dots\right\} \left(\frac{\p V_b}{\p\vec{r}_a}\right)^i\\
                                       &\approx& P_a \left(\frac{\p}{\p \vec{r}_a} \sum_b V_b\right)^i + (\nabla P)_a^k  \sum_b (\vec{r}_b - 
                                                           \vec{r}_a )^k \left(\frac{\p V_b}{\p\vec{r}_a}\right)^i\\
                                       &\equiv& P_a e_0^i -  (\nabla P)_a^k  V_a D^{ki}\\
                                       &\equiv& f_{\rm reg}^i + f_{\rm hyd}^i.
\eea
The first term is the ``regularization force'' that is responsible for driving particles into a good distribution. It only 
vanishes, $\vec{e}_0=0$, if the sum of the volumes is constant. The second term, $f_{\rm hyd}^i$, is the approximation to the hydrodynamic 
force. For a good particle distribution, $e_0^i \rightarrow 0$ and $D^{ki} \rightarrow \delta^{ki}$ and thus (using $\rho_a= m_a/V_a$) the Euler 
equation (\ref{eq:Newt_Euler}) is reproduced.

\subsubsection{SPH with integral-based derivatives}
\label{sec:SPH_with_integral_gradients}

One can find SPH equations based on the more accurate 
integral approximation derivatives by using the formal replacement 
Eq.~(\ref{eq:nabla_W_to_G}). By replacing the kernel gradients, one finds
\bea
\frac{d\vec{v}_a}{dt} &=& - \frac{1}{m_a} \sum_{b} X_a X_b \left\{  \frac{P_a}{\kappa_a^2} \vec{G}_a  
+ \frac{P_b}{\kappa_b^2} \vec{G}_b  \right\} \\
\frac{du_a}{dt}       &=& \frac{P_a X_a}{m_a \kappa_a^2} \sum_b X_b \vec{v}_{ab} \cdot \vec{G}_a,
\eea
with
\be
\left(\vec{G}_{a}\right)^k=  C^{kd}(\vec{r}_a,h_a) (\vec{r}_b - \vec{r}_a)^d W_{ab}(h_a)
\ee
and 
\be
\left(\vec{G}_{b}\right)^k=  C^{kd}(\vec{r}_b,h_b) (\vec{r}_b - \vec{r}_a)^d W_{ab}(h_b),
\ee
where a summation over $d$ is implied and the density can be calculated via Eq.~(\ref{eq:gen_vol_N}).
For the conventional choice $X= m$ this reproduces (up to the grad-h terms) the original equation set 
derived in \cite{garcia_senz12} from a Lagrangian. The experiments in \cite{garcia_senz12,cabezon12a,rosswog15b} 
clearly show that the use of this gradient prescription substantially improves the accuracy of SPH.

\subsubsection{Treatment of  shocks}
\label{sec:Newtonian_shocks}

The equations of gas dynamics allow for discontinuities to emerge even
from perfectly smooth initial conditions \citep{landau59}. At discontinuities,
the differential form of the fluid equations is no longer valid, and
their integral form needs to be used, which, at shocks, translates into
the Rankine--Hugoniot conditions, which relate the upstream and downstream 
properties of the flow. They show in particular, that the entropy increases
in shocks, i.e., that dissipation occurs inside the shock front. 
For this reason the inviscid SPH equations need to be augmented
by further measures that become active near shocks.

Another line of reasoning that suggests using artificial viscosity is the following.
One can think of the SPH particles as (macroscopic) fluid elements that follow 
streamlines, so in this sense SPH is a method of characteristics. Problems can 
occur when particle trajectories cross since in such a case fluid properties
at the crossing point can become multi-valued. The term linear in the velocities
in Eq.~(\ref{eq:basic:PI_AV}) was originally also introduced as a measure to avoid 
particle interpenetration where it should not occur \citep{hernquist89} and to 
damp particle noise. \cite{read12} 
designed special dissipation  switches to avoid particle properties becoming 
multi-valued at trajectory crossings.

\subsubsection*{Common form of the dissipative equations}

There are a number of successful Riemann solver implementations (see for example
\cite{inutsuka02,cha03,cha10,murante11,iwasaki11,puri14}), but the most widespread 
approach is the use ``artificial'' dissipation terms in SPH. Such terms are not necessarily 
meant to mimic physical  dissipation, their only purpose is to ensure that the 
Rankine--Hugoniot relations are fulfilled, though on a numerically resolvable 
rather than a microscopic scale.
The most commonly chosen approach is to add the following terms
\be
\left( \frac{d\vec{v}_a}{dt} \right)_{\rm diss} = - \sum_b m_b  \Pi_{ab} 
\nabla_a W_{ab} \quad {\rm and} \quad
\left( \frac{du_a}{dt}\right)_{\rm diss}=  \frac{1}{2} \sum_b m_b \Pi_{ab} \vec{v}_{ab} \cdot \nabla_a W_{ab}
\label{eq:basic_AV}
\ee
to the momentum and energy equation,
where $\Pi_{ab}$ is the artificial viscosity tensor. As long as $\Pi_{ab}$ is symmetric
in $a$ and $b$, the conservation of energy, linear and angular momentum is ensured
by the form of the equation and the anti-symmetry of $\nabla_a W_{ab}$ with respect to
the exchange of $a$ and $b$.
There is some freedom in choosing $\Pi_{ab}$, but the most commonly
used form is \citep{monaghan83}
\be
\Pi_{ab} =  \left\{\begin{array}{cl}
           \frac{- \alpha \bar{c}_{ab} \mu_{ab} + \beta
             \mu_{ab}^2}{\bar{\rho}_{ab}} \quad{\rm for} \; \;
           \vec{r}_{ab} \cdot \vec{v}_{ab} < 0\\  
           0 \quad \quad \quad \quad  {\rm otherwise}\\
           \end{array}\right. ,{\rm where} \quad
         \mu_{ab}= \frac{\bar{h}_{ab} \; \vec r_{ab}\cdot \vec v_{ab}}{r_{ab}^2 + \epsilon \bar{h}_{ab}^2},
\label{eq:basic:PI_AV}
\ee
where the barred quantities are arithmetic averages of particle $a$ and $b$ and $\vec{v}_{ab}= \vec{v}_a - \vec{v}_b$.
This is a SPH-translation of a bulk and a von-Neumann--Richtmyer viscosity,\epubtkFootnote{A 
step-by-step motivation of the term can be found in Section~2.7 of \cite{rosswog09b}.}
with viscous pressures of $P_B \propto - \rho l \nabla \cdot \vec{v}$
and $P_{\rm NR} \propto  \rho l^2 (\nabla \cdot \vec{v})^2$, respectively, where  $l$ 
is the local resolution length. This form has been chosen because it is Galilean
invariant, vanishes for rigid rotation and ensures exact conservation. The parameters
$\alpha$ and $\beta$ set the strength of the dissipative terms and are usually chosen
so that good results are obtained in standard benchmark tests.

Comparison with Riemann solver approaches \citep{monaghan97} suggests an alternative form 
that involves signal velocities and jumps in variables across 
characteristics. The main idea of these ``discontinuity capturing terms'' is that 
for any conserved scalar variable $A$ with $\sum_a m_a dA_a/dt=0$ a dissipative term 
of the form
\be
\left( \frac{dA_a}{dt}\right)_{\rm diss}= \sum_b m_b \frac{\alpha_{A,b} v_{\rm sig}}{\bar{\rho}_{ab}} 
(A_a-A_b)\hat{e}_{ab} \cdot \nabla_a W_{ab}
\ee
should be added, where the parameter $\alpha_{A,b}$ determines the exact amount of dissipation and $v_{\rm sig}$ 
is a signal velocity between particle $a$ and $b$. Applied to the velocity and the thermokinetic energy
$e= u + v^2/2$, this yields
\bea
\left( \frac{d \vec{v}_a}{dt} \right)_{\rm diss}&=& \sum_b m_b \frac{\alpha v_{\rm sig}(\vec{v}_a-\vec{v}_b)\cdot 
\hat{e}_{ab}}{\bar{\rho}_{ab}} \nabla_a W_{ab}\label{basic:eq:v_diss}\\
\left( \frac{d \hat{e}_a}{dt} \right)_{\rm diss}&=& \sum_b m_b \frac{e^\ast_a - e^\ast_b}{\bar{\rho}_{ab}} 
\hat{e}_{ab}\cdot \nabla_a W_{ab},\label{basic:eq:e_diss}
\eea
where, following \cite{price08a}, the energy $e^\ast$ includes velocity components along the line 
joining particles $a$ and $b$, $e^\ast_a= \frac{1}{2} \alpha v_{\rm sig} (\vec{v}_a\cdot\hat{e}_{ab})^2 
+ \alpha_u v_{\rm sig}^u u_a$. Note that in this equation different signal velocities and dissipation
parameters can be used for the velocities and the thermal energy terms. Using
$du_a/dt= d\hat{e}_a/dt -\vec{v}_a \cdot d\vec{v}_a/dt$, this translates into 
\be
\left( \frac{du_a}{dt} \right)_{\rm diss}= - \sum_b \frac{m_b}{\bar{\rho}_{ab}} \left[ \alpha v_{\rm sig} \;  
\frac{1}{2} (\vec{v}_{ab}\cdot\hat{e}_{ab})^2 + \alpha_u v^u_{\rm sig} (u_a-u_b) \right] \hat{e}_{ab} \cdot \nabla_a W_{ab}
\ee
for the thermal energy equation.
The first term in this equation bears similarities with the ``standard'' artificial viscosity prescription, see
Eq.~(\ref{eq:basic:PI_AV}).  The second one expresses the exchange of thermal energy between particles
and therefore represents an artificial thermal conductivity which smoothes discontinuities in the specific energy. Such 
artificial conductivity had been suggested earlier to cure the so-called ``wall heating problem'' \citep{noh87}. 
Tests have shown that artificial conductivity substantially improves SPH's performance
in simulating Sedov blast waves \citep{rosswog07c} and in the treatment of Kelvin--Helmholtz instabilities \citep{price08a}.
Note that the general strategy suggests to use dissipative terms also for the continuity equation so that particles
can exchange mass in a conservative way. This has, so far, been rarely applied, but \cite{read12} find good results
with this strategy in  multi-particle-mass simulations.

\subsubsection*{Time dependent dissipation parameters}

The choice of the dissipation parameters is crucial and simply using fixed parameters 
applies dissipation also where it is not needed and actually unwanted.\epubtkFootnote{For a discussion
of artificial dissipation effects in an astrophysical context (gravitationally unstable protoplanetary disks), 
see, for example, \cite{mayer04}}
 As a result, one simulates some type of viscous fluid rather than the intended inviscid medium. Therefore,
\cite{morris97} suggested an approach with time-dependent parameters 
for each particle and with triggers that indicate where dissipation is needed. Using
$\beta_a= 2\alpha_a$ in Eq.~(\ref{eq:basic:PI_AV}) they evolved $\alpha_a$ according 
to 
\be
\frac{d\alpha_a}{dt}= \mathcal{A}_a^+ - \mathcal{A}_a^- \quad {\rm with} \quad 
\mathcal{A}_a^+= \max \left(-(\nabla\cdot\vec{v})_a,0\right) \quad {\rm and} \quad
\mathcal{A}_a^- = \frac{\alpha_a(t) - \alpha_{\min}}{\tau_a},
\label{eq:alpha_steering_MM}
\ee
where $\alpha_{\min}$ represents a minimum, ``floor'' value 
for the viscosity parameter and $\tau_a\sim h_a/c_{{\rm s},a}$ is the individual decay time scale.
This approach (or slight modifications of it) has been shown to substantially 
reduce unwanted effects in practical simulations \citep{rosswog00,dolag05,wetzstein09}
but, as already realized in the original publication, triggering on the velocity
divergence also raises the viscosity in a slow compression with $(\nabla\cdot\vec{v})=$ const
where it is actually not needed.

\subsubsection*{New shock triggers}

More recently, some improvements of the basic Morris and Monaghan idea were suggested by \cite{cullen10}.
First, the authors argued that the floor value can be safely set to zero, provided 
that $\alpha$ can grow fast enough. To ensure the latter, the current value of $\alpha_a$ 
is compared to values indicated by triggers and, if necessary, $\alpha_a$ is increased 
instantaneously rather than by solving Eq.~(\ref{eq:alpha_steering_MM}). Second, 
instead of using $(\nabla\cdot\vec{v})_a$ one triggers on its time derivative 
to find the locally desired dissipation parameter:
\be
A_a= \xi_a \; \max \left[-\frac{d (\nabla\cdot\vec{v})_a}{dt},0\right] \quad {\rm and} \quad 
\alpha_{a, \rm des}= \alpha_{\max} \frac{A_a}{A_a + c^2_a/h_a^2},
\label{eq:trigger_CD}
\ee
where $c_a$ is a signal velocity, $\xi_a$ a limiter, see below, and $\alpha_{\max}$ a maximally
admissible dissipation value.
If $\alpha_{a, \rm des} > \alpha_{\rm a}$ then $\alpha_{\rm a}$ is raised immediately to 
$\alpha_{a, \rm des}$, otherwise it decays according to the $\mathcal{A}_a^-$-term 
in Eq.~(\ref{eq:alpha_steering_MM}). Apart from overall substantially reducing dissipation, 
this approach possesses the additional virtue that $\alpha$ peaks $\sim 2$ smoothing lengths ahead 
of a shock front and decays immediately after. Tests show that this produces much less 
unwanted dissipation than previous approaches.

\cite{read12} suggest a similar strategy, but trigger on the \emph{spatial} change of the compression
\be
A_a= \xi_a \; h_a^2 |\nabla (\nabla \cdot \vec{v})|_a \quad {\rm and} \quad 
\alpha_{a,\rm des}= \alpha_{\max} \frac{A_a}{A_a + h_a |\nabla \cdot \vec{v}|_a + 0.05 \; c_a},
\label{eq:trigger_RH}
\ee 
where $\nabla \cdot \vec{v} < 0$ and $\alpha_{\rm des}= 0$ otherwise. 
The major idea is to detect convergence \emph{before} it actually occurs. Particular care is taken
to ensure that all fluid properties remain single valued as particles approach each other and
higher-order gradient estimators are used. With this approach they 
find good results with only very little numerical noise.

\subsubsection*{Noise triggers}

The time scale $\tau_a$ on which dissipation is allowed to decay is usually chosen in
a trial-and-error approach. If chosen too large, the dissipation may be higher than desired,
if decaying too quickly, velocity noise may build up again after a shock has passed.
Therefore, a safer strategy is allow for a fast decay, but devise additional triggers that 
switch on if noise is detected. The main idea for identifying noise  is to monitor 
sign fluctuations of $\nabla \cdot \vec{v}$ in the vicinity of a particle. A particle is 
considered ``noisy'' if some of its neighbours are being compressed while others expand.\\
For example, the ratio 
\be
\frac{S_{1,a}}{S_{2,a}}\equiv \frac{\sum_b (\divv)_b}{\sum_b |\divv|_b}
\label{eq:noise_trigg}
\ee
can be used to construct a noise trigger. It deviates from $\pm 1$  in a noisy region since contributions of 
different sign are added up
in $S_{1,a}$ and therefore such deviations can be used as a noise indicator:
\be
\mathcal{N}_a^{(1)}= \left| \frac{\tilde{S}_{1,a}}{S_{2,a}} - 1\right|,
\label{eq:N_trigg_1}
\ee
where the quantity
\be
\tilde{S}_{1,a}= \left\{\begin{array}{ll}  -S_{1,a} & {\rm if} \; (\nabla \cdot \vec{v})_{a}\
 < 0 \\
         \quad S_{1,a} & {\rm else}\end{array}\right. .
\ee
If all particles in the neighborhood are either compressed or expanding, $\mathcal{N}_a^{(1)}$ 
vanishes. This trigger only reacts on sign changes, but does not account for the strengths
of the (de-)compressions. A second noise trigger that accounts for this uses
\be
\mathcal{S}^+_a =   \frac{1}{ N^+} \; \sum_{b, \divv_b>0}^{N^+}  \divv_b \quad {\rm and} \quad
\mathcal{S}^-_a = - \frac{1}{N^-} \; \sum_{b, \divv_b<0}^{N^-}  \divv_b,
\ee
where $N^+/N^-$ is number of neighbor particles with positive/negative $\divv$. 
The trigger then reads
\be
\mathcal{N}_a^{(2)}= \sqrt{\mathcal{S}^+_a \mathcal{S}^-_a} .
\label{eq:N_trigg_2}
\ee
If there are sign fluctuations in $\nabla \cdot \vec{v}$, but they are small
compared to $c_s/h$, the product is very small, if we have a uniform expansion or compression
one of the factors will be zero. So only for sign changes {\em and} significantly large compressions/expansions
will the product have a substantial value. Like in Eq.~(\ref{eq:trigger_CD}) the noise triggers $\mathcal{N}_a^{(1)}$ 
and $\mathcal{N}_a^{(2)}$ can be compared against reference values to steer the dissipation parameters.
For more details on noise triggers see \cite{rosswog15b}.\\

Note that in all these triggers \citep{cullen10,read12,rosswog15b} the gradients can 
straight forwardly be calculated from accurate expressions such as Eq.~(\ref{eq:full_IA_gradient})
or (\ref{eq:lin_exact_gradient}) rather than from kernel gradients.

\subsubsection*{Limiters}

Complementary to these triggers one can also try to actively suppress dissipation
where it is unwanted (with ideally working triggers this should, of course, not be necessary). 
For example, \cite{balsara95} suggested to distinguish between shock and shear motion based 
on the ratio
\be
\xi_a^{\rm B}= \frac{|\nabla \cdot \vec{v}|_a}{|\nabla \cdot \vec{v}|_a
+ |\nabla \times \vec{v}|_a + 0.0001 c_{s,a}/h_a}.
\label{eq:Balsara}
\ee
Dissipation is suppressed, $\xi_{a}^{\rm B} \rightarrow 0$, where $|\nabla \times \vec{v}|_a \gg |\nabla \cdot \vec{v}|_a$, 
whereas $\xi_{a}^{\rm B}$ tends to unity in the opposite limit. 
If symmetrized limiters are applied, $\Pi_{ab}'= \Pi_{ab} \; \bar{\xi}_{ab}$, exact conservation is ensured.
The Balsara limiter has been found very useful in many applications \citep{steinmetz96,navarro97,rosswog00}, but 
it can be challenged if shocks occur in a shearing environment like an accretion disk \citep{owen04}. Part of this
challenge comes from the finite accuracy of the standard SPH-derivatives. It is, however,
straight forward \citep{cullen10,read12,rosswog15b} to use more accurate derivatives such as,
for example,  Eq.~(\ref{eq:lin_exact_gradient}), in the limiters. Suppression of dissipation in shear flows can
then be obtained by simply replacing the SPH-gradient operators in Eq.~(\ref{eq:Balsara}) by more accurate
expressions, or by simply adding terms proportional to (accurate estimates for) $|\nabla \times \vec{v}|$ 
in the denominators of Eqs.~(\ref{eq:trigger_CD}) and (\ref{eq:trigger_RH}) (instead of multiplying $\alpha_{\rm des}$
with a limiter). Another alternative is the limiter proposed in \cite{cullen10} that also makes use 
of more accurate gradient estimates:
\be
\xi_{a}^{\rm C}= \frac{|2(1 - R_a)^4 (\nabla \cdot \vec{v})_a|^2}{|2(1 - R_a)^4 
           (\nabla \cdot \vec{v})_a|^2 + tr({\bf S}_a \cdot {\bf S}_a^t)},
\ee
where, in our notation and for generalized volume elements,
\be
R_a= \frac{V_a}{X_a} \sum_b {\rm sign}(\divv)_b X_b W_{ab}.
\ee
Note that $R_a$ is simply the ratio of a density summation where each term is weighted by the sign of $\divv$
and the normal density, Eq.~(\ref{eq:density}). Therefore, near a shock $R_a \rightarrow -1$. The matrix
${\bf S}$ is the traceless symmetric  part of the velocity gradient matrix $(\p_i v_j)$ and a measure for
the local shear. Similar to the Balsara factor, $\xi^{\rm C}$ approaches unity if compression clearly dominates over
shear and it vanishes in the opposite limit.

\subsection{Special-relativistic SPH}
\label{sec:SR_SPH}

The special-relativistic SPH equations can --like the Newtonian ones-- be elegantly derived from
a variational principle \citep{monaghan01,rosswog09b,rosswog10a,rosswog10b}. We discuss here
a formulation \citep{rosswog15b} that uses generalized volume elements, see Eq.~(\ref{eq:gen_vol_element}).
It is assumed that space-time is flat, that the metric, $\eta_{\mu \nu}$, has the signature 
(-,+,+,+) and units in which the speed of light is equal to unity, $c=1$, are adopted. We use
the Einstein summation convention and reserve 
Greek letters for space-time indices from 0\dots3 with 0 being the temporal component, while $i$ 
and $j$ refer to spatial components and SPH particles are labeled by $a,b$ and $k$.

The Lagrangian of a special-relativistic perfect fluid can then be written as \citep{fock64} 
\be
L= - \int T^{\mu \nu} U_\mu U_\nu dV,
\ee
where the energy-momentum tensor of an ideal fluid reads
\be
T^{\mu \nu}= (e + P) U^{\mu} U^{\nu} + P \eta^{\mu \nu}
\ee
with $e$ being the energy density, $P$ the pressure and $U^\lambda$ the four-velocity of a fluid element.
One can write the energy density as a sum of a contribution from rest mass density and one from the internal energy 
\be
e= \rho_{\rm rest} c^2 + u \rho_{\rm rest}= n m_0 c^2 (1 + u/c^2),
\label{eq:energy_density}
\ee
where  the speed of light was, for clarity, shown explicitly. The baryon number density $n$ is measured in the 
local fluid rest frame and the average baryon mass is denoted by $m_0$. With the conventions that 
all energies are written in units of $m_0c^2$ and $c=1$  we can use the normalization
of the four-velocity, $U_\mu U^\mu= - 1$, to simplify the Lagrangian to
\be
L= - \int n(1+u) dV.
\ee
The calculation will be performed in an a priori chosen ``computing frame'' (CF) and 
--due to length contraction-- the baryon number density measured in this frame, $N$, is 
increased by a Lorentz factor $\gamma$ with respect to the local fluid rest frame
\be 
N= \gamma n \label{eq:N_vs_n}.
\ee
Therefore, the Lagrangian can be written as
\be
L= - \int dV N \left(\frac{1+u}{\gamma}\right)
\quad {\rm or} \quad
L\simeq  - \sum_b V_b N_b \frac{1+u_b}{\gamma_b}= - \sum_b \nu_b \frac{1+u_b}{\gamma_b},\label{eq:SR_Lagrangian}
\ee 
where the baryon number carried by particle $b$, $\nu_b$, has been introduced.
If volume elements of the form Eq.~(\ref{eq:gen_vol_element}) are introduced, one can calculate 
CF baryon number densities as in the Newtonian case (see Eq.~(\ref{eq:density}))
\be
N_b= \frac{\nu_b}{V_b}= \frac{\nu_b}{X_b} \sum_k X_k W_{bk}(h_b),
\label{eq:CF_dens}
\ee
which reduces for the choice $X=\nu$ to the equivalent of the standard SPH sum, Eq.~(\ref{eq:dens_sum}), but
with $\rho/m$ being replaced by $N/\nu$.
To obtain the equations of motion, $\nabla_a N_b$ and $d N_a/dt$ are needed, which follow 
from Eq.~(\ref{eq:gen_vol_element}) as
\be
\nabla_a N_b= \frac{\nu_b}{X_b \tilde{\Omega}_b} \sum_k X_k \nabla_a W_{bk}(h_b)
\quad
{\rm and}
\quad
\frac{d N_a}{dt}  =  \frac{\nu_a}{X_a \tilde{\Omega}_a} \sum_k X_k \vec{v}_{bk} \cdot \nabla_b 
W_{bk}(h_b)\label{eq:dNdt_SR}
\ee
and which contain the generalized ``grad-h'' terms 
\be
\tilde{\Omega}_b= 1 - \frac{\nu_b}{X_b} \frac{\p h_b}{\p N_b} \sum_k X_k \frac{\p W_{kb}(h_b)}{\p h_b}.
\ee
It turns out to be beneficial to use the canonical momentum per baryon
\be
\vec{S}_a \equiv \frac{1}{\nu_a}\frac{\p L}{\p \vec{v}_a}=  
\gamma_a \vec{v}_a \left ( 1+u_a + \frac{P_a}{n_a}\right)
\label{eq:S_a}
\ee
as a numerical variable. Its evolution equation follows directly from the Euler--Lagrange equations as \citep{rosswog15b}
\be
\frac{d \vec{S}_a}{dt}
= -  \frac{1}{\nu_a}  \sum_{b} \left\{ \frac{P_a V_a^2}{\tilde{\Omega}_a} \frac{X_b}{X_a} \nabla_a W_{ab}(h_a) +
                                                                    \frac{P_b V_b^2}{\tilde{\Omega}_b} \frac{X_a}{X_b} \nabla_a W_{ab}(h_b) \right\}.
\label{eq:gen_mom_SR}
\ee
For the choice $V_k= \nu_k/N_k$, this reduces to the momentum equation given in \cite{rosswog10b}.

The canonical energy 
\be
E\equiv \sum_a \frac{\p L}{\p \vec{v}_a} \cdot \vec{v}_a - L = \sum_a \nu_a \epsilon_a, 
\ee
suggests to use
\be
\epsilon_a = \vec{v}_a \cdot \vec{S}_a + \frac{1+u_a}{\gamma_a} = \gamma_a \left(1 + u_a + \frac{P_a}{n_a} \right) - \frac{P_a}{N_a}
= \gamma_a \enth_a - \frac{P_a}{N_a},
\label{eq:en_a}
\ee
as numerical energy variable. Here the specific, relativistic enthalpy was abbreviated as
\be
\enth_a= 1 + u_a + \frac{P_a}{n_a}.
\label{eq:enthalpy}
\ee
The subsequent derivation is identical to the one in \cite{rosswog09b} up to their Eq.~(165),
\be
\frac{d \epsilon_a}{dt}= \vec{v}_a \cdot \frac{d \vec{S}_a}{dt} + \frac{P_a}{N_a^2} \frac{d N_a}{dt},
\ee
which, upon using Eqs.~(\ref{eq:dNdt_SR}) and (\ref{eq:gen_mom_SR}), yields the 
special-relativistic energy equation
\be
\frac{d \epsilon_a}{dt}= - \frac{1}{\nu_a} \sum_b \left\{ \frac{P_a V_a^2}{\tilde{\Omega}_a} \frac{X_b}{X_a} \vec{v}_b \cdot \nabla_a W_{ab}(h_a)
                                                                                            + \frac{P_b V_b^2}{\tilde{\Omega}_b} \frac{X_a}{X_b} \vec{v}_a \cdot \nabla_a W_{ab}(h_b)\right\}.
\label{eq:gen_ener_SR}
\ee
Again, for $V_k= \nu_k/N_k$, this reduces to the energy equation given in \cite{rosswog10b}. Of course,
the set of equations needs to be closed by an equation of state which in the simplest case can be a polytrope.

Note that by choosing the canonical energy and momentum as numerical variables,
one avoids complications such as time derivatives of Lorentz factors, that have plagued earlier SPH formulations \citep{laguna93a}.
The price one has to pay is that the physical variables (such as $u$ and $\vec{v}$) need to be recovered at every time
step from $N$, $\epsilon$ and $\vec{S}$ by solving a non-linear equation. For this task approaches very similar to what is used
in Eulerian relativistic hydrodynamics can be applied \citep{chow97,rosswog10b}.

As in the non-relativistic case, these equations need to be augmented by extra measures to deal with shocks. The 
dissipative terms  \citep{chow97}
\be
\left(\frac{d\vec{S}_a}{dt}\right)_{\rm diss}= - \sum_b \nu_b \Pi_{ab} 
\overline{\nabla_a W_{ab}} \;\; {\rm with} \;\; \Pi_{ab}= - 
\frac{K v_{\rm sig}}{\bar{N}_{ab}} (\vec{S}_a^\ast-\vec{S}_b^\ast) \cdot\hat{e}_{ab}
\label{eq:diss_mom}
\ee
\be
\left(\frac{d\epsilon_a}{dt}\right)_{\rm diss}=  - \sum_b \nu_b \vec{\Omega}_{ab} \cdot
\overline{\nabla_a W_{ab}} \;\; {\rm with} \;\; \vec{\Omega}_{ab} = -
\frac{K v_{\rm sig}}{\bar{N}_{ab}} (\epsilon_a^\ast-\epsilon_b^\ast)\hat{e}_{ab}
\label{eq:diss_en}
\ee
with symmetrized kernel gradients
\be
\overline{\nabla_a W_{ab}} = \frac{1}{2}\left[\nabla_a W_{ab}(h_a) +  \nabla_a W_{ab}(h_b) \right],
\ee
\be
\gamma_k^\ast= \frac{1}{\sqrt{1-(\vec{v}_k\cdot \hat{e}_{ab})^2}},
\vec{S}_k^\ast= \gamma^\ast_k \left(1+u_k+\frac{P_k}{n_k}\right) \vec{v}_k
\quad 
{\rm and} 
\quad
\vec{\epsilon}_k^\ast= \gamma^\ast_k \left(1+u_k+\frac{P_k}{n_k}\right) - \frac{P_k}{N_k},
\ee
where the asterisk denotes the projection to the line connecting two particles,
give good results, even in very challenging shock tests \citep{rosswog10b}.
A good choice for $v_{\rm sig}$ is \citep{rosswog10b} 
\be
v_{\rm sig,ab}= \max (\alpha_a,\alpha_b),\label{eq:vsig}
\ee
where
\be
\alpha_k^{\pm}= \max (0,\pm \lambda^\pm_k)
\ee
with $\lambda^\pm_k$ being the extreme local eigenvalues of the Euler equations, see e.g., \cite{marti03},
\be
\lambda^\pm_k= \frac{v_\parallel(1-c_{\rm s}^2) \pm c_{\rm s} \sqrt{(1-v^2)(1-v_\parallel^2 - 
v_\perp^2 c_{\rm s}^2)}}{1-v^2 c_{\rm s}^2}
\ee
and $c_{{\rm s},k}$ is the relativistic sound velocity of particle $k$,   
$c_{s,k}= \sqrt{\frac{(\Gamma-1) (\enth-1)}{\enth}}$. 
In 1 D, this simply reduces
to the usual velocity addition law, $\lambda^\pm_k= (v_k\pm c_{{\rm s},k})/(1\pm v_k c_{{\rm s},k}) $.
As in the non-relativistic case, the challenge lies in designing triggers that switch on where needed,
but not otherwise. The strategies that can be applied here are straight forward translations of those described in 
Section~\ref{sec:Newtonian_shocks}. We refer to \cite{rosswog15b} for more details on this topic.

\subsubsection*{Integral approximation-based special-relativistic SPH}

Like in the Newtonian case, alternative SPH equations can be obtained \citep{rosswog15b}
by replacing the kernel gradients by the functions $\vec{G}$, see Eq.~(\ref{eq:IA_gradient}):
\be
\frac{d\vec{S}_a}{dt}= - \frac{1}{\nu_a} \sum_b \left\{ P_a V_a^2 \frac{X_b}{X_a} \vec{G}_{a} +
                                         P_b V_b^2 \frac{X_a}{X_b} \vec{G}_{b} \right\}
\label{eq:momentum_eq_no_diss_integral}
\ee
and
\be
\frac{d \epsilon_a}{dt} = 
- \frac{1}{\nu_a} \sum_b \left\{ P_a V_a^2 \frac{X_b}{X_a} \vec{v}_b\cdot \vec{G}_{a} +
                                         P_b V_b^2 \frac{X_a}{X_b}  \vec{v}_a\cdot \vec{G}_{b} \right\},
\label{eq:ener_eq_no_diss_integral}
\ee 
where (sum over $d$)
\be
\left(\vec{G}_{a}\right)^k=  C^{kd}(\vec{r}_a,h_a) (\vec{r}_b - \vec{r}_a)^d W_{ab}(h_a)
\ee
and 
\be
\left(\vec{G}_{b}\right)^k=  C^{kd}(\vec{r}_b,h_b) (\vec{r}_b - \vec{r}_a)^d W_{ab}(h_b).
\ee
The density calculation remains unchanged from Eq.~(\ref{eq:CF_dens}).
The same dissipative terms as for the kernel-gradient-based approach can be used, but it is
important to replace $\overline{\nabla_a W_{ab}}$ by
\be
\overline{\vec{G}}_{ab}= \frac{1}{2}\left[\vec{G}_a +  \vec{G}_b \right],
\ee
since otherwise numerical instabilities can occur. This form has been extensively tested \citep{rosswog15b}
and shown to deliver significantly more accurate results than the kernel-gradient-based approach.

\subsection{General-relativistic SPH}
\label{sec:GR_SPH}

For binaries that contain a neutron or a black hole general-relativistic effects are important. 
The first relativistic SPH formulations were developed by \cite{kheyfets90}
and \cite{mann91,mann93}.
Shortly after, \cite{laguna93a} developed a 3D, general-relativistic SPH code that
was subsequently applied to the tidal disruption of stars by massive black holes \citep{laguna93b}.
Their SPH formulation is complicated by several issues: the continuity 
equation contains a gravitational source term that requires SPH kernels
for curved space-times. Moreover, owing to their choice of variables, the equations contain 
time derivatives of Lorentz factors that are treated by finite difference approximations and restrict
the ability to handle shocks to only moderate Lorentz factors. The Laguna et al. formulation has been 
extended by \cite{rantsiou08} and applied to neutron star black hole binaries, see 
Section~\ref{sec:appl_NSBH}. We focus here on SPH in a fixed background metric, approximate GR 
treatments are briefly discussed in Section~\ref{sec:appl_NSNS_NSBH}.

In smooth continuation of the Newtonian and special-relativistic approaches,
we focus here on the derivation from a Lagrangian.
To this end, we  assume that a prescribed metric $g_{\mu \nu}$ is known as a function of the coordinates
and that the perturbations that the fluid induces to the space-time geometry can be safely
neglected. Again, $c=1$ and signature ($-,+,+,+$) are assumed and the same index conventions
as in the special relativistic section are adopted. Contravariant spatial indices of a vector 
quantity $w$ at particle $a$ are denoted as $w^i_a$, while covariant ones will be written as 
$(w_i)_a$. The line element and proper time are given by $ds^2= g_{\mu \nu} \, dx^\mu \, dx^\nu$ and 
$d\tau^2= - ds^2$ and the proper time is related to a coordinate time $t$ by 
\be
\Theta d\tau = dt,
\ee
where a generalization of the Lorentz-factor 
\be
\Theta\equiv \frac{1}{\sqrt{-g_{\mu\nu} v^\mu v^\nu}} \quad {\rm with} \quad v^\alpha=\frac{dx^\alpha}{dt}
\ee
was introduced. This relates to the four-velocity $U^\nu$, normalized to $U^\mu U_\mu= -1$, by
\be
v^\mu= \frac{dx^\mu}{dt}= \frac{dx^\mu}{d\tau} \frac{d\tau}{dt}= \frac{U^\mu}{\Theta}= \frac{U^\mu}{U^0}.
\label{eq:v_mu}
\ee
The Lagrangian of an ideal relativistic fluid  can be written as \citep{fock64}
\be
L= - \int T^{\mu \nu} U_\mu U_\nu \sqrt{-g} dV,
\ee
where $g= {\rm det}(g_{\mu \nu})$ and $T^{\mu\nu}$ denotes the energy-momentum tensor of an ideal 
fluid without viscosity and conductivity
\be
T^{\mu \nu}= (e+P)U^\mu U^\nu + P g^{\mu \nu}
\ee
with the energy density $e$ given in Eq.~(\ref{eq:energy_density}). With these conventions 
the Lagrangian can be written, similar to the special-relativistic case, as
\be
L= - \int n(1+u)\sqrt{-g} dV\label{eq:Lagrangian_cont}.
\ee
As before, for practical simulations we give up general covariance and choose a particular
``computing frame''. 
To find a SPH discretization in terms of a suitable density variable one can express
local baryon number conservation, $(U^\mu n);_\mu= 0$, as \citep{siegler00a}
\be 
\p_\mu (\sqrt{-g} U^\mu n)= 0,
\ee
or, more explicitly, as
\be 
\p_t (N) + \p_i(N v^i)= 0, \label{eq:continuity_N}
\ee
where Eq.~(\ref{eq:v_mu}) was used and the computing frame baryon number density
\be
N= \sqrt{-g} \Theta n\label{eq:N_n}
\ee
was introduced.
The total conserved baryon number can then be expressed as a sum over fluid parcels
with volume $\Delta V_b$ located at $\vec{r}_b$, where each parcel carries a baryon number $\nu_b$
\be 
\mathcal{N}= \int N dV \simeq \sum_b N_b \Delta V_b = \sum_b \nu_b.\label{eq:parcel_volumes}
\ee
Eq.~(\ref{eq:continuity_N}) looks like the Newtonian continuity equation which suggests to use it
in the SPH discretization process of a quantity $f$:
\be
\tilde{f}(\vec{r}) \simeq \sum_b f_b \frac{\nu_b}{N_b} W(\vec{r}-\vec{r}_b,h),\label{eq:SPH_discretization}
\ee
where the subscript $b$ indicates that a quantity is evaluated at a position $\vec{r}_b$ 
and $W$ is the smoothing kernel. If all $\nu_b$ are kept constant in time, exact baryon 
number conservation is guaranteed and no continuity equation needs to be solved (this 
can be done, though, if desired).
If Eq.~(\ref{eq:SPH_discretization}) is applied to the baryon number density $N$ at the 
position of particle $a$, one finds
\be
N_a= N(\vec{r}_a)= \sum_b \nu_b W(\vec{r}_a-\vec{r}_b,h_a)\label{eq:N_r}.
\ee
Note that the locally smoothed quantities are evaluated with flat-space kernels
which assumes that the local space-time curvature radius is large in comparison 
to the local fluid resolution length. Such an approach is very convenient, but 
(more involved) alternatives to this approach exist \citep{kheyfets90,laguna93a}. 
It is usually desirable to adjust the smoothing length locally to fully exploit the natural adaptivity 
of a particle method. As before, one can adopt the smoothing length according to
$
h_a= \eta (\nu_a/N_a)^{1/3}.
$ 
Due to the mutual dependence an iteration is required at each time step to obtain 
consistent values for both density and smoothing length, similar to the Newtonian and
special-relativistic case.
Motivated by Eqs.~(\ref{eq:N_n}) and (\ref{eq:parcel_volumes}), one can re-write
the fluid Lagrangian, Eq.~(\ref{eq:Lagrangian_cont}),  in terms of our computing 
frame number density $N$,
\be
L= -\int \frac{1+u}{\Theta} N dV,
\ee
or, in  discretized form,
\be
L_{\rm SPH}= - \sum_b \nu_b \left( \frac{1+u}{\Theta} \right)_b.
\ee
This Lagrangian has the same form as in the special-relativistic case, see Eq.~(\ref{eq:SR_Lagrangian}),
with the Lorentz factor $\gamma$ being replaced with its generalized form $\Theta$.
Using the Euler--Lagrange equations and taking care of the velocity dependence of both 
the generalized Lorentz-factor $\Theta$ and the internal energy, 
$\frac{\p u_b}{\p v^i_a}= \frac{\p u_b}{\p n_b} \frac{\p n_b}{\p v^i_a}$, 
one finds the canonical momentum per baryon 
\be
(S_i)_a\equiv \frac{1}{\nu_a} \frac{\p L}{\p v^{i}_{a}}= \Theta_a \left(1+u_a+\frac{P_a}{n_a} \right) 
(g_{i\mu} v^\mu)_a= \left(1+u_a+\frac{P_a}{n_a} \right) (U_i)_a.
\ee
which generalizes Eq.~(\ref{eq:S_a}). The Euler--Lagrange equations \citep{rosswog10a} deliver
\bea
\frac{d (S_i)_a}{dt}
&=& - \sum_b \nu_b 
\left\{ 
\frac{P_a \sqrt{-g}_a}{\Omega_a N_a^2} \frac{\p W_{ab}(h_a)}{\p x^i_a} +
\frac{P_b \sqrt{-g}_b}{\Omega_b N_b^2} \frac{\p W_{ab}(h_b)}{\p x^i_a}
\right\}
+ \left(\frac{\sqrt{-g}}{2 N} T^{\mu\nu} 
\frac{\p g_{\mu\nu}}{\p x^i}\right)_a,
\label{eq:GR_momentum_evolution}
\eea
again very similar to the special-relativistic case.
Like before,  the canonical energy
\be
E\equiv \sum_a \frac{\p L}{\p v^i_a} v^i_a - L= \sum_a \nu_a \left( v^i_a (S_i)_a + \frac{1+u_a}{\Theta_a}\right)
\ee
suggests to use
\be
e_a \equiv v^i_a (S_i)_a + \frac{1+u_a}{\Theta_a}
\ee
as numerical variable whose evolution equation follows from straight forward 
differentiation \citep{rosswog10a} as
\bea
\frac{de_a}{dt}&=& - \sum_b \nu_b \left\{ 
\frac{P_a \sqrt{-g}_a v^i_b}{\Omega_a N_a^2} \; \frac{\p W_{ab}(h_a)}{\p x^i_a} + 
\frac{P_b \sqrt{-g}_b v^i_a}{\Omega_b N_b^2} \; \frac{\p W_{ab}(h_b)}{\p x^i_a}  
\right\}
- \left( \frac{\sqrt{-g}}{2 N} T^{\mu\nu} \p_t g_{\mu\nu}\right)_a.
\label{eq:GR_energy_evolution}
\eea
Together with an equation of state, the equations (\ref{eq:N_r}), (\ref{eq:GR_momentum_evolution}) and
(\ref{eq:GR_energy_evolution}) represent a complete and self-consistently derived set of SPH equations.
The gravitational terms are identical to those of \cite{siegler00a}
and \cite{monaghan01}, but the hydrodynamic terms
differ in both the particle symmetrization and the presence of the grad-h terms
\be
\Omega_b= 1- \frac{\p h_b}{\p N_b} \sum_k \nu_k \frac{\p W_{bk}(h_b)}{\p h_b}.
\ee
Note that
the only choices in our above derivation were the $h$-dependence in Eq.~(\ref{eq:N_r}) and how to adapt 
the smoothing length. The subsequent calculation contained no arbitrariness concerning the symmetry in
particle indices, everything followed stringently from the first law of thermodynamics and the 
Euler--Lagrange equations. Another important point to note is that the derived energy equation, 
Eq.~(\ref{eq:GR_energy_evolution}), does not contain destabilizing  time derivatives of Lorentz 
factors \citep{norman86} on the RHS -- in contrast to  earlier SPH formulations \citep{laguna93a}.
The above SPH equations can also be recast in the language of the 3+1 formalism, this can be 
found in \cite{tejeda12a}.

\subsubsection{Limiting cases}

It is a straight forward exercise to show that, in the limit of vanishing hydrodynamic terms (i.e.,
$u$ and $P$ = 0), the evolution equations (\ref{eq:GR_momentum_evolution}) and (\ref{eq:GR_energy_evolution})
reduce to the equation of geodesic motion \citep{tejeda12a}.
If, in the opposite limit, we are neglecting the gravitational terms in 
Eqs.~(\ref{eq:GR_momentum_evolution}) and (\ref{eq:GR_energy_evolution}) and assume
flat space-time with Cartesian coordinates one has
$\sqrt{-g} \rightarrow 1$ and $\Theta \rightarrow \gamma$, and Eq.~(\ref{eq:N_n}) becomes 
$N= \gamma n$. The momentum and energy equations reduce in this limit  to 
the previous equations (\ref{eq:gen_mom_SR}) and (\ref{eq:gen_ener_SR}) (for $X= \nu$).

\subsection{Frequently used SPH codes}

A number of SPH codes are regularly used throughout various areas of astrophysics and it is beyond the scope
of this review to give a detailed account of each of them. In Table~\ref{tab:SPH_codes} we briefly summarize 
some of the basic features of Newtonian SPH codes that have been used in the context of compact 
object simulations and that will be referred to in the subsequent sections of this review. For a more detailed 
code description we refer to the original papers. SPH approaches that use GR or various approximations to it
are further discussed in Section~\ref{sec:appl_NSNS_NSBH}.

\begin{landscape}
\begin{table}
\caption{Frequently used Newtonian SPH codes}
\label{tab:SPH_codes}
\renewcommand{\arraystretch}{1.8}
{\small
\begin{tabular}{ >{\RaggedRight\noindent}p{3.8cm}lllllllll }
               reference        & name/group & SPH equations  & self-gravity  & AV  steering   & EOS, burning  & \\
\hline\\
             \cite{springel01a,springel05a}   & \textsc{Gadget}      & entropy equation, Eq.~(\ref{eq:momentum_gadget}) & Oct-tree & fixed parameters & polytrope\\
             \cite{starcrash}                          &  \textsc{StarCrash}  & ``vanilla ice''& FFT on grid & fixed parameters & polytrope\\
             \cite{rosswog02a}                     & Leicester   &  ``vanilla ice''  & binary tree \citep{benz90b} & time-dep., Balsara-switch & Shen \\
             \cite{rosswog08b}                     & Bremen     &  ``vanilla ice''   & binary tree \citep{benz90b} & time-dep., Balsara-switch & Helmholtz  + network         \\
             \cite{rosswog07c}                     &     \textsc{Magma}    &   hydro \& self-gravity                    & binary tree   & time-dep., Balsara-switch,    &  Shen &\\
                                                              &                     &   from Lagrangian                           &                     & conductivity     & \\
             \cite{guerrero04}                       &  Barcelona     &  ``vanilla ice'' & oct-tree & fixed parameters,&  ions, electrons, photons&\\
                                                               &                      &                        &               &  Balsara-switch   &   + network    &\\
             \cite{fryer06}                             & \textsc{SNSPH}          & ``vanilla ice''   & oct-tree & fixed parameters & polytrope,\\
                                                               &                     &                              &                                          &  & Helmholtz + network \\
             \cite{wadsley04}                        & \textsc{Gasoline}       & ``vanilla ice''   & K-D tree & fixed parameters,   & polytrope &  \\
                                                               &                      &                     &               & Balsara-switch \\
\end{tabular}}
``Shen'' EOS: \cite{shen98a,shen98b}, ``Helmholtz'' EOS: \cite{timmes99}
\end{table}
\end{landscape}

\subsection{Importance of initial conditions}
\label{sec:IC}

It is not always sufficiently appreciated  how important the initial conditions are for the accuracy of SPH simulations.
Unfortunately, it can become rather non-trivial to construct them. One obvious requirement is the regularity 
of  a particle distribution so that the quality
indicators $\mathcal{Q}_1 - \mathcal{Q}_4$, Eqs.~(\ref{eq:quality_int}) and (\ref{eq:gradient_quality}), 
are fulfilled with high accuracy. This suggests placing the particles initially on some type of lattice.
However, these lattices should ideally possess at least two more properties: a) they should not contain preferred directions
b) they should represent a \emph{stable} minimum energy configuration for the applied SPH formulation.
Condition a) is desirable since otherwise shocks propagating along a symmetry axis of a lattice will 
continuously ``collect'' particles in this direction and this can lead to ``lattice ringing effects''. Condition b)
is important since a regular lattice does by no means guarantee that the configuration actually 
represents an equilibrium for the SPH particles. As briefly outlined in Section~\ref{sec:SPH_from_variational_principle},
the momentum equation derived from a Lagrangian also contains a ``regularization force'' contribution that 
strives to achieve a good particle distribution. If the initial lattice does not represent such a configuration,
the  particles will start to move ``off the lattice'' and this
will introduce noise. In Figure~\ref{fig:noise} we show the result of a numerical experiment.
10K particles are placed either on a quadratic or hexagonal lattice in the domain $[-1,1] \times [-1,1]$ with particles
at the edges being fixed. If the configurations are allowed to evolve without dissipation (here we use the
common cubic spline kernel), the particles in the quadratic lattice case soon begin to move off the grid
and gain average velocities of $\langle v \rangle \approx 0.015 \; c_{\rm s}$, while the remain in the initial
configuration when initially placed on a hexagonal lattice ($\langle v \rangle \approx 10^{-4} \; c_{\rm s}$). 
This is discussed in more detail in \cite{rosswog15b}. From a theoretical point of view, the relation between 
kernel, particle configuration and stability/noise is to date  only incompletely understood and must be addressed
in careful future studies. 

\epubtkImage{}{%
  \begin{figure}[htb]
  \centerline{\includegraphics[angle=-90,width=\textwidth]{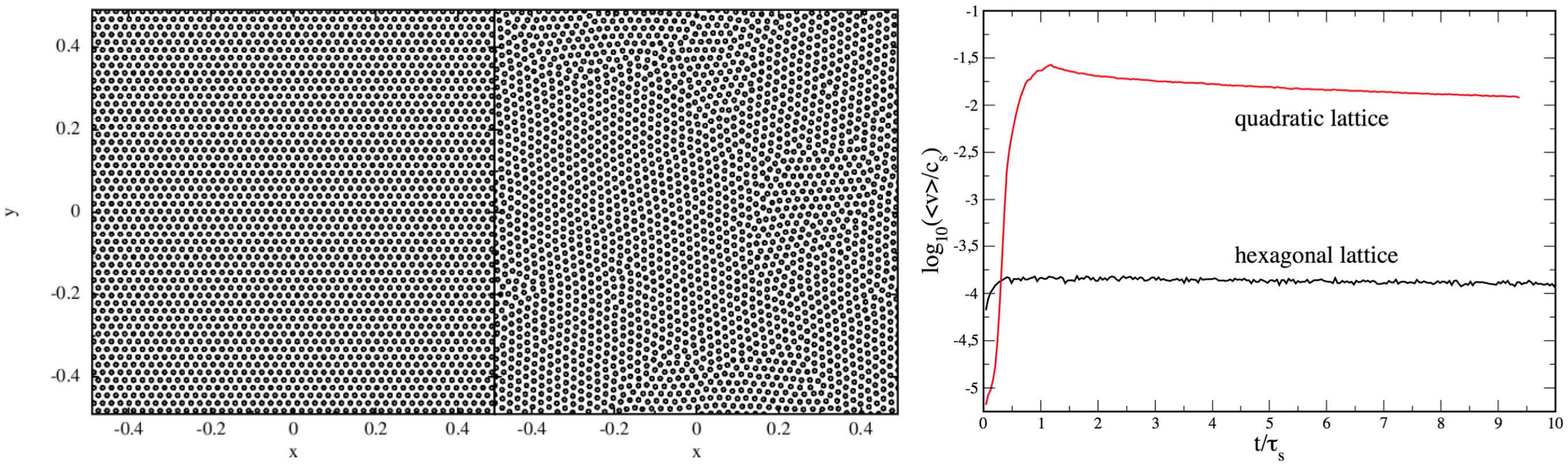}}

  \vspace*{-2.5cm}

    \caption{Numerical experiment to illustrate the importance of the initial particle lattice. In this experiment particles
               are placed on  either a quadratic or a hexagonal lattice and they are subsequently allowed to evolve (no dissipation,
               cubic spline kernel). In the quadratic lattice case, the particles begin to move around and rearrange themselves 
               (middle panel after 10 sound crossing times), while they stay on the original lattice for the hexagonal case 
               (left panel; same time).
               The logarithm of the average particle velocities for both cases (velocities in units of the sound speed, times in 
               sound crossing times) are shown in the right panel.}
   \label{fig:noise}
\end{figure}}

In experiments \citep{rosswog15b} different kernels show a very different noise
behavior and this seems to be uncorrelated with the accuracy properties of the kernels: kernels
that are rather poor density and gradient estimators may be excellent in producing very little noise (see for example
the QCM$_6$ kernel described in \cite{rosswog15b}). On the other hand, very accurate kernels (like W$_{\rm H,9}$) may be
still allow for a fair amount of noise in a dynamical simulation. The family of Wendland kernels \citep{wendland95}
has a number of interesting properties, among them the stability against particle pairing despite having
a vanishing central derivative \citep{dehnen12} and their reluctance to tolerate sub-resolution particle motion 
\citep{rosswog15b}, in other words noise. In those experiments where particle noise is relevant for the overall
accuracy, for example in the Gresho--Chan vortex test, see Section~\ref{sec:gresho}, the  Wendland kernel, Eq.~(\ref{eq:wend33}), 
performs substantially better than any other of the explored kernels.

A heuristic approach to construct good, low-noise initial conditions for the actually used kernel function is 
to apply a velocity-dependent damping term, $\vec{f}_{\rm damp} \propto - \vec{v}_{\rm damp,a}/\tau_a$, 
with a suitably chosen damping time scale $\tau_a$
to the momentum equation. This ``relaxation process'' can be applied until some convergence criterion is met (say, the kinetic
energy or some density oscillation amplitude has dropped below some threshold). 
This procedure becomes, however, difficult to apply for more complicated initial conditions
and it can become as time consuming as the subsequent production simulation. For an interesting 
recent suggestion on how to construct more general initial conditions see \cite{diehl12}.

\subsection{The performance of SPH}

Here we cannot give an exhaustive overview over the performance of SPH in general.
We do want to address, however, a number of issues that are of particular relevance in astrophysics.
Thereby we put particular emphasis on the new improvements of SPH that were discussed in the
previous sections.
Most astrophysical simulations, however, are still carried out with older methodologies. This is natural since, 
on the one hand, SPH is still a relatively young numerical method and improvements are 
constantly being suggested and, on the other hand, writing a well-tested and robust production
code is usually a rather laborious endeavor. Nevertheless, efforts should be taken to ensure
that latest developments find their way into production codes. In this sense the below shown
examples are also meant as a motivation for computational astrophysicists to keep their 
simulation tools up-to-date.

The accuracy of SPH with respect to commonly used techniques can be enhanced by using
\begin{itemize}
\item higher-order kernels, for example the $W_{\rm H,9}$ or the Wendland kernel $W_{3,3}$, see Section~\ref{sec:kernel_choice}
\item different volume elements, see Section~\ref{sec:volume_elements}
\item more accurate integral-based derivatives, see Section~\ref{sec:integral_gradients}
\item modern dissipation triggers, see Section~\ref{sec:Newtonian_shocks}.
\end{itemize}
In the examples shown here, we use a special-relativistic SPH code (``SPHINCS\_SR'', \cite{rosswog15b})
to demonstrate SPH's performance in a few astrophysically relevant examples. Below, we  
refer several times to  the ``$\mathcal{F}_1$ formulation'': it consists of baryon number density calculated via
Eq.~(\ref{eq:CF_dens}) with volume weight $X= P^{0.05}$, the integral approximation-based 
form of the SPH equations, see Eqs.~(\ref{eq:momentum_eq_no_diss_integral}) and 
(\ref{eq:ener_eq_no_diss_integral}), and shock trigger Eq.~(\ref{eq:trigger_CD}) 
and noise triggers, see \cite{rosswog15b} for more details.

\epubtkImage{}{%
  \begin{figure}[htb]
    \centerline{\includegraphics[width=0.6\textwidth,angle=-90]{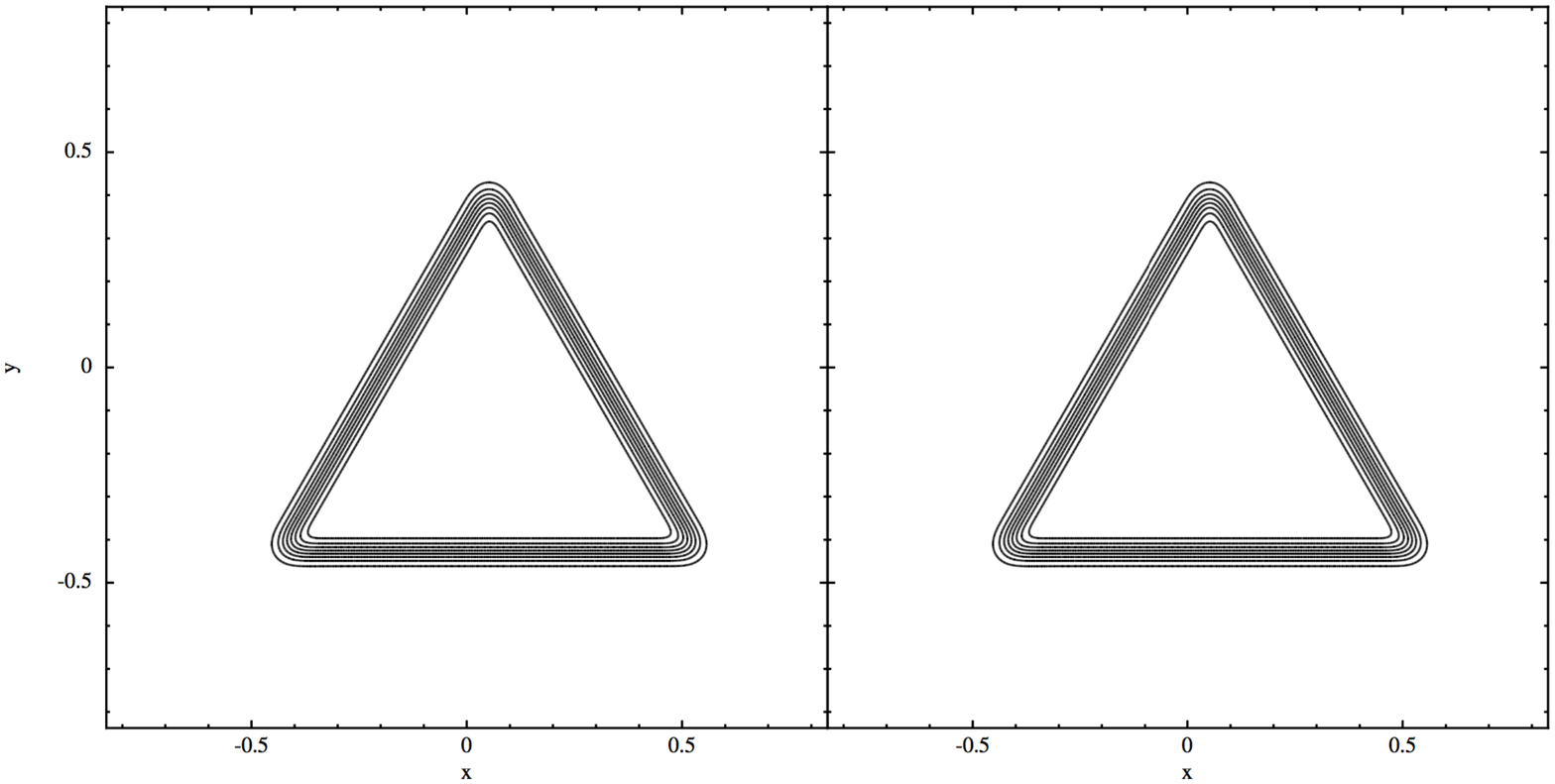}}

\vspace*{-1.5cm}

    \caption{Demonstration of SPH's superb advection properties. A high-density wedge ($N=2$) 
in pressure equilibrium with the background ($N=1$) is advected with velocity $v_x= 0.9999$, 
corresponding to a Lorentz factor of $\Gamma= 70.7$, through a box with periodic boundaries. 
For this test only 7000 particles are used. After crossing the box 10 times (or more than 23\,000 
time steps; right panel) no deterioration of the shape with respect to the initial condition (left panel) 
is noticeable.}
    \label{fig:advection}
\end{figure}} 

\subsubsection{Advection}
\label{sec:advection_test}

SPH is essentially perfect in terms of advection: if a particle carries a certain property,
say some electron fraction, it simply transports this property with it (unless the property is changed
by physical processes, of course). The advection accuracy is, contrary to Eulerian schemes,
independent of the numerical resolution and just governed by the integration accuracy of
the involved ordinary differential equations.
We briefly want to illustrate the advection accuracy  in an example that is a very serious challenge
for Eulerian schemes, but essentially a ``free lunch'' for SPH: the highly supersonic advection of a density
pattern through the computational domain with periodic boundary conditions. 
To this end, we set up  7K SPH particles inside the domain
$[-1,1] \times [-1,1]$. Each fluid element is assigned a velocity in the x-direction of 
$v_x= 0.9999$ corresponding to a Lorentz factor of $\gamma\approx 70.7$ and 
periodic boundary conditions are applied. The result of this experiment is shown in 
Figure~\ref{fig:advection}: after crossing the computational domain ten times (more than 
23 000 time steps; right panel) the shape of the triangle has not changed in any 
noticeable way with respect to the initial condition (left panel). 

\subsubsection{Shocks}
\label{sec:shock_test}

Since shocks play a crucial role in many areas of astrophysics we want to briefly discuss
SPH's performance in shocks. As an illustration, we show the result of a 2D  relativistic shock tube test
where the left state is given by $[P,v_x,v_y,N]_L= [40/3,0,0,10]$ and the right state by
$[P,v_x,v_y,N]_R= [10^{-6},0,0,1]$. The polytropic exponent is $\gamma = 5/3$ and the results
are shown at $t= 0.25$. In the experiment  80K particles were
initially placed on a hexagonal lattice in region $[-0.4,0.4] \times [-0.02,0.02]$.
Overall, the exact solution (red solid line) is well captured, see Figure~\ref{fig:shock}. The 
discontinuities, however, are less sharp than those obtained with modern High Resolution 
Shock Capturing methods with the same resolution. Characteristic for SPH is that the particles that were initially 
located on a lattice get compressed in the shock and, in the post-shock region, they need to 
``re-grid'' themselves into a new particle distribution. This also means that the particles 
have to pass each other and therefore also acquire a small velocity component in y-direction.
As a result, there is some unavoidable ``re-meshing noise'' behind the shock front.
This can be tamed by using smoother kernels and, in general, the details of the results depend on the
exactly chosen dissipation parameters and initial conditions. Note that the noise trigger that is used 
in the shown simulation applies dissipation
in the re-gridding region behind the shock, therefore the contact discontinuity has been somewhat broadened. Reducing the
amount of dissipation in this region (i.e., raising the tolerable reference value for noise) would sharpen the contact 
discontinuity, but at the price of increased re-meshing noise in the velocity.

\epubtkImage{}{%
  \begin{figure}[htb]
    \centerline{\includegraphics[width=10cm,angle=-90]{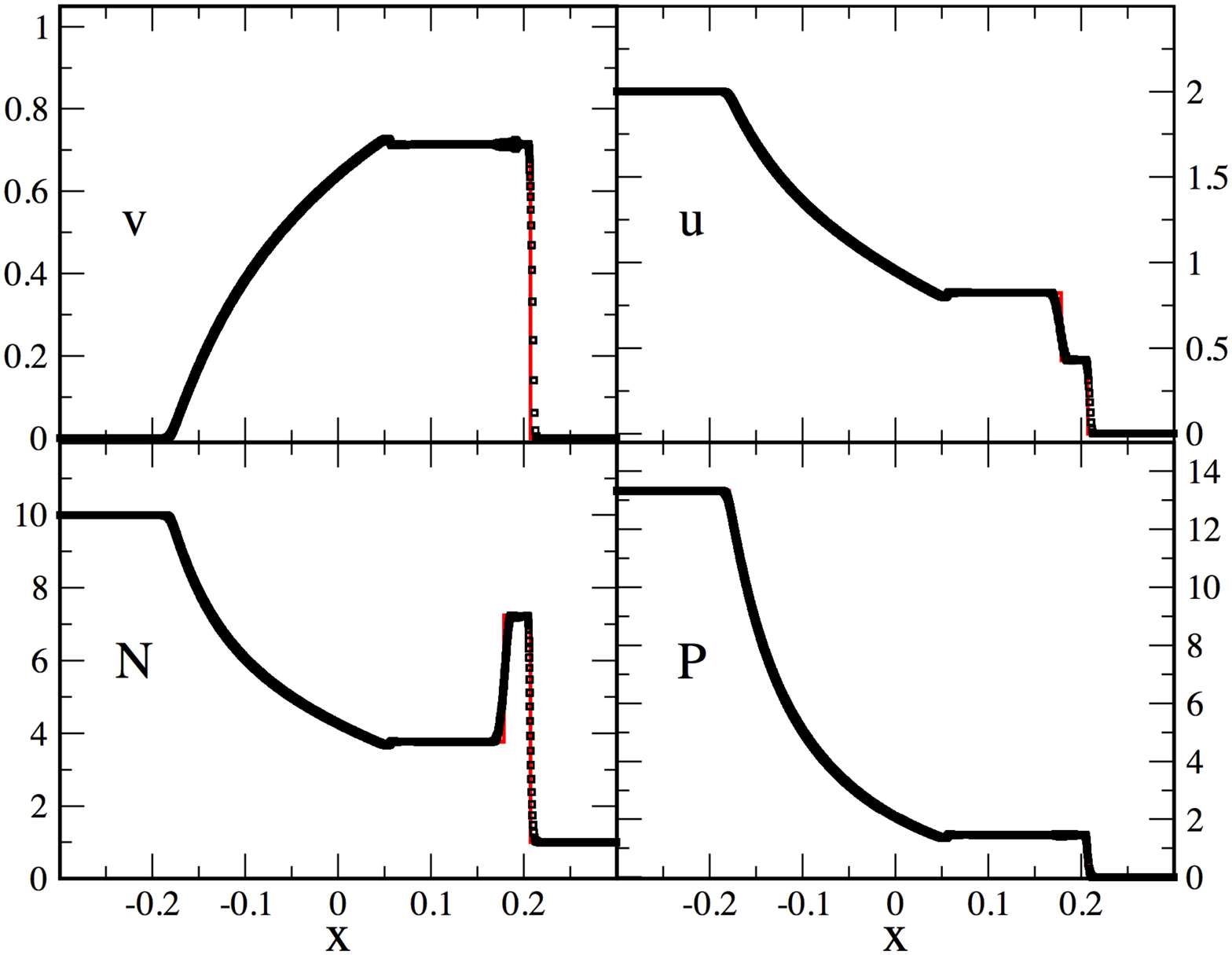}}
    \caption{Two-dimensional relativistic shock tube test. The exact solution is shown as red, solid line, the SPH results as open squares. Each SPH particle is shown.}
    \label{fig:shock}
\end{figure}}

\subsubsection{Fluid instabilities}
\label{sec:KH}

The standard formulation of SPH has recently been criticized \citep{thacker00,agertz07,springel10a,read10} 
for its performance in handling of fluid instabilities. As discussed in Section~\ref{sec:volume_elements}, this is related to
the different smoothness of density and internal energy which can lead to spurious pressure
forces acting like a surface tension. We want to briefly illustrate that a modern formulation can capture
Kelvin--Helmholtz instabilities accurately.\epubtkFootnote{The formulation $\mathcal{F}_1$ from \cite{rosswog15b}
is used for this test.}  Three stripes of hexagonal
lattices are placed in the domain $[-1,1] \times [-1,1]$ so that a density of $N=2$ is realized in the middle
stripe ($|y|<0.3$) and $N=1$ in the surrounding stripes. The high-density stripe moves with  $v_x=0.2$,
the other two with $v_x=-0.2$, the pressure is $P_0=10$ everywhere and the polytropic exponent is $\Gamma=5/3$.
No perturbation mode is triggered explicitly, we wait until small perturbations that occur as particles at the interface 
pass each other grow into a healthy Kelvin--Helmholtz instability. Figure~\ref{fig:KH} shows snapshots at 
$t=$ 2, 3.5 and 5.0. Note that ``standard methods'' (fixed, high dissipation parameters, volume element $\nu/N$ or $m/\rho$
and direct gradients of the M$_4$ kernel) do, for this setup, not lead to a noticeable instability on the shown time 
scale \citep{rosswog15b}.

\epubtkImage{}{%
  \begin{figure}[htb]
    \centerline{\includegraphics[angle=-90,width=\textwidth]{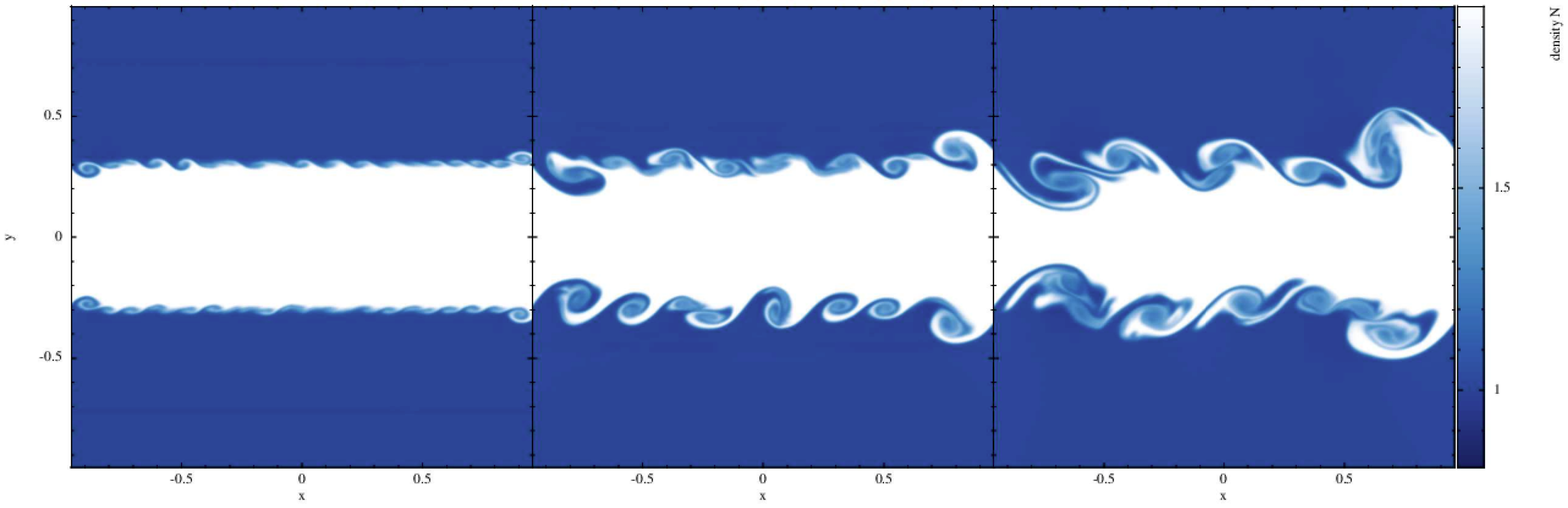}}

\vspace*{-2.5cm}

 \caption{Example of an untriggered Kelvin--Helmholtz instability. The central density stripe 
               has a density of $N=2$ and moves with velocity $v_x=0.2$, the surrounding stripes have
               density $N=1$ and $v_x=-0.2$. No particular mode was excited, the instability is triggered by
               small numerical irregularities at the interface that grow into Kelvin-Helmholtz billows. The shown
               snapshots are taken at $t=$ 2, 3.5 and 5.0, 1000K SPH particles were used. The simulation has been
               performed with the $\mathcal{F}_1$ SPH formulation of \cite{rosswog15b}, note that with
               ``standard techniques'' (constant, high dissipation, standard volume elements and direct 
              gradients of the M$_4$ kernel) no Kelvin--Helmholtz billows develop on the shown time scale.}
   \label{fig:KH}
\end{figure}}

\subsubsection{The Gresho--Chan vortex}
\label{sec:gresho}

The Gresho--Chan vortex \citep{gresho90} is considered as a particularly difficult test,
in general and in particular for SPH. As shown in \cite{springel10a}, standard SPH 
shows very poor convergence in this test.  The test deals with a stationary vortex 
that should be in stable equilibrium. Since centrifugal forces and pressure gradients 
balance exactly, any deviation from the initial configuration that develops over time 
is spurious and of purely numerical origin. The azimuthal component of the velocity 
in this test rises linearly up to a maximum value of $v_0$ which is reached at $r= R_1$ 
and subsequently decreases linearly back to zero at 2$R_1$
\be
v_\varphi (r)=  v_0 \left\{
  \begin{array}{ l l l}
     u \hspace*{1cm} {\rm for \quad } u \le 1\\
     2 - u  \hspace*{0.4cm} {\rm for \quad} 1 < u \le 2,\\
     0   \hspace*{1.05cm} {\rm for \quad} u > 2\\
   \end{array} \right.
\ee
where $u= r/R_1$. We require that centrifugal and pressure accelerations balance, 
therefore the pressure becomes
\be
P(r)= P_0 + \left\{
  \begin{array}{ l l l}
     \frac{1}{2} v_0^2 u^2  \hspace*{3.0cm} {\rm for \quad } u \le 1\\
     4 v_0^2 \left(\frac{u^2}{8} - u + \ln{u} + 1 \right)  \hspace*{0.3cm} {\rm for \quad } 1 < u \le 2\\
     4 v_0^2 \left(\ln 2 - \frac{1}{2}\right) \hspace*{1.8cm} {\rm for \quad } u > 2.\\
   \end{array} \right.
\ee
In the literature on non-relativistic hydrodynamics \citep{liska03,springel10a,read12,dehnen12} usually $v_0= 1$ 
is chosen together with $R_1= 0.2$, a uniform density $\rho= 1$ and a polytropic exponent of 5/3. 
Since we perform this test with the special-relativistic code SPHINCS\_SR in the Newtonian limit,
we use most of these values, but choose $R_1= 2 \times 10^{-4}$  and  $v_0= 10^{-3}$ to be 
safely in the non-relativistic regime.  We use this test here to illustrate which 
difference modern concepts can make in this challenging test, see Figure~\ref{fig:gresho}. 
The left panel shows the result from a modern SPH
formulation (formulation $\mathcal{F}_1$ from \citealp{rosswog15b}) and one that applies ``traditional''
approaches (fixed, high dissipation parameters, volume element $\nu/N$ or $m/\rho$
 and direct gradients of the M$_4$ kernel). The traditional approach fails completely and 
actually does not converge to the correct solution \citep{springel10a,rosswog15b} while the new, 
more sophisticated approach yields very good results.

\epubtkImage{}{%
\begin{figure}[htbp]
  \centerline{\includegraphics[angle=0,width=\textwidth]{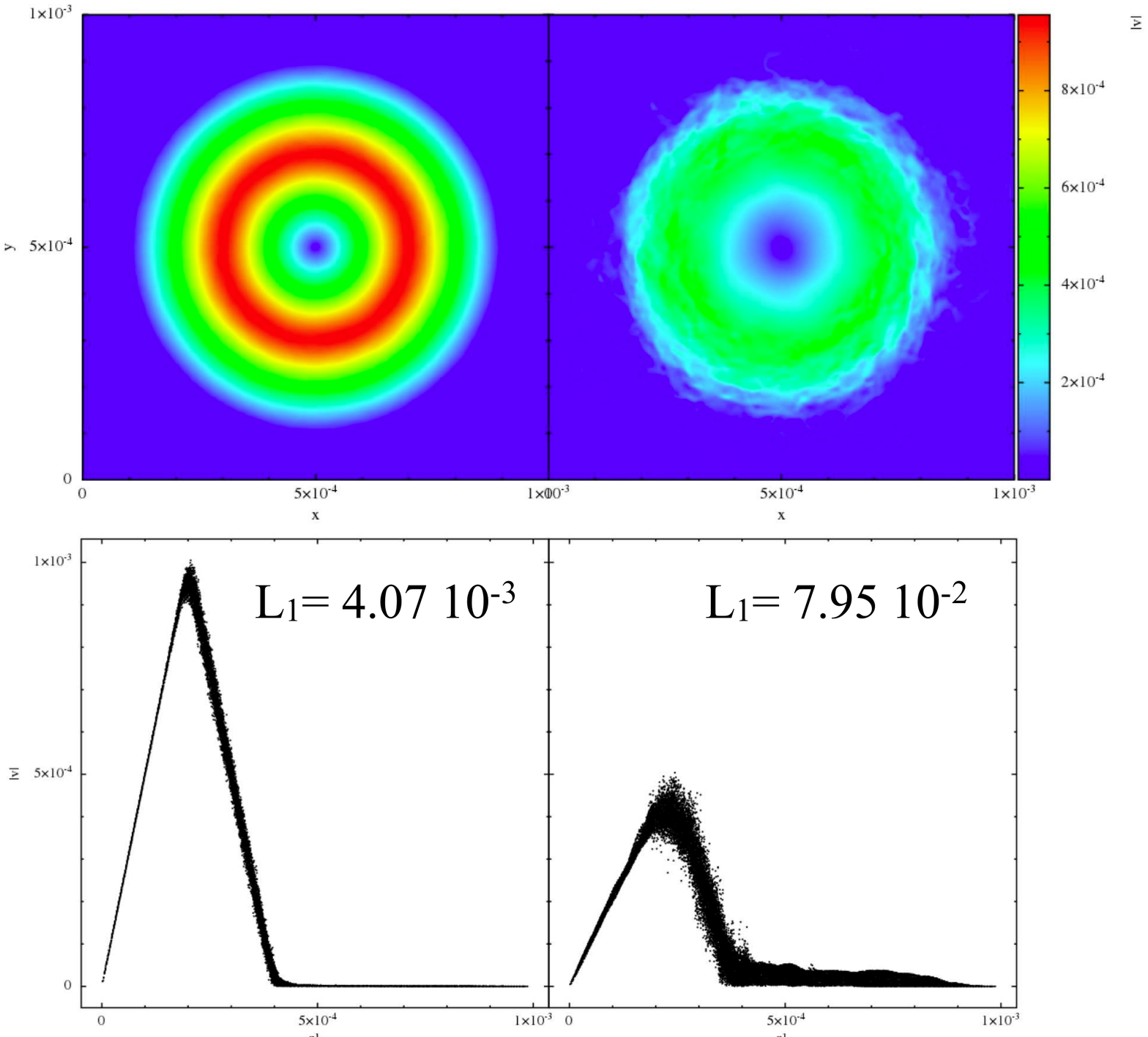}}
  \caption{Comparison of the performance of a modern (good dissipation triggers, high-order kernel, 
integral-based gradient calculation; this is the $\mathcal{F}_1$ formulation from \citealp{rosswog15b}) 
and a more traditional SPH formulation (constant high dissipation, cubic spline kernel, gradients 
from kernel derivatives; this is the $\mathcal{F}_4$ formulation from the same paper) in the 
challenging Gresho--Chan vortex  problem at $t=1$. At this time, the combination of noise 
and high dissipation have, for the second case, already very seriously deteriorated the ideally stationary 
solution. The modern formulation, in contrast, has stayed very close to the initial condition. Also 
indicated is the $L_1$ error of the velocity.}
  \label{fig:gresho}
\end{figure}}

\subsection{Summary: SPH}

The most outstanding property of SPH is its exact numerical conservation. This can straight forwardly
be achieved via symmetries in the particle indices of the SPH equations together with anti-symmetric gradient
estimates. The most elegant and least arbitrary strategy to obtain a conservative SPH formulation
is to start from a fluid Lagrangian and to derive the evolution equations via a variational principle. This approach can 
be applied in the Newtonian, special- and general relativistic case, see Sections~\ref{sec:Newt_SPH}, \ref{sec:SR_SPH} 
and \ref{sec:GR_SPH}. Advection of fluid properties is essentially perfect in SPH and it is in particular
not dependent on the coordinate frame in which the simulation is performed.

SPH robustly captures shocks, but they are, at a given resolution, not as sharp as those from
state-of-the-art high-resolution shock capturing schemes. Moreover, standard SPH has been 
criticized for its (in)ability to resolve fluid instabilities under certain circumstances. Another issue 
that requires attention when performing SPH simulations are initial conditions. When not perpared 
carefully, they can easily lead to noisy results, since the regularization force discussed in 
Sec.~\ref{sec:self_regularization} leads for poor particle distribution to a fair amount of particle
velocity fluctuations. This issue is particular severe when poor kernel functions and/or low neighbour 
numbers are used, see Sec.~\ref{chap:kernel_approx}.

Recently, a number of improvements to SPH techniques have been suggested. These include a) more accurate gradient estimates,
see Sec~\ref{sec:SPH_derivs}, b) new volume elements which eliminate spurious surface tension effects, see 
Section~\ref{sec:volume_elements}, c) higher-order kernels, see Section~\ref{sec:kernel_choice} and d) more sophisticated dissipation 
switches, see Section~\ref{sec:Newtonian_shocks}.  As illustrated, for example by the Gresho--Chan vortex test in
Section~\ref{sec:gresho}, enhancing SPH with these improvements can substantially 
increase its accuracy with respect to older SPH formulations.


\section{Astrophysical Applications}
\label{sec:appl}

The remaining part of this review is dedicated to  actual applications of SPH
to astrophysical studies of compact objects.  We focus on encounters of
\begin{itemize}
\item two white dwarfs (Section~\ref{sec:appl_WDWD}), 
\item two neutron stars (Section~\ref{sec:appl_NSNS}) and 
\item a neutron star with a black hole  (Section~\ref{sec:appl_NSBH}).
\end{itemize}
In each case the focus is on gravitational wave-driven binary mergers.
In locations with large stellar number densities, e.g., globular clusters,
dynamical collisions between stars occur frequently and encounters between
two neutron stars and a neutron star with a stellar-mass black hole may
yield very interesting signatures. Therefore such encounters are also briefly 
discussed.

In each of these fields a wealth of important results have been achieved with a number of different
methods. Naturally, since the scope of this review are SPH methods, we will focus our attention here
to those studies that are at least partially based on SPH simulations. For further studies that are based
on different methods we have to refer to the literature.

\subsection{Double white-dwarf encounters}
\label{sec:appl_WDWD}

\subsubsection{Relevance}

White Dwarfs (WDs) are the evolutionary end stages of most stars in the Universe,
for every solar mass of stars that forms $\sim 0.22$ WDs will be produced on average. As a result, the Milky Way 
contains $\sim 10^{10}$ WDs \citep{napiwotzki09} in total and  $\sim 10^8$ double WD systems
\citep{nelemans01a}. About half of these systems have separations that are small enough (orbital
periods $<10$ hrs) so that gravitational wave emission will bring them into contact 
within a Hubble time, making them a major target for the eLISA mission \citep{elisa13}. 
Once in contact, in almost all cases the binary system will merge, in the remaining
small fraction of cases mass transfer may stabilize the orbital decay and lead to long-lived 
interacting binaries such as AM CVn systems \citep{paczynski67,warner95,nelemans01b,nelemans05,solheim10}.

Those systems that merge may have a manifold of interesting possible outcomes. The  merger 
of two He WDs may produce a low-mass He star \citep{webbink84,iben86,saio00,han02}, He-CO 
mergers may form hydrogen-deficient giant or R CrB stars \citep{webbink84,iben96,clayton07} and 
if two CO WDs merge, the outcome may be a more massive, possibly hot and high B-field WD 
\citep{bergeron91,barstow95,segretain97}. A good fraction of the CO-CO merger remnants probably transforms into
ONeMg WDs which finally, due to electron captures on Ne and Mg, undergo an accretion-induced collapse (AIC)
to a neutron star \citep{saio85,nomoto91,saio98}.
Given that the nuclear binding energy that can still be released by burning to iron group elements 
(1.6 MeV from He, 1.1 MeV from C and 0.8 MeV from O) is large, it is not too surprising
that there are also various pathways to thermonuclear explosions. The ignition of helium
on the surface of a WD may lead to weak thermonuclear explosions \citep{bildsten07,foley09,perets10}, 
sometimes called ``.Ia'' supernovae. The modern view is that WDWD mergers might also trigger
type Ia supernovae (SN~Ia) \citep{webbink84,iben84} and, in some cases,
even particularly bright ``super-Chandrasekhar'' explosions, e.g., \cite{howell06,hicken07}. 

SN~Ia  are important as cosmological distance indicators, as factories for intermediate mass 
and iron-group nuclei, as cosmic ray accelerators, kinetic energy sources for galaxy evolution or 
simply in their own right as end points of  binary stellar evolution. After having been the second-best 
option behind the ``single degenerate'' model for decades, it now seems entirely possible that double degenerate mergers 
are behind a sizeable fraction of  SN~Ia. It seems that with the re-discovery of double 
degenerates as promising type Ia progenitors an interesting time for supernovae research has begun.
See \cite{howel11} and \cite{maoz14} for two excellent recent reviews on this topic.

Below, we will briefly summarize the challenges in a numerical simulation of a WDWD merger 
(Section~\ref{sec:challenges_WDWD}) and then discuss recent results concerning mass transferring 
systems (Section~\ref{sec:WDWD_MT}) and, closely related, to the final merger of a WDWD binary 
and possibilities to trigger SN~Ia (Section~\ref{sec:WDWD_SNIa}). We will also briefly discuss dynamical
collisions of WDs (Section~\ref{sec:WDWD_collisions}). For SPH studies that explore the 
gravitational wave signatures of WDWD mergers we refer to the literature 
\citep{loren05,dan11,vandenbroek12}.

Note that in this section we explicitly include the constants $G$ and $c$ in the equations to 
allow for a simple link to the astrophysical literature.

\subsubsection{Challenges}
\label{sec:challenges_WDWD}

WDWD merger simulations are challenging for a number of reasons not the least of which are
the onset of mass transfer and the self-consistent triggering of thermonuclear explosions.

While two white dwarfs revolve around their common centre of mass, gravitational wave 
emission reduces the separation $a$ of a circular binary orbit at a rate of \citep{peters63,peters64}
\be
\dot{a}_{\rm GW}= -\frac{64 G^3}{5 c^5} \frac{m_1 m_2 M}{a^3},
\ee
where $m_1$ and $m_2$ are the component masses.
Although it is gravitational wave emission that drives the binary towards mass transfer/merger
in the first place,  its dynamical impact at the merger stage is completely negligible since
\be
\frac{\tau_{\rm GW}}{P_{\rm orb}}= \frac{a/|\dot{a}_{\rm GW}|}{2\pi/\omega_{\rm K}}=
6.6 \cdot 10^8 \left( \frac{a}{2 \times 10^9 {\rm cm}}\right)^{5/2} \left(\frac{0.6\,M_{\odot}}{m_1} \right) 
\left( \frac{0.6\,M_{\odot}}{m_2}\right) \left( \frac{1.2\,M_{\odot}}{M}\right)^{1/2}.
\ee
Mass transfer will set in once the size of the Roche lobe of one of the stars has become comparable to the 
enclosed star. Due to the inverted mass-radius relationship of WDs, it is always the less massive WD (``secondary'') that fills 
its Roche lobe first. From the Roche lobe size and Kepler's third law, the average density 
$\bar{\rho}$ of the donor star can be related to the orbital period \citep{paczynski71,frank02}
\be
\bar{\rho} \approx \frac{115\mathrm{\ g\ cm}^{-3}}{P^2_{\mathrm{hr}}},
\ee
where $P_{\rm hr}$ is the orbital period measured in  hours. In other words: the shorter the
orbital period, the higher the density of the mass donating star. For periods below 1 hr the donor 
densities exceed those of  main sequence stars which signals that a compact star is involved. If
one uses the typical dynamical time scale of a WD, $\tau_{\rm dyn}\approx (G \bar{\rho})^{-1/2}$, one finds
\be
\frac{P_{\rm orb}}{\tau_{\rm dyn}}\approx 10,
\ee
so that a single orbit would already require $\sim 10\,000 (n_{\rm dyn}/1000)$ numerical time steps, if $n_{\rm dyn}$
denotes the number of numerical time steps per stellar dynamical time. This demonstrates that long-lived mass transfer
over tens of orbital periods can become quite computationally expensive and may place limits on
the numerical resolution that can be afforded in such a simulation. On the other hand, when
numerically resolvable mass transfer sets in, it already has a rate of
\begin{equation}\label{eq:mmin}
\dot M_{\rm lim}\sim \frac{1\ {\rm particle \; mass}}{{\rm orbital \; period}}
\sim 2 \times 10^{-8}
\frac{M_{\odot}}{\rm s}\left(\frac{10^6}{\rm npart}\right)
\left(\frac{M}{1\,M_{\odot}}\right)^{3/2}
\left(\frac{2\cdot 10^9\ {\rm cm}}{a_0}\right)^{3/2},
\end{equation}
where ${\rm npart}$ is the total number of SPH particles, ${\rm M_{tot}}$ is the
total mass of the binary and $a_{0}$ is the separation between the stars
at the onset of mass transfer. This limit due to finite numerical resolution is several
orders of magnitude above the Eddington limit of WDs, therefore sub-Eddington 
accretion rates are hardly ever resolvable within a global, 3D SPH simulation. The 
transferred matter comes initially
from the tenuous WD surface which, in SPH, is the poorest resolved region of
the star. Following this matter is also a challenge for Eulerian methods since it
needs to be disentangled from the ``vacuum'' background and, due to the 
resolution-dependent angular momentum conservation, it is difficult to obtain the 
correct feedback on the orbital evolution. In other words: the consistent simulation
of mass transfer and its feedback on the binary dynamics is a serious challenge for 
every numerical method. 

The onset of mass transfer represents a juncture in the life of WDWD binary, since now the stability
of mass transfer decides whether the binary can survive or will inevitably merge. 
The latter depends sensitively on the internal structure of the donor star, the binary mass 
ratio and the angular momentum transport mechanisms \citep[e.g.,][]{marsh04,gokhale07}.
Due to the inverse mass-radius relationship of WDs, the secondary will expand on mass loss and 
therefore tendentially speed up the mass loss further. On the other hand, since the mass is transferred 
to the higher mass object, momentum/centre of mass conservation will tend to 
widen the orbit and therefore tendentially reduce mass transfer. If the circularization radius 
of the transferred matter is smaller than the primary radius it will directly impact on the stellar 
surface and tend to spin up the accreting star. In this way, orbital angular momentum 
is lost to the spin of the primary which, in turn, decreases the orbital separation and accelerates 
the mass transfer. If, on the other hand, the circularization radius is larger than the primary radius 
and a disk can form, angular momentum can, via the large lever arm of the disk, be fed back into the orbital 
motion and stabilize the binary system \citep{iben98,piro11}. To make things even more complicated, if tidal interaction 
substantially heats up the mass donating star it may impact on its internal structure and therefore change its 
response to mass loss.

To capture these complex angular momentum transfer mechanisms reliably in a simulation
requires a very accurate numerical angular momentum conservation. We want to 
briefly illustrate this point with a small numerical experiment. A $0.3+0.6\,M_{\odot}$ WD binary system 
is prepared in a Keplerian orbit, so that mass transfer is about to set in. To mimic the effect of numerical 
angular momentum loss in a controllable way, we add small artificial forces similar to those emerging 
from gravitational wave emission \citep{peters63,peters64,davies94} and adjust the overall value so 
that 4\% or 0.5\% of angular momentum per orbit are lost. These results are compared to a 
simulation without artificial loss terms where the angular momentum is conserved to better 
than 0.01\% per orbit, see Figure~\ref{fig:conservation_GWs}. The effect on the mutual separation 
$a$ (in $10^9\,{\rm cm}$) is shown in the upper panels and the gravitational wave amplitude, 
$h_+$ ($r$ is the distance to the observer) as calculated in the quadrupole formalism, are shown 
the lower panels. 

\epubtkImage{}{%
\begin{figure}[htb]
  \centerline{\includegraphics[angle=90,width=\textwidth]{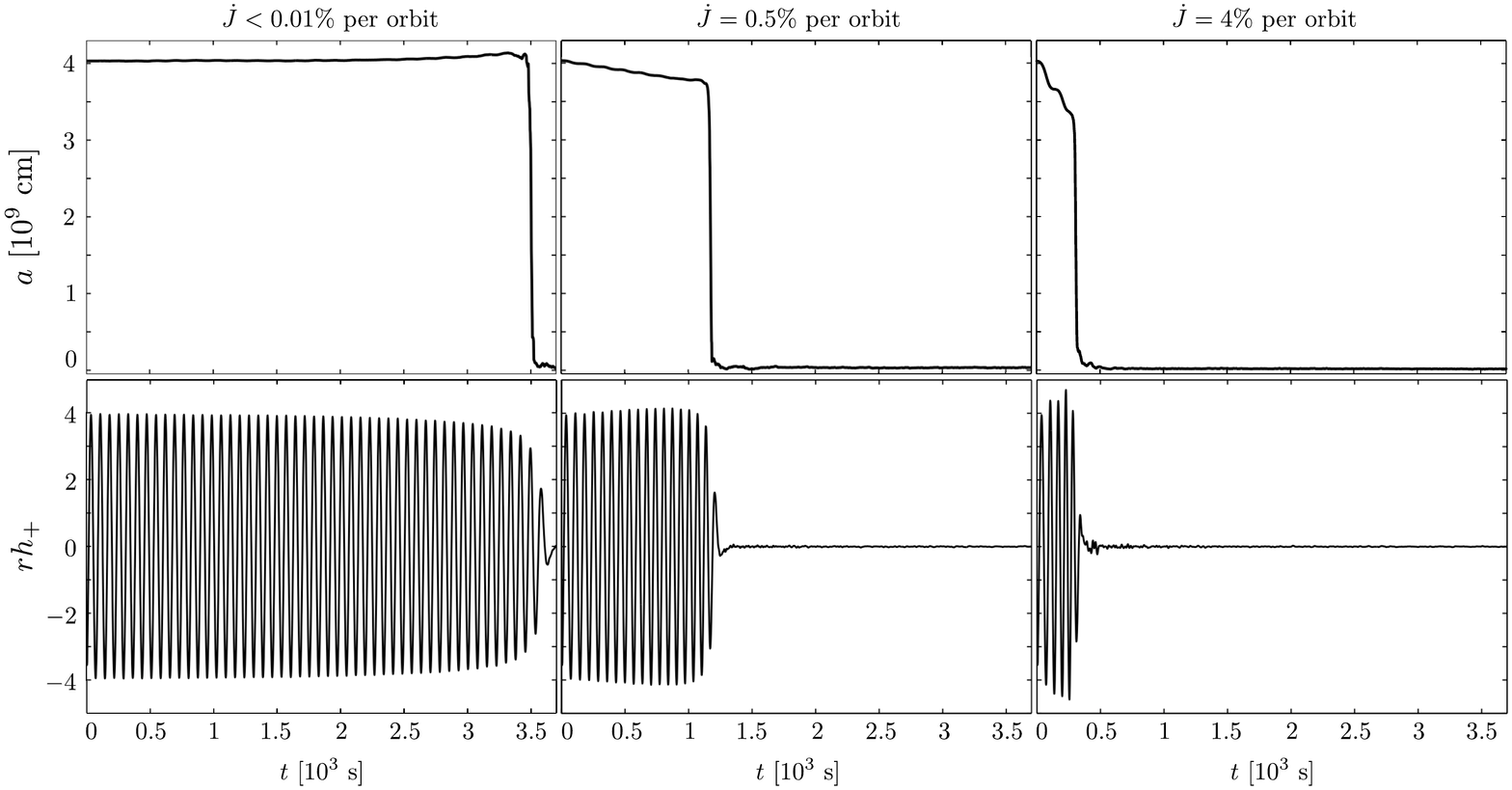}}

\vspace*{-1cm}

  \caption{Numerical experiment to illustrate the sensitivity of a binary system to the non-conservation 
of angular momentum. A WDWD binary system (0.3 and 0.6 \msun) is adiabatically relaxed to the point 
where mass transfer is about to set in and then its orbital evolution is followed in a 3D hydrodynamic 
simulation. To mimic numerical loss of angular momentum, an artificial force is applied that removes 
angular momentum at a rate of 0.5\% (middle) and 4\% per orbit (right panel). The simulation shown in 
the leftmost panel  conserves angular momentum to better than 0.01\% per orbit. Shown is the binary 
separation $a$ (upper row) and the gravitational wave amplitude $h_+$ times to distance to the source 
$r$. Figure courtesy of Marius Dan.}
  \label{fig:conservation_GWs}
\end{figure}}

Even the moderate loss of 0.5\% angular momentum per orbit leads to a quick artificial merger and a mass transfer
duration that is reduced by more than a factor of three. These conservation requirements make SPH a natural 
choice for WDWD merger simulations and it has indeed been  the first  method used for these type of problems.

As outlined above, one of the most exciting possibilities is the triggering of thermonuclear 
explosions during the interaction of two WDs. Such explosions can  either be triggered by a shock wave
where the thermonuclear energy generation behind the shock wears down possible dissipative effects
or, spontaneously, if the local conditions for burning are favorable enough so that it occurs 
faster than the star can react by expanding. \cite{seitenzahl09} have studied detonation conditions
in detail via local simulations and found that critical detonation conditions can require that length 
scales down to centimeters are resolved which is, of course,  a serious challenge for global, 3D simulations of
objects with radii of $\sim 10^9$ cm.

There is also a huge disparity in terms of time scales. Whenever nuclear burning is important for
the dynamics of the gas flow, the nuclear time scales are many orders of magnitude shorter than
the admissible hydrodynamic time steps. Therefore, nuclear networks are usually implemented via
operator splitting methods, see e.g., \cite{benz89,rosswog09a, raskin10,garcia_senz13}. Because of 
the exact advection in SPH the post-processing of hydrodynamic trajectories with larger nuclear
networks to obtain detailed abundance patterns is straight forward.  For burning processes in tenuous
surface layers, however, SPH is seriously challenged since here the resolution is poorest. For
such problems, hybrid approaches that combine SPH with, say, AMR methods \citep{guillochon10,dan15}
seem to be the best strategies.


\subsubsection{Dynamics and mass transfer in white dwarf binaries systems}
\label{sec:WDWD_MT}

Three-dimensional simulations of WDWD mergers were pioneered by \cite{benz90b}. Their major motivation was to understand the merger dynamics and the possible
role of double degenerate systems as SN~Ia progenitors. 
They used an SPH formulation as described in Section~\ref{sec:Newt_vanilla} (``vanilla ice'') together with  7000 SPH 
particles, an equation of state for a non-degenerate ideal gas with a completely
degenerate, fully relativistic electron component and they restricted themselves to the study 
of a 0.9 - 1.2 \Msun system. No attempts were undertaken to include nuclear burning in this 
study (but see \cite{benz89}). Each star was relaxed in isolation, see Section~\ref{sec:IC}, and subsequently placed in 
a circular Keplerian orbit so that the secondary was overfilling its critical lobe by $\sim8\%$.
Under these conditions the secondary star was disrupted within slightly more than two orbital periods, 
forming a three-component system of a rather unperturbed primary, a hot pressure 
supported spherical envelope and a rotationally supported outer disk. About 0.6 \% of a  
solar mass were able to escape, the remaining $\sim 1.7$ \msun, supported mainly by 
pressure gradients, showed no sign of collapse.

\cite{rasio95} were more interested in the equilibrium and the (secular,
dynamical and mass transfer) stability properties of close binary systems. They studied
systems both with stiff ($\Gamma> 5/3$; as models for neutron stars) and soft ($\Gamma= 5/3$)
polytropic equations of state, as approximations for the EOS of  (not too massive) WDs 
and low-mass main sequence star binaries.
They put particular emphasis on constructing accurate, \emph{synchronized} initial conditions 
\citep{rasio94}. These were obtained by relaxing the binary system in a corotating frame where,
in equilibrium, all velocities should vanish. The resulting configurations satisfied the virial 
theorem to an accuracy of about one part in $10^3$. With these initial 
conditions they found a more gradual increase in the mass transfer rate in comparison to 
\cite{benz90b}, but nevertheless the binary was disrupted after only a few  orbital periods.

\cite{segretain97} focussed on the question whether 
particularly massive and hot WDs could be the result of binary mergers \citep{bergeron91}. They applied
a simulation technology similar to \cite{benz90b} and concentrated on a binary system
with non-spinning WDs of 0.6 and 0.9 \msun.  They showed, for example, that such a merger remnant would need
to lose about 90\% of its angular momentum in order to reproduce  properties of the observed 
candidate WDs.

Although \cite{rasio95}  had already  explored the construction of accurate initial
conditions, essentially all subsequent simulations 
\citep{guerrero04,loren05,yoon07a,loren09,pakmor10,pakmor11,zhu13a} 
were carried out with rather approximate initial conditions consisting of spherical stars placed
in orbits where, according to simple Roche-lobe geometry estimates, mass transfer should
set in. \cite{marsh04} had identified in  a detailed orbital stability analysis
definitely stable regions (roughly for primary masses substantially larger than the companion mass), 
definitely unstable (mass ratios between 2/3 and 1) and an intermediate region where the 
stability of mass transfer is less clear. 
Motivated by large discrepancies in the mass transfer 
duration that had been observed between careful grid-based \citep{motl02,dsouza06,motl07} 
and earlier SPH simulations \citep{benz90b,rasio95,segretain97,guerrero04,yoon07a,pakmor10}
\cite{dan09,dan11} focussed on the mass transfer in this unclear 
regime.  They  very carefully relaxed the binary system in a corotating frame 
and thereby adiabatically reduced the mutual separation until the first particle climbed up 
to saddle point $L_1$ in the effective potential, see Figure~\ref{fig:effect_potential}. 

\epubtkImage{}{%
  \begin{figure}[htb]
    \centerline{\includegraphics[width=8cm,angle=0]{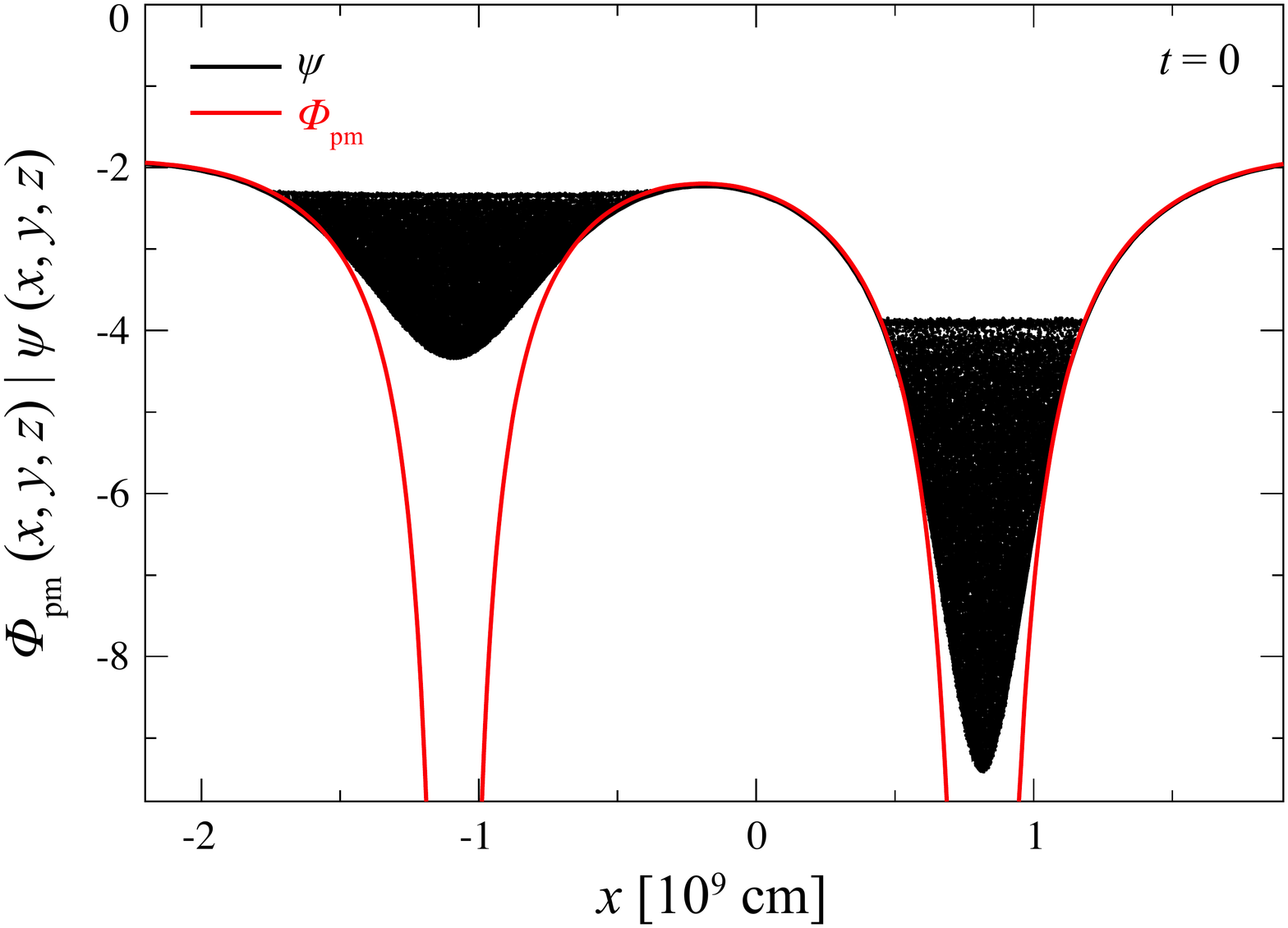}}
    \caption{Initial conditions for a synchronized WDWD binary system at the onset of resolvable mass transfer. 
The  red line is the point mass Roche potential, the SPH particle values are shown as filled black circles. 
Once the first SPH particle has crossed the $L_1$ point, the system is transformed from the corotating 
binary frame to a space-fixed frame where it is hydrodynamically evolved. Figure from \cite{dan11}.}
    \label{fig:effect_potential}
\end{figure}}

In their study such carefully constructed initial conditions were compared to the previously 
commonly used approximate initial conditions.  Apart from the inaccuracies inherent to the 
analytical Roche lobe estimates, approximate initial conditions also
neglect the tidal deformations of the stars and therefore seriously underestimate 
the initial separation at the onset of mass transfer. For this reason, such initial conditions have  
up to 15\% too little angular momentum and, as a result, binary systems with inaccurate initial 
conditions merge too  violently on a much too short time scale. As a result, temperatures 
and densities in the final remnant are over- and the size of tidal tails are underestimated.
The carefully prepared binary systems all showed dozens of orbits of numerically resolvable 
mass transfer. Given that, due to the finite resolution, the mass transfer is already highly 
super-Eddington when it starts being resolvable all the results on mass transfer duration 
have to be considered as strict lower limits. One particular
example, a 0.2 \Msun He-WD and a 0.8 CO-WD, merged with approximate initial conditions within
two orbital periods (comparable to earlier SPH results), but only after painfully long
84 orbital periods when the initial conditions were prepared carefully. This particular
example also illustrated the suitability of SPH for such investigations: during the orbital
evolution, which corresponds to $\approx$~17\,000 dynamical time scales, energy and angular
momentum were conserved to better than 1\%! All of the investigated (according to the Marsh
et al. analysis unstable) binary systems merged in the end although only after several dozens 
of orbital periods. Some systems showed a systematic widening of the orbits after the onset
of mass transfer. Although they were still disrupted in the end, this indicated that
systems in the parameter space vicinity of this 0.2 - 0.8 \Msun system may evolve into short-period
AM CVn systems.

In two recent studies, \cite{dan12,dan14a} systematically explored the  parameter space 
by simulating 225 different binary systems with masses ranging from 0.2 to 1.2 \msun. All of
the initial conditions were prepared carefully as \cite{dan11}. Despite the only moderate resolution 
(40 K particles) that could be afforded in such a broad study, they found 
excellent agreement with the orbital evolution predicted by mass transfer stability analysis 
\citep{marsh04,gokhale07}.

\subsubsection{Double white dwarf mergers and possible pathways to thermonuclear supernovae}
\label{sec:WDWD_SNIa}

\epubtkImage{}{%
  \begin{figure}[htbp]
    \centerline{
      \includegraphics[width=0.3\textwidth]{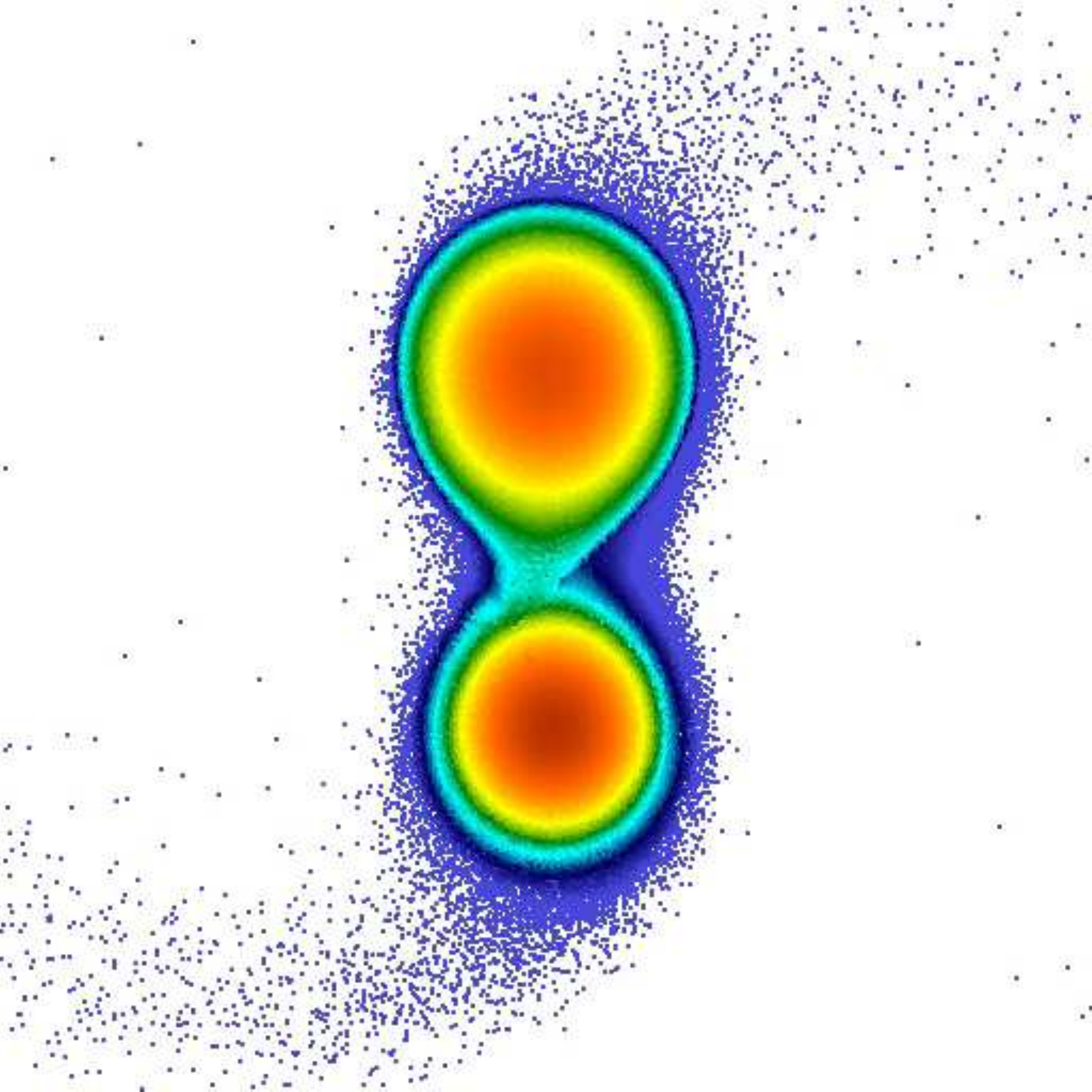}
      \includegraphics[width=0.3\textwidth]{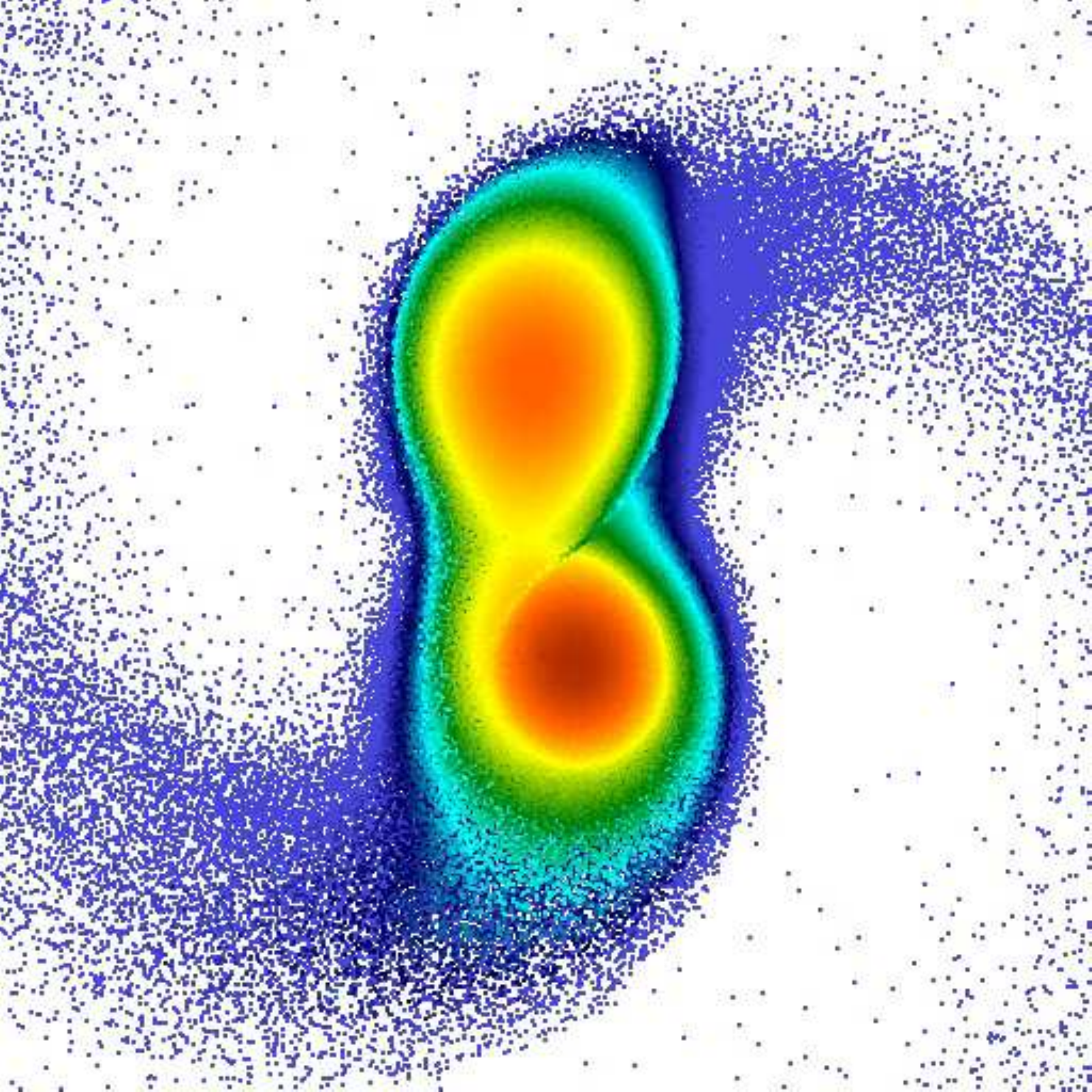}
    }
     \centerline{
      \includegraphics[width=0.3\textwidth]{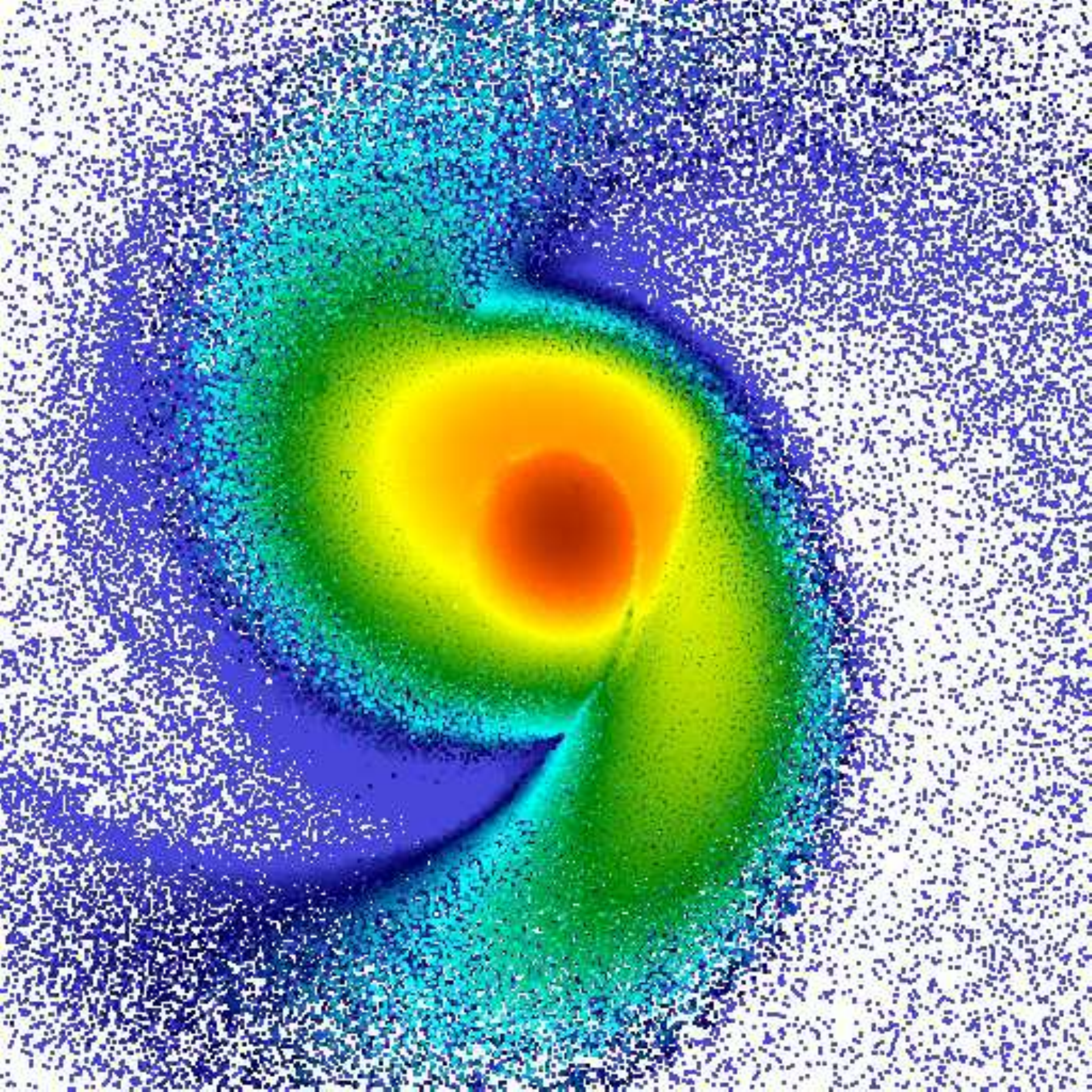}
      \includegraphics[width=0.3\textwidth]{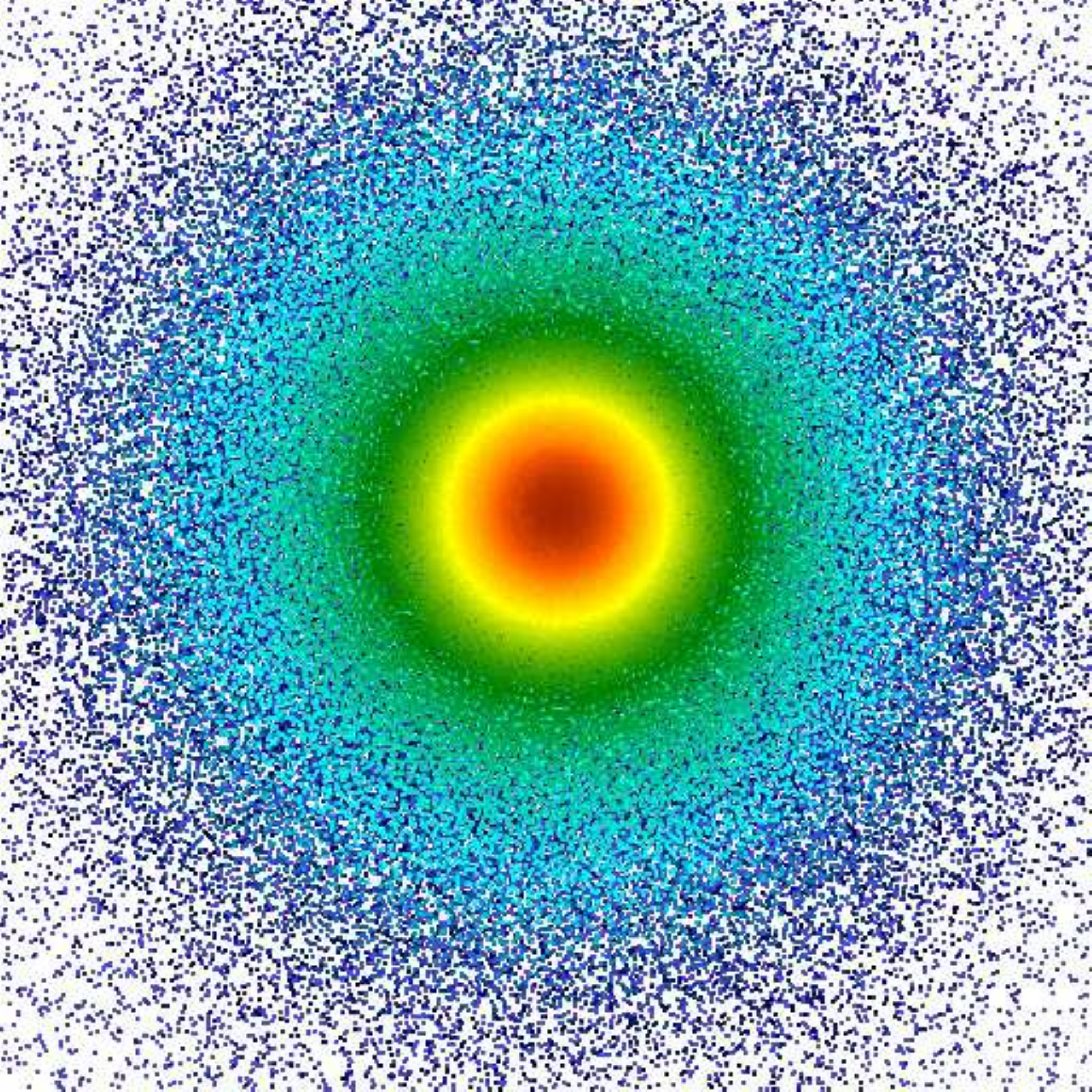}
    }
    \caption{Illustration of the morphology of a WDWD merger (mass ratio  $q= 0.78$). 
                  From \cite{diehl08}.}
    \label{fig:WDWD_merger}
\end{figure}}

The merger of two white dwarfs, the so-called ``double degenerate scenario'', had already 
been suggested relatively early on \citep{webbink84,iben84} as a promising type Ia progenitor channel.
It was initially modelled as CO-rich matter being accreted from a thick disk onto a central, 
cold WD \citep{nomoto85,saio85,saio98,piersanti03a,piersanti03b,saio04}.
Since for such thick disks accretion rates close to the Eddington limit ($\dot{M} \sim 10^{-5}$ \Msun yr$^{-1}$) 
are expected, most studies concluded that carbon ignition would start in the envelope of the central WD
and, as the burning flame propagates inwards within $\sim$ 5000 years, it would transform the WD 
from CO into ONeMg \citep{saio85,saio98}. When approaching the Chandrasekhar mass, Ne and Mg 
would undergo electron captures and the final result would be an accretion-induced collapse to a 
neutron star rather than a SN~Ia. Partially based on these studies, the double degenerate
model was long regarded as only the second best model that had some good motivation (consistent
rates, lack of hydrogen in SN~Ia spectra, Chandrasekhar mass as motivation for the uniformity of 
type Ia properties), but lacked a convincing pathway to an explosion.

The Barcelona group were the first to explore the effect of nuclear burning 
during a WD merger event \citep{guerrero04}. They implemented the reduced 14-isotope
$\alpha$-network of \cite{benz89} with updated reaction rates into a ``vanilla-ice'' SPH
code with artificial viscosity enhanced by the Balsara factor, see Sections~\ref{sec:Newt_vanilla} 
and \ref{sec:Newtonian_shocks}. They typically used 40 K particles, approximate initial conditions
as described above and explored six different combinations of masses/chemical compositions.
They found an orbital dynamics similar to \cite{benz89,segretain97} and, although in the outer,
partially degenerate layers of the central core temperatures around $10^9$ K were encountered,
no dynamically important nuclear burning was observed. Whenever it set in, the remnant had time 
to quench it by expansion, both for the He and CO accreting systems. Therefore they concluded 
that direct SN~Ia explosions were unlikely, but some remnants could evolve into subdwarf B objects 
as suggested in \cite{iben90}.

\cite{yoon07a} challenged the ``classical picture'' of the cold WD accreting from a 
thick disk as an oversimplification. Instead, they suggested that the subsequent 
secular evolution of the remnant would be better studied by treating the central 
object as a differentially rotating CO star with a central, slowly rotating,
cold core engulfed by a rapidly rotating hot envelope, which, in turn, is embedded 
and fed by a centrifugally supported Keplerian accretion disk. The further evolution 
of such a system is then governed by the thermal cooling of the hot envelope and 
the redistribution of angular momentum inside of the central remnant and the 
accretion of the matter from the disk into the envelope.
They based their study of the secular remnant evolution on a dynamical merger 
calculation of two CO WDs with 0.6 and 0.9 \msun. To this end they used an SPH code originally
developed for neutron star merger calculations \citep{rosswog99,rosswog00,rosswog02a,rosswog03a,rosswog08b} 
extended by the Helmholtz EOS \citep{timmes99} and a quasi-equilibrium reduced $\alpha$-network
\citep{hix98}. Particular care was taken to avoid artefacts from the artificial viscosity treatment
and time-dependent viscosity parameters \citep{morris97} and a Balsara-switch \citep{balsara95}, see
Section~\ref{sec:Newtonian_shocks}, were used in the simulation. As suggested by the work
of \cite{segretain97}, they assumed non-synchronized stars and started the simulations from
approximate initial conditions, see Section~\ref{sec:WDWD_MT}. Once a stationary remnant had 
been formed, the results were mapped into a 1D hydrodynamic stellar evolution code \citep{yoon04a} and its 
secular evolution was followed including the effects of rotation and angular momentum transport. 
They found that the growth of the stellar core is controlled by the neutrino 
cooling at the interface between the core and the envelope and that carbon ignition could be avoided 
provided that a) once the merger reaches a quasi-static equilibrium temperatures are below the 
carbon ignition threshold, b) the angular momentum loss occurs on a time scale longer than the neutrino 
cooling time scale and c) the mass accretion from the centrifugally supported disk is low enough
($\dot{M} \le 5 \times 10^{-6} - 10^{-5}$ \Msun yr$^{-1}$). From such remnants an explosion may be
triggered $\sim 10^5$ years after the merger. Such systems, however, may need unrealistically low 
viscosities.

A more recent study \citep{shen12a} started from two remnants of CO WD mergers \citep{dan11} and followed 
their viscous longterm evolution. Their conclusion was more in line with earlier studies: they
expected that the long-term result would be ONe or an accretion-induced collapse to a neutron star rather
than a SN~Ia.

In recent years, WD mergers have been extensively explored as possible pathways to SN~Ia. Not too
surprisingly, a number of pathways have been discovered that very likely lead to a thermonuclear
explosion directly prior to or during the merger. Whether these explosions are responsible for 
(some fraction of) normal SNe Ia or for peculiar subtypes needs to be further explored 
in future work. Many of the recent studies used a number of different numerical tools to 
explore various aspects of WDWD mergers. We focus here on those studies where SPH simulations 
were involved.

\subsubsection*{Explosions prior to merger}

\cite{dan11} had carefully studied the impact of mass transfer on the orbital dynamics.
In these SPH simulations the feedback on the orbit is accurately modelled, but due to SPH's automatic 
``refinement  on density'' the properties of the transferred matter are not well resolved.
Therefore,  a ``best-of-both-worlds'' approach was followed in \cite{guillochon10}/\cite{dan11} where 
the  impact of the mass transfer on the orbital dynamics was simulated with SPH, while recording
the orbital evolution and the mass transfer rate. This information was used in a second set of 
simulations that was performed with the FLASH code \citep{fryxell00}. This second study 
focussed on the detailed hydrodynamic interaction of the transferred mass with the accretor star. For 
He-CO binary systems where helium directly impacts on a primary of a mass $>$ 0.9 \msun,
they found that helium surface explosions can be self-consistently triggered via Kelvin--Helmholtz (KH)
instabilities. These instabilities occur at the interface between the incoming helium stream 
and an already formed helium torus around the primary. ``Knots'' produced by the KH instabilities can
lead to local ignition points once the triple-alpha time scale becomes shorter than the 
local dynamical time scale. The resulting detonations travel around the primary surface 
and collide on the side opposite to the ignition point. Such helium surface detonations
may resemble weak type Ia SNe \citep{bildsten07,foley09,perets10} and they may
drive shock waves into the CO core which concentrate in one or more focal 
points, similar to what was found in the 2D study of \cite{fink07}. This 
could possibly lead to an explosion via a ``double-detonation'' mechanism.
In a subsequent large-scale parameter study \citep{dan12,dan14a} found,
based on a comparison between nuclear burning and hydrodynamical time scales,
that a large fraction of the helium-accreting systems do produce explosions early on: 
all dynamically unstable systems with primary masses $< 1.1$ \Msun together with 
secondary masses $>0.4$ \Msun triggered helium-detonations at surface contact. A 
good fraction of these systems could also produce in addition KH-instability-induced 
detonations as described in detail in \cite{guillochon10}. 
There was no definitive evidence for explosions prior to  contact for any of the 
studied CO-transferring systems.

\subsubsection*{Explosions during merger}

\cite{pakmor10} studied double degenerate mergers, but
-- contrary to earlier
studies-- they focussed on very massive WDs with masses close to 0.9 \msun.
They used the \textsc{Gadget} code \citep{springel05a}  with some modifications \citep{pakmor12a}, for example, the 
Timmes EOS \citep{timmes99} and a 13 isotope network were implemented for their study.  
In order to facilitate the network implementation, the energy equation (instead of,
as usually in \textsc{Gadget}, the entropy equation) was evolved. No efforts were undertaken
to reduce the constant, untriggered artificial viscosity. They placed the stars on orbit with the
approximate initial conditions described above and found the secondary to be disrupted 
within two orbital periods. In a second step, several hot ($>2.9 \times 10^9$K) particles
were identified and the remnant was artificially ignited in these hot spots. The explosion
was followed with a grid-based hydrodynamics code \citep{fink07,roepke07} that had
been used in earlier SN~Ia studies. In a third step, the nucleosynthesis was post-processed
and synthetic light curves were calculated \citep{kromer09}. The explosion resulted in 
a moderate amount of \Nifs (0.1 \msun), large amounts (1.1 \msun) of intermediate 
mass elements and oxygen (0.5 \msun) and less than 0.1 \Msun of unburnt
carbon. The kinetic energy of the explosion ($1.3 \times 10^{51}$ erg) was typical for 
a SN~Ia, but the resulting velocities were relatively small, so that the explosion
resembled a sub-luminous 1991bg-like supernova. An important condition for reaching ignition 
is  a mass ratio close to unity. Some variation in total mass is expected, but cases with less 
than 0.9 \Msun of a primary mass would struggle to reach the ignition
temperatures and -- if successful -- 
the lower densities would lead to even lower resulting \Nifs   masses and therefore lower luminosities. 
For substantially higher masses, in contrast, the burning would proceed at larger densities
and therefore result in  much larger amounts of \nifs. Based on population synthesis
models \citep{ruiter09} they estimated that mergers of this type of system could account for
2\,--\,11 \% of the observed SN~Ia rate.
In a follow up study \citep{pakmor11} the sensitivity of the proposed model to the mass ratio
was studied. The authors concluded that binaries with a primary mass near 0.9 \Msun ignite
a detonation immediately at contact, provided that the mass ratio $q$ exceeds 0.8. Both the 
abundance tomography  and the lower-than-standard velocities provided support for
the idea of this type of merger producing sub-luminous, 1991bg-type supernovae.

As a variation of the theme, \cite{pakmor12b} also explored the merger of a higher 
mass system with 0.9 and 1.1 \msun. Using the
same assumption about the triggering of detonations as before, they found a substantially
larger mass of \Nifs (0.6 \msun), 0.5 \Msun of intermediate mass elements, 0.5 \Msun of Oxygen 
and about 0.15 \Msun of unburnt carbon. Due to its higher density, only the primary is able
to burn \Nifs and therefore the brightness of the SN~Ia would be closely related to the primary
mass. The secondary is only incompletely burnt and thus provides the bulk of the intermediate mass 
elements. Overall, the authors concluded that such a merger reproduces the observational 
properties of normal SN~Ia reasonably well.
In \cite{kromer13}  the results of a 0.9 and 0.76 \Msun CO-CO merger were analyzed
and unburned oxygen close to the centre of the ejecta was found which produces narrow emission
lines of [O1] in the late-time spectrum, similar to what is observed in the sub-luminous 
SN~2010lp \citep{leibundgut93}.

\cite{fryer10} applied a sequence of computational tools to study the spectra that 
can be expected from a supernova triggered by a double-degenerate merger. Motivated by population synthesis
calculations, they simulated a CO-CO binary of 0.9 and 1.2 \Msun  with the SNSPH
code \citep{fryer06}. They assumed that the remnant would explode at the Chandrasekhar
mass limit, into a gas cloud consisting of the remaining merger debris. They found a density
profile with $\rho \propto r^{-4}$,   inserted an explosion
\citep{meakin09} into such a matter distribution and calculated signatures with the radiation-hydrodynamics 
code \textsc{Rage} \citep{fryer07,fryer09}.  They found that in such ``enshrouded''  SNe Ia the debris extends 
and delays the X-ray flux from a shock breakout and produces a signal that is closer to a SN Ib/c. 
Also the  V-band peak was extended and much broader than in a normal SN~Ia  with
the early spectra being dominated by CO lines only.  They concluded that, within their model, 
a CO-CO merger with a total mass $>1.5$ \Msun would not produce spectra and light curves 
that resemble normal type Ia supernovae.

One of the insights gained from the study of \cite{pakmor10} was that (massive) mergers 
with similar masses are more likely progenitors of SNe Ia than the mergers with larger mass differences that
were studied earlier. This motivated \cite{zhu13a} to perform a large parameter study where they
systematically scanned the parameter space from 0.4-0.4 \Msun up to 1.0-1.0 \msun. In their study,
they used the \textsc{Gasoline} code  \citep{wadsley04} together with the Helmholtz EOS \citep{timmes99,timmes00a}, 
no nuclear reactions were included. All their simulations were performed with non-spinning stars
and approximate initial conditions as outlined above. Mergers with ``similar'' masses produced a 
well-mixed, hot central core while ``dissimilar'' masses produced a rather unaffected cold core 
surrounded by a hot envelope and disk, consistent with earlier studies.  They found that the central density 
ratio of the accreting and donating star of $\rho_a/\rho_d > 0.6$ is a good criterion for those 
systems that produce hot cores (i.e., to define ``similar'' masses).

\cite{dan14a}  also performed a very broad scan of the parameter space. They studied 
the temperature distribution inside the remnant for different stellar spins: tidally locked initial conditions 
produce hot spots (which are the most likely locations for detonations to be initiated) in the 
outer layers of the core, while irrotational
systems produce them deep inside of the core,  consistent with the results of \cite{zhu13a}. 
Thus, the spin state of the WDs may possibly 
be decisive for the question where the ignition is triggered which may have its bearings on the resulting
supernova. Dan et al. found essentially no chemical mixing between the stars for mass ratios below
$q \approx 0.45$, but maximum mixing for a mass ratio of unity. Contrary to \cite{zhu13a},
nowhere complete mixing was found, but this difference can be convincingly attributed to
the different stellar spins that were investigated (tidal locking in \citealp{dan14a}, no spins in \citealp{zhu13a}).
In addition to the helium-accreting systems that likely undergo a detonation for a total mass beyond 
1.1 \Msun, they also found CO binaries with total masses beyond 2.1 \Msun to
be prone to a CO explosion. Such systems may be candidates for the so-called super-Chandrasekhar
SN~Ia explosions.  They also  discussed the possibility of ``hybrid supernovae''
where a ONeMg core with a significant helium layer  collapses and forms a neutron star. While
technically being a (probably weak) core-collapse supernova 
\citep{podsiadlowski04,kitaura06}, most of the explosion energy may come from helium
burning. Such hybrid supernovae may be candidates for the class of ``Ca-rich'' SNe Ib
\citep{perets10} as the burning conditions seem to favor the production of intermediate mass
elements.

\cite{raskin12a} explored 10 merging WDWD binary systems with total masses between 
1.28 and 2.12 \Msun with  the SNSPH code  \citep{fryer06} coupled to the Helmholtz EOS and a 13-isotope 
nuclear network. They used a heuristic procedure to construct tidally deformed, synchronized binaries
by letting the stars fall towards each other in free fall, and repeatedly set the fluid velocities to zero.
They had coated their CO WDs with atmospheres of helium (smaller than 2.5\% of the total mass) and found 
that for all cases where the primary had a mass of 1.06 \msun, the helium detonated,  no Carbon detonation 
was encountered, though.

Given the difficulty to identify the central engine of a SNe~Ia on purely theoretical grounds, \cite{raskin13a} explored possible observational signatures stemming from the tidal tails of a WD 
merger. They followed the ejecta from a SNSPH simulation
\citep{fryer06} with n-body methods and -- 
assuming spherical symmetry -- they explored by means of a 1D Lagrangian code how a supernova would
interact with such a medium. Provided the time lag between merger and supernova is short enough ($<100$ s), 
detectable shock emission at radio, optical, and/or X-ray wavelengths is expected. For delay times 
between $10^8$~s and 100 years one expects broad NaID absorption features, and, since this has not been 
observed to date, they concluded that if (some) type Ia supernovae are indeed caused by WDWD mergers, the delay 
times need to be either short ($<100$ s) or rather long ($>100$ years). If the tails can expand and cool for 
$\sim 10^4$ years, they produce the observable narrow NaID and Ca II K\& K lines which are seen in some 
fraction of type Ia supernovae.

An interesting study from a Santa Cruz--Berkeley collaboration \citep{moll14a,raskin14a} combined again
the strengths of different numerical methods. The merger process was calculated with the SNSPH code
\citep{fryer06}, the subsequent explosion with the grid-based code 
\textsc{Castro} \citep{almgren10,zhang11} and synthetic light curves and spectra were calculated with \textsc{Sedona}  \citep{kasen06}.
For some cases without immediate explosion the viscous remnant evolution was followed further with ZEUS-MP2 \citep{hayes06}. 
The first part of the study \citep{moll14a} focussed on prompt detonations, while the second part explored 
the properties of detonations that emerge in later phases, after the secondary has been completely disrupted. 
For the first part, three mergers (1.20 - 1.06, 1.06 - 1.06 and 0.96 - 0.81 \msun) were simulated with
 the SPH code and subsequently mapped into \textsc{Castro}, where soon after simulation start ($<0.1$ s)
detonations emerged. They found the best agreement with common SNe Ia  (0.58 \Msun of \nifs) for the
binary with 0.96 - 0.81 \msun. More massive systems  lead to 
more \Nifs and therefore unusually bright SNe Ia. The remnant asymmetry at the moment of detonation leads
to large asymmetries in the elemental distributions and therefore to strong viewing angle effects for the resulting supernova. 
Depending on viewing angle, the peak bolometric luminosity varied by a factor of two and the flux in the ultraviolet 
even varied by an order of magnitude. All of the three models approximately fulfilled the width-luminosity 
relation (``brighter means broader''; \citealp{phillips99}).

The companion study \citep{raskin14a} explored cases where the secondary has become completely disrupted before a detonation
sets in, so that the primary explodes into a disk-like CO structure. To initiate detonations, the SPH simulations were
stopped once a stationary structure had formed, all the material with $\rho>10^6$ \Gcc was burnt in a separate simulation 
and, subsequently, the generated energy was deposited back as thermal energy into the merger remnant and the SPH simulation 
was resumed. As a double-check of the robustness of this approach, two simulations were also mapped into \textsc{Castro} and detonated
there as well. Overall there was good agreement both in morphology and the nucleosynthetic yields. The explosion 
inside the disk produced an hourglass-shaped  remnant geometry with strong viewing angle effects. The 
disk scale height, initially set by the mass ratio and the  different merger dynamics and burning processes,
turned out to be an important factor for the viewing angle dependence of the later supernova. The other 
crucial factor was the primary mass that  determines the resulting amount of
\nifs. Interestingly, the location of the detonation spot, whether at the surface or in the core of the primary, 
had a relatively small effect compared to the presence of an accretion disk. While qualitatively in agreement
with the width-luminosity relation, the lightcurves lasted much longer than 
standard SNe Ia. The surrounding CO disk from the secondary remained essentially unburnt, but, impeding the expansion,
it lead to relatively small intermediate mass element absorption velocities. The large asymmetries in the abundance distribution
could lead to a large overestimate of the involved \Nifs masses if spherical symmetry is assumed in interpreting observations.
The lightcurves and spectra were peculiar with weak features 
from intermediate mass elements but relatively strong carbon absorption. The study also explored how 
longer-term viscous evolution before a detonation sets in would affect the supernova. Longer delay times 
were found to produce likely larger \Nifs masses and more symmetrical remnants.  Such systems might be 
candidates for super-Chandra SNe Ia. 

\subsubsection{Simulations of white dwarf\,--\,white dwarf collisions}
\label{sec:WDWD_collisions}

A physically interesting alternative to gravitational wave-driven binary mergers are dynamical 
collisions between two WDs as they are expected in globular clusters and galactic cores.
They have first been explored, once more, by \cite{benz89}, probably
in one of the first hydrodynamic simulations of WDs that included a nuclear reaction 
network. At that time they found that nuclear burning could, in central collisions,
help to undbind a substantial amount of matter, but this amount was found to be
irrelevant for the chemical evolution of galaxies. More recently, the topic has been taken up 
again by \cite{rosswog09c} and, independently, by 
\cite{raskin09}. Both of these studies employed SPH simulations ($2 \times 10^6$
SPH particles in \cite{rosswog09c}, $8 \times 10^5$ particles in \cite{raskin09})
coupled to small nuclear reactions networks. Both groups concluded that an amount of radioactive
\Nifs could be produced that is comparable to what is observed in normal type Ia supernovae
($\sim 0.5$ \msun). In \cite{rosswog09c} one of the simulations was repeated 
within the \textsc{Flash} code \citep{fryxell00} to judge the robustness of the result and
given the enormous sensitivity of the nuclear reactions to temperature
overall good agreement was found, see Figure~\ref{fig:WDWD_collision}. 
Rosswog et al. also post-processed the nucleosynthesis with larger nuclear networks
and calculated synthetic light curves and spectra  with the \textsc{Sedona} code \citep{kasen06}.
Interestingly, the resulting light-curves and spectra \citep{rosswog09c}
look like normal type Ia supernova, even the width-luminosity relationship \citep{phillips99}
is fulfilled to good accuracy. These results have been confirmed recently in further studies
\citep{raskin10,hawley12,kushnir13,garcia_senz13}.

\epubtkImage{}{%
  \begin{figure}[htb]
   \centerline{\includegraphics[width=13cm,angle=0]{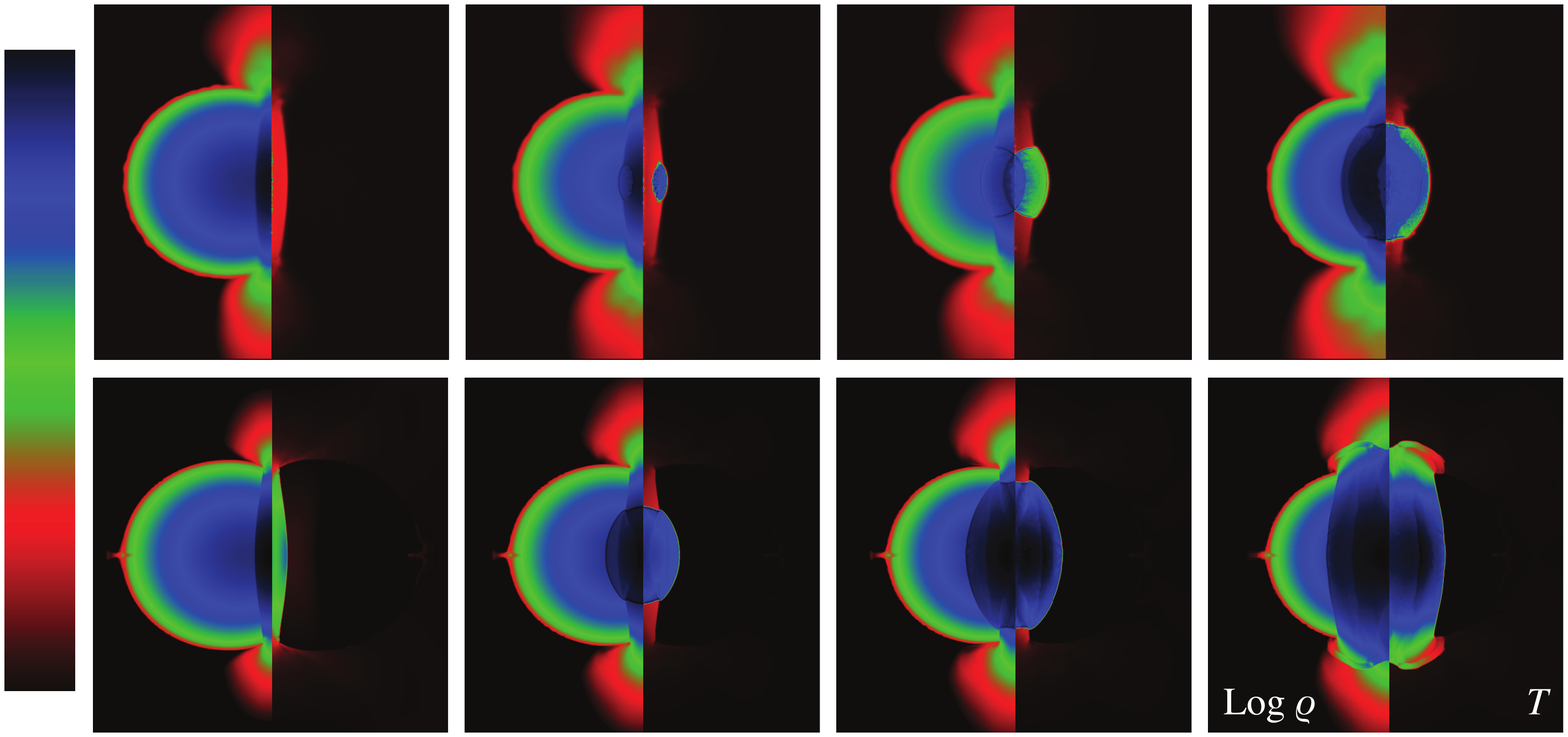}}

\vspace*{-1cm}

    \caption{Comparison of the head-on collision of two WDs once calculated with an SPH code \citep{rosswog08b} (upper row) and once with the Adaptive Mesh Refinement code \textsc{Flash} \citep{fryxell00} (lower row). Each of the panels is split into density (left) and temperature (right). Given the enormous temperature sensitivity of explosive nuclear burning the agreement is remarkably good, the produced energy from the SPH calculation is $\log(E_{\rm erg})= 51.21$ and  $\log(E_{\rm erg})= 51.11$ for the \textsc{Flash} simulation. Figure taken from \cite{rosswog09c}.}
   \label{fig:WDWD_collision}
\end{figure}}

This result is interesting for a number of reasons. First, the detonation mechanism
is parameter-free and extremely robust: the free-fall velocity between WDs naturally
produces relative velocities in excess of the WD sound speeds, therefore strong shocks
are inevitable. Moreover, the most likely involved WD masses are near the peak of the mass
distribution, $\sim 0.6$ \msun, or, due to mass segregation effects, possibly slightly larger, but well
below the Chandrasekhar mass. This has the benefit that the nuclear burning occurs
at moderate densities ($\rho \sim 10^7$ \gcc) and thus produces naturally the observed
mix of $\sim 0.5$ \Msun  \Nifs and intermediate-mass elements, without any fine-tuning
such as the deflagration-detonation transition that is required in the single-degenerate
scenario \citep{hillebrandt00}. However, based on simple order of magnitude estimates,
the original studies \citep{raskin09,rosswog09c} concluded that, while being very interesting
explosions, the rates would likely be too low to make a substantial contribution to the observed 
supernova sample. More recently, however, there have been claims \citep{thompson11,kushnir13}
that the Kozai--Lidov mechanism in triple stellar systems may substantially boost the rates of 
WDWD collisions so that they could constitute a sizeable fraction of the SN~Ia rate. Contrary
to these claims, a recent study by \cite{hamers13} finds  that the 
contribution from the triple-induced channels to SN~Ia is small.  Here further studies are
needed to quantify how relevant collisions really are for explaining normal SNe Ia.

\subsubsection{Summary double white dwarf encounters}

SPH has very often been used to model mergers of, and later on also collisions between, two WDs.
This is mainly since SPH is not restricted by any predefined geometry and has excellent conservation
properties. As illustrated in the numerical experiment shown in Figure~\ref{fig:conservation_GWs},
even small (artificial) losses of angular momentum can lead to very large errors in the prediction
of the mass transfer duration and the gravitational wave signal. SPH's tendency to ``follow the density'' 
makes it ideal to predict, for example, the gravitational wave signatures of WD mergers. But it is 
exactly this tendency which makes it very difficult for SPH to study
thermonuclear ignition processes in low-density regions that are very important, for example, for the 
double-detonation scenario.  This suggests to apply in such cases a combination of numerical tools: 
SPH for bulk motion and orbital dynamics and Eulerian (Adaptive Mesh) hydrodynamics for 
low-density regions that need high resolution. As 
outlined above, there have recently been a number of successful studies that have followed such strategies.

SPH simulations have played a pivotal role in ``re-discovering'' the importance of white dwarf mergers
(and possibly, to some extent, collisions) as progenitor systems of SNe Ia. In the last few years, a number
of new possible pathways to thermonuclear explosions prior to or during a WD merger have been discovered.
There is, however, not yet a clear consensus whether they produce just ``peculiar'' SN~Ia-like events or whether
they may be even responsible for the bulk of ``standard'' SN~Ia. Here, a lot of progress can be expected within
the next few years, both from the  modelling and the observational side.

\subsection{Encounters between neutron stars and black holes}
\label{sec:appl_NSNS_NSBH}

\subsubsection{Relevance}

The relevance of compact binary systems consisting of  two neutron stars (NS) or a neutron star
and a stellar mass black hole (BH) is hard to overrate. The first observed double neutron star system,
PSR~1913+16, proved, although indirectly, the existence of gravitational waves (GWs): the orbit decays in
excellent agreement with the predictions from general relativity (GR) \citep{taylor89,weisberg10}. By
now there are 10 binary systems that are thought to consist of two neutron stars \citep{lorimer08}, but
to date no NSBH system has been identified, although both types of systems should form in 
similar evolutionary processes.  This is usually attributed to  possibly smaller numbers in 
comparison to NS-NS systems \citep{belczynski09} and to a lower detection probability in current surveys. 
Once formed, gravitational wave  emission drives compact binary systems towards coalescence on a time
scale approximately given by \cite{lorimer08}
\be
\tau_{\mathrm{GW}}= 9.83 \times 10^6\mathrm{\ years} \left(\frac{P_{\rm orb}}{\rm hr}\right)^{8/3} \left(\frac{M}{M_{\odot}} 
\right)^{-2/3} \left(\frac{\mu}{M_{\odot}} \right)^{-1}     
           (1-e^2)^{7/2},
\ee 
where $P_{\rm orb}$ is the current orbital period, $M$ the total and $\mu$ the reduced mass and $e$ the eccentricity.
This implies that the initial orbital period must be $\leq 1$ day for a coalescence to occur within a Hubble
time. As the inspiral time depends sensitively on the orbital period and eccentricity, which are set by individual
evolutionary histories, one expects a large spread in inspiralling times. 
Near its innermost stable circular orbit (ISCO) a compact binary system emits gravitational waves at a frequency and amplitude 
of
\bea
f &\approx& 594\mathrm{\ Hz} \left( \frac{6 G M}{c^2 a}\right)^{3/2} \left( \frac{7.4\,M_{\odot}}{M} \right)\label{eq:GW_frequ}\\
h &\approx& 3.6 \times 10^{-22} \left( \frac{m_{\rm 1}}{6\,M_{\odot}} \right) \left( \frac{m_{\rm 2}} {1.4\,M_{\odot}} \right) 
                                                       \left( \frac{6 G M}{c^2 a} \right) \left( \frac{100\mathrm{\ Mpc}}{r} \right),\label{eq:GW_ampl}
\eea
where $a$ is the binary separation, $r$ the distance to the GW source and
the scalings are oriented at a  NSBH system. Thus the emission from the  late inspiral stages will sweep through
the frequency bands of the advanced detector facilities from $\sim 10$ to $\sim 3000$ Hz \citep{ligo,virgo}
making compact binary mergers (CBMs) the prime targets for the ground-based gravitational wave detections. The estimated
coalescence rates are rather uncertain, though, they range from 1 - 1000 Myr$^{-1}$ MWEG$^{-1}$ for NSNS binaries and from
0.05 - 100 Myr$^{-1}$ MWEG$^{-1}$ for NSBH binaries \citep{abadie10}, where MWEG stands for 
``Milky Way Equivalent Galaxy''.

In addition, CBMs have long been suspected to be the engines of short Gamma-ray Bursts (sGRBs) 
\citep{paczynski86,goodman86,eichler89,narayan92}. 
Although already their projected distribution on the sky and their fluence distribution pointed to a 
cosmological source \citep{goodman86,paczynski86,schmidt88,meegan92,piran92,schmidt01,guetta05},
their cosmological origin was only firmly established by the first afterglow observations for short bursts in 2005 \citep{hjorth05,bloom06}.
This established the scale for both distance and energy and proved that sGRBs occur in both early- and 
late-type galaxies. CBMs are natural candidates for sGRBs since accreting compact objects are very 
efficient converters of gravitational energy into electromagnetic radiation, they occur 
at rates that are consistent with those of sGRBs \citep{guetta05,nakar06,guetta06} and CBMs are expected 
 to occur in both  early and late-type galaxies.  Kicks imparted at birth provide a natural 
explanation for the observed projected offsets of $\sim$ 5 kpc from their host galaxy \citep{fong13}, and, 
with dynamical time scales of $\sim1$ ms (either orbital at the ISCO or the neutron star oscillation
time scales), they naturally provide the observed short-time fluctuations. Moreover, for cases where an accretion torus
forms, the expected viscous lifetime is  roughly comparable with a typical sGRB duration ($\sim 0.2$ s).
While this picture is certainly not without open questions, see \cite{piran04,lee07,nakar07,gehrels09,berger11,berger14a}
for recent reviews, it has survived the confrontation with three
decades of observational results and -- while 
competitors have emerged -- it is still the most commonly accepted model for the engine of short GRBs.

\cite{lattimer74,lattimer76} and \cite{lattimer77} suggested that the decompression 
of initially cold neutron star matter could lead to rapid neutron capture, or ``r-process'',  
nucleosynthesis so that CBMs may actually also be an important source of heavy elements. Although discussed 
convincingly in a number of subsequent publications \citep{symbalisty82,eichler89}, this idea kept
the status of an exotic alternative to the prevailing idea that the heaviest elements are
formed in core-collapse supernovae, see \cite{arcones13a} for a recent review. Early SPH 
simulations of NSNS mergers \citep{rosswog98c,rosswog99} showed that of order $\sim$~1\% of 
neutron-rich material is ejected per merger event, enough to be a substantial or even the major source of
r-process. A nucleosynthesis post-processing of these SPH results \citep{freiburghaus99b} 
confirmed that this material  is indeed a natural candidate for the robust, heavy r-process \citep{sneden08}. 
Initially it was doubted \citep{qian00,argast04} that CMBs as main r-process source are consistent 
with galactic chemical evolution, but a recent study based on a detailed chemical evolution model 
\citep{matteucci14} finds room for a substantial contribution of neutron star mergers and recent
hydrodynamic galaxy formation studies \citep{shen14a,vandevoort14a} even come to the conclusion 
that neutron star mergers are consistent with being the dominant r-process source in the Universe.
Moreover,  state-of-the-art supernova models seem unable to provide the conditions that are needed to 
produce heavy elements with nucleon numbers in excess of $A=90$ \citep{arcones07,fischer10,
huedepohl10,roberts10}.\epubtkFootnote{A possible exception may be magnetically driven explosion
of rapidly rotating stars \citep{winteler12}.} On the other hand, essentially all recent studies agree
that CBMs eject enough mass to be at least a major r-process source 
\citep{oechslin07a,bauswein13a,rosswog14a,hotokezaka13a,kyutoku13} and that the resulting abundance 
pattern beyond $A\approx 130$ resembles the solar-system 
distribution \citep{metzger10b,roberts11,korobkin12a,goriely11a,goriely11b,eichler15}.
Recent studies \citep{wanajo14,just14} suggest that compact binary mergers with their different ejection channels for
neutron-rich matter could actually even be responsible for the whole range of r-process.

In June 2013 the SWIFT satellite detected a relatively nearby ($z=0.356$) sGRB, GRB~130603B, \citep{melandri13} for 
which the HST \citep{tanvir13a,berger13a} detected 9 days after the burst a nIR point source with properties
close to what had been predicted \citep{kasen13a,barnes13a,grossman14a,tanaka13a,tanaka14a} by models for 
``macro-'' or ``kilonovae'' \citep{li98,kulkarni05,rosswog05a,metzger10a,metzger10b,roberts11}, 
radioactively powered transients from the decay of freshly produced r-process elements. If this 
interpretation is correct, GRB~130603B would provide the first observational confirmation of the 
long-suspected link between CBMs, nucleosynthesis and Gamma-Ray Bursts.

\subsubsection{Differences between double neutron and neutron star black hole mergers}

Both NSNS and NSBH systems share the same three stages of the merger: a) the \emph{secular inspiral} where the mutual
separation is much larger than the object radii and the orbital evolution can be very accurately  described by
Post-Newtonian methods \citep{blanchet06}, b) the \emph{merger phase} where relativistic hydrodynamics is important and
c) the subsequent \emph{ringdown phase}. Although similar in their formation paths and in their relevance, there are a 
number of differences between NSNS and NSBH systems. For example, when the surfaces of two neutron stars come 
into contact the neutron star matter can heat up and radiate neutrinos. Closely related, if matter from the interaction 
region is dynamically ejected, it may -- due to the large temperatures
-- have the chance to change its electron
fraction due to weak interactions. Such material may have  a different nucleosynthetic signature than the matter
that is ejected during a NSBH merger. Moreover, the interface between two neutron stars is prone to hydrodynamic
instabilities that can amplify existing neutron star magnetic fields \citep{price06,anderson08b,rezzolla11,zrake13}.
The arguably largest differences between the two types of mergers, however,  are the total binary mass and its mass ratio
$q$. For neutron stars, the deviation from unity $|1-q|$ is small (for masses that are known to better than 0.01 \msun,
J1807-2500B has the largest deviation with a mass ratio of $q=0.88$ \citep{lattimer12a}) while for NSBH systems a broad
range  of total masses is expected. Since the merger dynamics is very sensitive
to the mass ratio, a much larger diversity is expected in the dynamical behavior of NSBH systems.
The larger bh mass has also as a consequence that the plunging phase of the orbit sets in at larger separations and 
therefore lower GW  frequencies, see Eq.~(\ref{eq:GW_frequ}). Moreover, for black holes the dimensionless spin parameter can be close to unity,
$a_{\mathrm{BH}}= c J/G M^2 \approx 1$, while for neutron stars it is restricted by the mass shedding limit to somewhat lower values 
$a_{\mathrm{NS}} < 0.7$ \citep{lo11}, therefore spin-orbit coupling effects could potentially be larger for NSBH systems and lead to 
observable effects e.g., \cite{buonanno03,grandclement04}.

\subsubsection{Challenges}

Compact binary mergers are challenging to model since in principle each of the fundamental interactions
enters: the strong (equation of state),  weak (composition, neutrino emission, nucleosynthesis), electromagnetic (transients,
neutron star crust) and of course (strong) gravity.
 Huge progress has been achieved during the last decade, but so far none of the approaches ``has
it all'' and, depending on the focus of the study, different compromises have to be accepted.

The compactness parameters, $\mathcal{C}\equiv GM/R c^2$, of $\sim 0.2$ for neutron
stars and $\sim 0.5$ for black holes suggest a general-relativistic treatment. 
Contrary to the WDWD case discussed in Section~\ref{sec:appl_WDWD} now the gravitational 
and orbital time scales can become comparable
\be
\frac{\tau_{\rm GW}}{P_{\rm orb}}= \frac{a/|\dot{a}_{\rm GW}|}{2\pi/\omega_{\rm K}}=
5.0 \left( \frac{a}{           6 G (M/2.8\,M_{\odot})/c^2                    }\right)^{5/2} \left(\frac{1.4\,M_{\odot}}{m_1} \right) 
\left( \frac{1.4\,M_{\odot}}{m_2}\right) \left( \frac{2.8\,M_{\odot}}{M}\right)^{1/2},
\label{eq:tau_GW_tau_orb}
\ee
so that backreaction from the gravitational wave emission on the dynamical evolution can no more be
safely neglected. And, unless one considers the case where the black hole completely dominates
the spacetime geometry, it is difficult to make admissible
approximations to dynamical GR gravity.

Of comparable importance is the neutron star equation of state (EOS). Together with gravity it determines the compactness
of the neutron stars which in turn impacts on peak GW frequencies. It also influences the torus masses, the amount of 
ejected matter, and, since it sets the $\beta$-equilibrium value of the electron fraction $Y_e$ in neutron star matter, also the resulting 
nucleosynthesis and the possibly emerging electromagnetic transients. Unfortunately, the EOS is not well known 
beyond $\sim$ 3 times nuclear density. On the other hand, since the EOS seriously impacts on a number of 
observable properties from a CBM, one can be optimistic and hope to conversely constrain the high-density EOS via 
astronomical observations.

Since the bulk of a CBM remnant consists of high density matter ($\rho>10^{10}$\gcc), photons are very efficiently trapped and
the only cooling agents are neutrinos. Moreover, with temperatures of order MeV, weak interactions become so fast
that they change the electron fractions $Y_e$ substantially on a dynamical time scale ($\sim$~1 ms). While such effects can be safely
ignored when gravitational wave emission is the main focus, these processes are crucial for the neutrino signature,  for the 
``engine physics'' of GRBs (e.g., via $\nu\bar{\nu}$ annihilation), and for nucleosynthesis, since the nuclear 
reactions are very sensitive to the neutron-to-proton ratio which is set by the weak interactions. Fortunately, the treatment of neutrino 
interactions is not as delicate as in the core-collapse supernova case where changes on the percent level can decide 
between a successful and a failed  explosion \citep{janka07}. Thus, depending on the exact focus, leakage schemes or 
simple transport approximations may be admissible in the case of compact binary mergers. But also approximate treatments
face the challenge that the remnant is intrinsically three-dimensional, that the optical depths change from $\tau \sim 10^4$ 
in the hypermassive neutron star (e.g., \cite{rosswog03a}), to essentially zero in the outer regions of the disk and that the 
neutrino-nucleon interactions are highly energy-dependent.

Neutron stars are naturally endowed with strong magnetic fields and a compact binary merger offers a wealth of possibilities 
to amplify them. They may be decisive for the fundamental mechanism to produce a GRB in the first place, but they may
also determine -- via transport of angular momentum -- when the central object produced in a NSNS merger collapses into a 
black hole or how accretion disks evolve further under the transport mediated via 
the magneto-rotational instability (MRI) \citep{balbus98}.

In addition to these challenges from different physical processes, there are also purely numerical challenges which are, however,
closely connected to the physical ones. For example, the length scales that need to be resolved  to follow the growth
of the MRI can be minute in comparison to the natural length scales of the problem. Or, another example, the viscous dissipation
time scale of an accretion disk,
\be
\tau_{\rm visc} \sim 0.3\mathrm{\ s} \;  \left(\frac{0.1}{\alpha}\right)  \left( \frac{r}{300\mathrm{\ km}} \right)^{3/2}  
\left(\frac{3\,M_{\odot}}{M}\right)^{1/2} \left( \frac{r/H}{2} \right)^2
\label{eq:tau_visc}
\ee
with $\alpha$ being the Shakura--Sunyaev viscosity parameterization \citep{shakura73}, $M$ the central mass, 
$r$ a typical disk radius and $H$ the disk scale height, may be challengingly long in comparison to the hydrodynamic 
time step that is allowed by the CFL stability criterion \citep{press92},  
\be
\Delta t < 10^{-5}\mathrm{\ s} \; \; \left(\frac{\Delta x}{1\mathrm{\ km}}\right) \; \left(\frac{0.3 \; {\rm c}}{c_{\rm s}}\right).
\label{eq:CFL}
\ee

\subsubsection{The current status of SPH- vs  grid-based simulations}
\label{sec:SPH_vs_Eulerian_NSNS_NSBH}

A lot of progress has been achieved in recent years in simulations of CBMs. This includes 
many different microphysical aspects as well as the dynamic solution of the Einstein equations.
The main focus of this review are SPH methods and therefore we will restrict the
detailed discussion to work that makes use of SPH. Nevertheless, it is worth briefly comparing 
the current status of SPH-based simulations with those that have been obtained with
grid-based methods. As will be explained in more detail below, SPH had from the beginning 
a very good track record with respect to the implementation of various microphysics ingredients. 
On the relativistic gravity side, however, it is lagging behind in terms of implementations 
that dynamically solve the Einstein equations, a task that has been achieved with Eulerian methods
already more than a decade ago \citep{shibata99,shibata00}. In SPH, apart from Newtonian gravity, 
Post-Newtonian and Conformal Flatness Approaches (CFA) exist, but up to now no coupling 
between SPH-hydrodynamics and a dynamic spacetime solver has been achieved.

Naturally, this has implications for the types of problems that have been addressed and there are
interesting questions related to nsns and nsbh mergers that have so far not yet (or only  approximately) 
been tackled with SPH approaches. One example with far-reaching astrophysical 
consequences is the collapse of a hypermassive neutron star (HMNS) that temporarily forms after a 
binary neutron star merger. Observations now indicate a lower limit on the maximum neutron star mass
of around 2.0 \Msun (1.97 $\pm$ 0.04 \Msun for PSR~J1614+2230, see \cite{demorest10} and 2.01 $\pm$ 0.04
\Msun for PSR~J0348+0432,  see \cite{antoniadis13}) and therefore it is very likely that the hot and 
rapidly differentially rotating central remnant of a nsns merger is at least temporarily stabilized against
a gravitational collapse to a black hole, see e.g., \cite{hotokezaka13b}. It appears actually entirely plausible that the low-mass end
of the nsns distribution may actually leave behind a massive but stable neutron star rather than a black 
hole. In SPH, the question when a collapse sets in, can so far only be addressed within the Conformal
Flatness Approximation (CFA), see below. Being exact for spherical symmetry, the CFA should be fairly 
accurate in describing the collapse itself. It is, however, less clear how accurate the CFA is during the last 
inspiral stages where the deviations from spherical symmetry are substantial. Therefore quantitative
statements about the HMNS lifetimes need to be interpreted with care. On the other 
hand, differential rotation has a major impact in the stabilisation and therefore hydrodynamic resolution is
also crucial for the question of the collapse time. Here, SPH with its natural tendency to refine on density
should perform very well once coupled to a dynamic spacetime solver.

Another question of high astrophysical significance is the amount of matter that becomes ejected into
space during a nsns or nsbh merger. It is likely one of the major sources of r-process in the cosmos and
thought to cause electromagnetic transients  similar
to the recently observed ``macronova'' event in the aftermath of GRB~130603B \citep{tanvir13a,berger13b}.
In terms of mass ejection, one could expect large differences between the results of fully relativistic, grid-based hydrodynamics
results and Newtonian or approximate GR SPH approaches. The expected differences are twofold. On the
one hand, Newtonian/approximate GR treatments may yield stars of a different compactness which in turn
would influence the dynamics and torques and therefore the ejecta amount.  But apart from gravity, SPH
has a clear edge in dealing with ejecta: mass conservation is exact, advection is exact (i.e., a composition 
only changes due to nuclear reactions but not due to numerical effects), angular momentum conservation
is exact and vacuum \emph{really} corresponds to the absence of matter. Eulerian schemes usually face here
the challenges that conservation of mass, angular momentum and the accuracy of advection are 
resolution-dependent and that vacuum most often is modelled as background fluid of lower density.
Given this rather long list of challenges, it is actually encouraging that the results of different groups with
very different methods agree reasonably well these days. For NSNS mergers Newtonian SPH simulations \citep{rosswog13b}
find a range from $8 \times 10^{-3} - 4 \times 10^{-2}$ \msun, approximate GR SPH calculations \citep{bauswein13a}
find a range from $10^{-3} - 2 \times 10^{-2}$ \Msun and full GR calculations (Hotokezaka et al. 2013) find 
$10^{-4} - 10^{-2}$ \msun.\epubtkFootnote{The fair comparison of these numbers is complicated by the fact that also
different equations of state and mass ratios have been used.} Even the results from Newtonian nsbh calculations 
agree quite well with the GR results (compare  Table~1 in \cite{rosswog13b} and the results of \cite{kyutoku13}).

Another advantage of SPH is that  one can also decide to just focus on the 
ejected matter. For example, a recent SPH-based study \citep{rosswog14a} has followed the evolution of the
dynamic merger ejecta for as long as 100 years, while Eulerian methods are usually restricted to very few 
tens of milliseconds. During this expansion the density was followed from supra-nuclear densities ($> 2 \times 10^{14}$
g/ccm) down to values that are below the interstellar matter density ($< 10^{-25}$ g/ccm).

For many of the topics that will be addressed below, there have been parallel efforts on the Eulerian side and
within the scope of this review we will not be able to do justice to all these parallel developments. As a starting 
point, we want to point to a number of excellent textbooks \citep{alcubierre08,baumgarte10,rezzolla13a} 
that deal with relativistic (mostly Eulerian) fluid dynamics and to various recent review articles 
\citep{duez10a,shibata11,faber12,pfeiffer12,lehner14a}.

\subsubsection{Neutron star \,--\,neutron star mergers}
\label{sec:appl_NSNS}

\subsubsection*{Early Newtonian calculations with polytropic equation of state}

The earliest NSNS merger calculations \citep{rasio92,davies94,zhuge94,rasio94,rasio95,zhuge96} were performed with Newtonian 
gravity and a polytropic equation of state, sometimes a simple gravitational wave backreaction force was added. While
these initial studies were, of course, rather simple, they set the stage for future efforts and settled questions about the 
qualitative merger dynamics and some of the emerging phenomena. For example, they established the emergence of
a Kelvin--Helmholtz unstable vortex sheet at the interface between the two stars, which, due to the larger shear, is 
more pronounced for initially non-rotating stars. Moreover, they confirmed the expectation that a relatively baryon-free 
funnel would form along the binary rotation axis \citep{davies94} (although this conclusion may need to be revisited in 
the light of emerging, neutrino-driven winds, see e.g., \cite{dessart09,perego14b,martin15}). These early simulations also established the 
basic morphological differences between tidally locked and irrotational binaries and between binaries of different 
mass ratios. In addition, these studies also drove technical developments that became very useful in later studies
such as the relaxation techniques  to construct synchronized binary systems \citep{rasio94} or the procedures to 
analyze the GW energy spectrum in the frequency band \citep{zhuge94,zhuge96}. See also \cite{rasio99} for a review
on earlier research.

\subsubsection*{Studies with focus on microphysics}

Studies with a focus on microphysics (in a Newtonian framework) were pioneered by \cite{ruffert96} 
who implemented a nuclear equation of state and a neutrino-leakage scheme into their Eulerian (PPM) hydrodynamics code.
In SPH, the effects of a nuclear equation of state (EOS)  were first explored in \cite{rosswog99}. The authors implemented
the Lattimer--Swesty EOS \citep{lattimer91} and neutrino cooling in the simple free-streaming limit to bracket the
effects that neutrino emission may possibly have.  To avoid artefacts from excessive artificial dissipation, they also included 
the time-dependent viscosity parameters suggested by \cite{morris97}, see Section~\ref{sec:Newtonian_shocks}. 
They found torus masses between 0.1 and 0.3 \msun, and, maybe most importantly, that between $4 \times 10^{-3}$ 
and $4 \times 10^{-2}$ \Msun of neutron star matter becomes ejected. A companion paper \citep{freiburghaus99b} 
post-processed trajectories from this study and found that all the matter undergoes r-process and yields an abundance 
pattern close to the one observed in the solar system, provided that the initial electron fraction is $Y_e\approx 0.1$. 
A subsequent study \citep{rosswog00} explored the effects of different initial stellar spins and mass ratios $q\neq 1$
on the ejecta masses.

The simulation ingredients were further refined in \cite{rosswog02a,rosswog03a,rosswog03c}. Here the \cite{shen98a,shen98b} EOS, extended down to very low densities, was used and a detailed multi-flavor neutrino leakage scheme was
developed \citep{rosswog03a} that takes particular care to avoid using average neutrino  energies. These studies were also performed
at a substantially higher resolution (up to $10^6$ particles) than previous SPH studies of the topic. The typical neutrino luminosities
turned out to be $\sim 2 \times 10^{53}$ erg/s  with typical average neutrino energies of 8/15/20 MeV for $\nu_e, \bar{\nu}_e$ and
the heavy lepton neutrinos. Since GRBs were a major focus of the studies,  neutrino annihilation was calculated in a post-processing
step and, barring possible complications from baryonic pollution, it was concluded that $\nu \bar{\nu}$ annihilation should lead to 
relativistic ouflows and could produce moderately energetic sGRBs. Simple estimates indicated, however, that strong 
neutrino-driven winds are likely to occur, that could, depending on the wind geometry, pose a possible threat for the emergence
of ultra-relativistic outflow/a sGRB. A more recent, 2D study of the neutrino emission from a merger remnant \citep{dessart09}
found indeed strong baryonic winds with mass loss rates $\dot{M} \sim 10^{-3}$ \msun/s emerging along the binary rotation axis. 
Recently, studies of neutrino-driven winds have been extended to 3D \citep{perego14b} and the properties of the blown-off material have been
studied in detail.
This complex of topics, $\nu-$driven winds, baryonic pollution, collapse of the central merger remnant will for sure
receive more attention in the future. Based on simple arguments, it was also argued that any initial seed magnetic fields 
should be amplified to values near equipartition, making magnetically launched/aided outflow likely. Subsequent studies 
\citep{price06,anderson08b,rezzolla11,zrake13} found indeed a strong amplification of initial seed magnetic fields.

In a recent set of studies the neutron star mass parameter space was scanned by exploring systematically mass
combinations from 1.0 to 2.0 \Msun \citep{korobkin12a,rosswog13a,rosswog13b}. The main focus here was the dynamically 
ejected mass and its possible observational signatures. One interesting result \citep{korobkin12a} was that the nucleosynthetic
abundance pattern is  essentially identical for the dynamic ejecta of all mass combinations and even NSBH systems yield
practically an identical pattern. While extremely robust
to a variation of the astrophysical parameters, the pattern showed some sensitivity to the involved nuclear physics, for example
to a change of the mass formula or the distribution of the fission fragments. The authors concluded that the dynamic ejecta of
neutron star mergers are excellent candidates for the source of the heavy, so-called ``strong r-process'' that is observed in 
a variety of metal-poor stars and shows each time the same relative abundance pattern for nuclei beyond barium \citep{sneden08}. 
Based on these results, predictions were made for the resulting ``macronovae'' (or sometimes called ``kilonovae'') 
\citep{li98,kulkarni05,rosswog05a,metzger10a,metzger10b,roberts11}. The first set of models assumed spherical symmetry
\citep{piran13a,rosswog13a}, but subsequent studies \cite{grossman14a} were based on the 3D remnant structure obtained by
hydrodynamic simulations of the expanding ejecta. This study included the nuclear energy release during the hydrodynamic
evolution \citep{rosswog14a}. The study substantially profited from SPH's geometric
flexibility and its treatment of vacuum as just empty (i.e., SPH particle-free) space. The ejecta expansion was followed for as many
as 40 orders of magnitude in density, from nuclear matter down to the densities of interstellar matter. Since from the 
latter calculations the 3D remnant structure was known, also viewing angle effects for macronovae could be explored 
\citep{grossman14a}. Accounting for the very large opacities of the r-process ejecta \citep{barnes13a,kasen13a}, 
\cite{grossman14a} predicted that the resulting macronova should peak, depending on the binary system, between 3 and 5 
days after the merger in the nIR, roughly consistent with what has been observed in the aftermath of GRB~130603B 
\citep{tanvir13a,berger13b}.

\subsubsection*{Studies with approximate GR gravity}

A natural next step beyond Newtonian gravity is the application of  Post-Newtonian expansions.
\cite{blanchet90} developed an approximate formalism for moderately relativistic, self-gravi\-ta\-ting 
fluids which allows to write all the equations in a quasi-Newtonian form and casts all relativistic non-localities in terms of Poisson equations
with compactly supported sources. The 1PN equations require the solution of eight Poisson equations and to account for the
lowest order radiation reaction terms requires the solution of yet
another Poisson equation. While -- with nine Poisson equations --
computationally already quite heavy, the efforts to implement the scheme into SPH by two groups \citep{faber00,ayal01,faber01,faber02b}
turned out to be not particularly useful, mainly since for realistic neutron stars with compactness $\mathcal{C}\approx 0.17$ the corrective
1PN terms are comparable to the Newtonian ones, which can lead to instabilities. As a result, one of the groups \citep{ayal01}
decided to study ``neutron stars'' of small compactness ($M < 1$ \msun, $R\approx 30$ km), while the other \citep{faber00,faber01,faber02b}
artificially downsized the 1PN effects by choosing a different speed of light for the corresponding terms. While both approaches
represented admissible first steps, the corresponding results are astrophysically difficult to interpret.

A second, more successful approach, was the resort to the so-called conformal flatness approximation (CFA)
\citep{isenberg08,wilson95a,wilson96,mathews97,mathews98}. Here the basic assumption is that the spatial part of the metric
is conformally flat, i.e.,  it can be written as a multiple (the prefactor depends on space and time and absorbs the overall 
scale of the metric) of the Kronecker symbol $\gamma_{ij}= \Psi^4 \delta_{ij}$, and that
it remains so during the subsequent evolution. The latter, however, is an assumption and by no means guaranteed. 
Physically this corresponds to gravitational wave-free space time. Consequently, the inspiral of a binary system has 
to be achieved by adding an ad hoc radiation reaction force. The CFA also cannot handle frame dragging effects.
On the positive side, for spherically symmetric space times the CFA coincides exactly with GR and for small deviation
from spherical symmetry, say for rapidly rotating neutron stars, it has been shown \citep{cook96} to deliver very accurate results.
For more general cases such as a binary merger, the accuracy is difficult to assess. Nevertheless, given how complicated the overall
problem is, the CFA is certainly a very useful approximation to full GR, in particular since it is computationally much more efficient
than solving Einstein's field equations. 

The CFA was implemented into SPH by \cite{oechslin02} and slightly later by \cite{faber04}. The 
major difference between the two approaches was that Oechslin et al. solved the set of six coupled, non-linear elliptic
field equations by means of a multi-grid solver \citep{press92}, while Faber et al. used spectral methods from the \textsc{Lorene}
library on two spherically symmetric grids around the stars. Both studies used polytropic equations of state (\citealp{oechslin02} used
$\Gamma= 2.0, 2.6$ and 3.0;  \citealp{faber04} used $\Gamma= 2.0$) and approximative radiation reaction  terms
based on the Burke--Thorne potential \citep{burke71,thorne69b}. Oechslin et al. used a combination of a bulk and a 
von-Neumann--Richtmyer artificial viscosity steered similarly as in the \cite{morris97} approach, while Faber et al.\
argued that shocks would not be important and artificial dissipation would not be needed.

In a subsequent study, \cite{oechslin04} explored how the presence of quark matter in neutron stars would 
impact on a NSNS merger and its gravitational wave signal. They combined a relativistic mean field model (above
$\rho = 2 \times 10^{14}$ \gcc) with a stiff polytrope as a model for the hadronic EOS and added in an MIT bag model
so that quark matter would appear at $5 \times 10^{14}$ \Gcc and would completely dominate the EOS
for high densities ($> 10^{15}$ \gcc). While the impact on the GW frequencies at the ISCO remained moderate ($<10\%$), 
the post-merger GW signal was substantially influenced in those cases where the central object did not collapse
immediately into a BH. In a subsequent study,
\citep{oechslin06} implemented the \cite{shen98a,shen98b} EOS and a range of NS mass ratios was explored,
mainly with respect to the question how large resulting torus masses would be and whether such merger remnants could 
likely power bursts similar to GRBs 050509b, 050709, 050724, 050813. The found range of disk masses from 1\,--\,9\% of the baryonic
mass of the NSNS binary was considered  promising and broadly consistent with CBM being the central engines of sGRB.

In \cite{oechslin07a} the same group also explored the Lattimer--Swesty EOS \citep{lattimer91}, the cold
EOS of \cite{akmal98} and ideal gas equations of state with parameters fitted to nuclear EOSs. The merger outcome
was  rather sensitive to the nuclear matter EOS: the remnant collapsed either immediately or very soon after merger 
for the soft Lattimer--Swesty EOS and for all other cases it did not show signs of collapse for tens of dynamical time scales.
Both ejecta and disk masses were found to increase with an increasing deviation of the mass ratio from unity. The ejecta
masses were in a range between $10^{-3}$ and $10^{-2}$ \msun, comparable, but slightly lower than the earlier, Newtonian
estimates \citep{rosswog99}. In terms of  their GW signature, it 
turned out that the peak in the GW wave energy spectrum that is related to the formation of the hypermassive merger remnant
has a frequency that is sensitive to the nuclear EOS \citep{oechslin07b}. In comparison, the mass ratio and neutron star spin 
only had a weak impact.

\epubtkImage{}{%
  \begin{figure}[htbp]
    \centerline{\includegraphics[width=8cm]{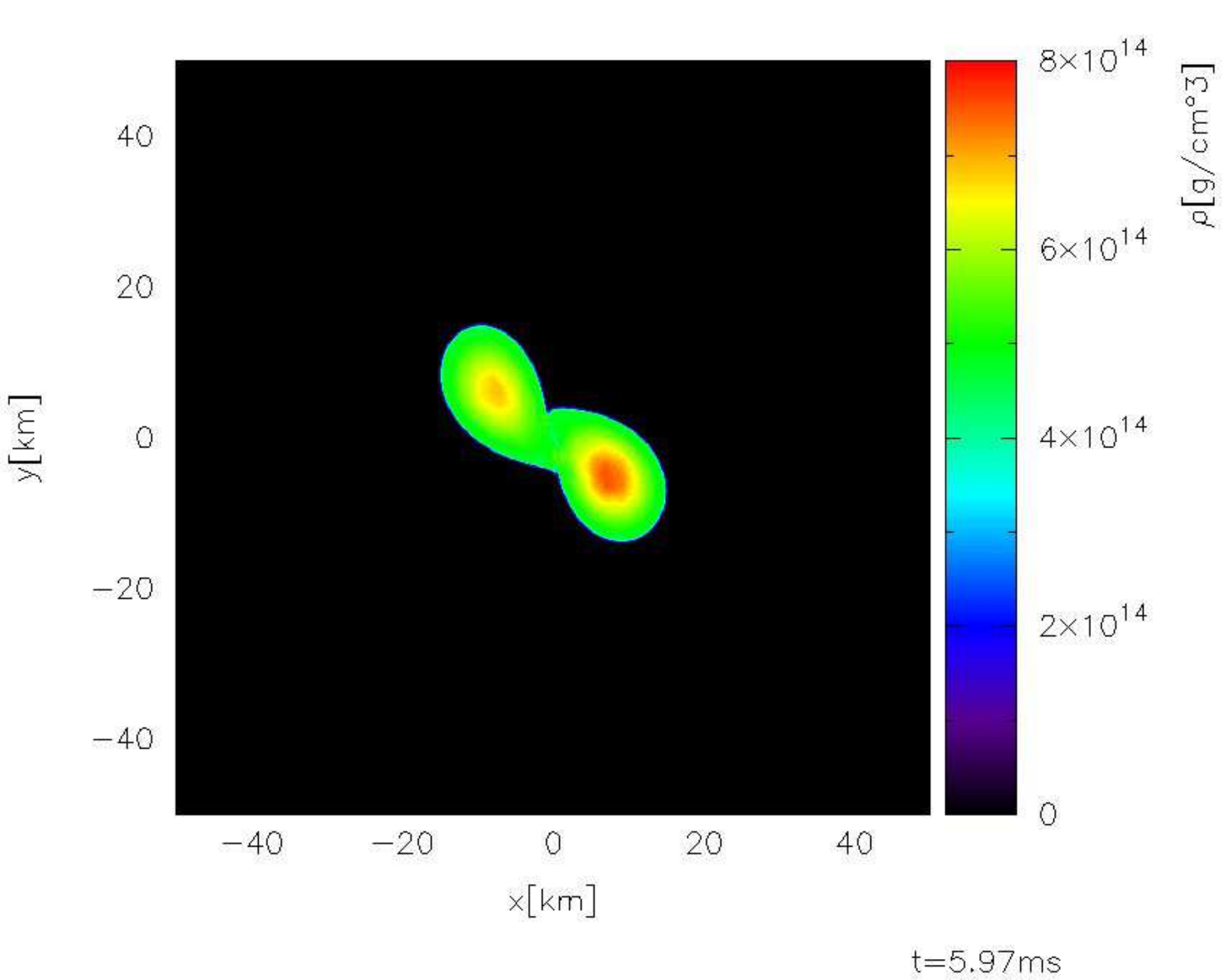}}
    \centerline{\includegraphics[width=8cm]{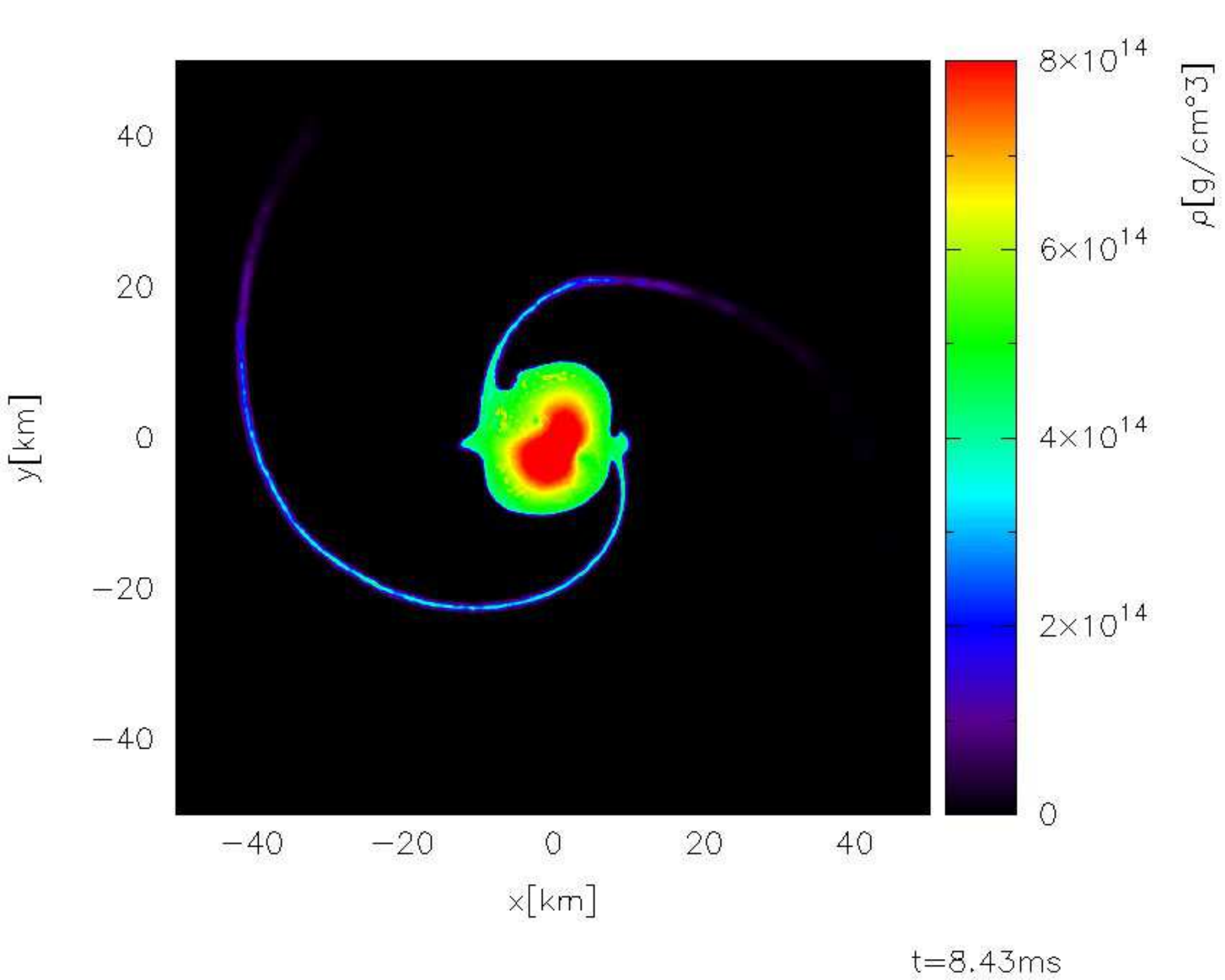}}
    \centerline{\includegraphics[width=8cm]{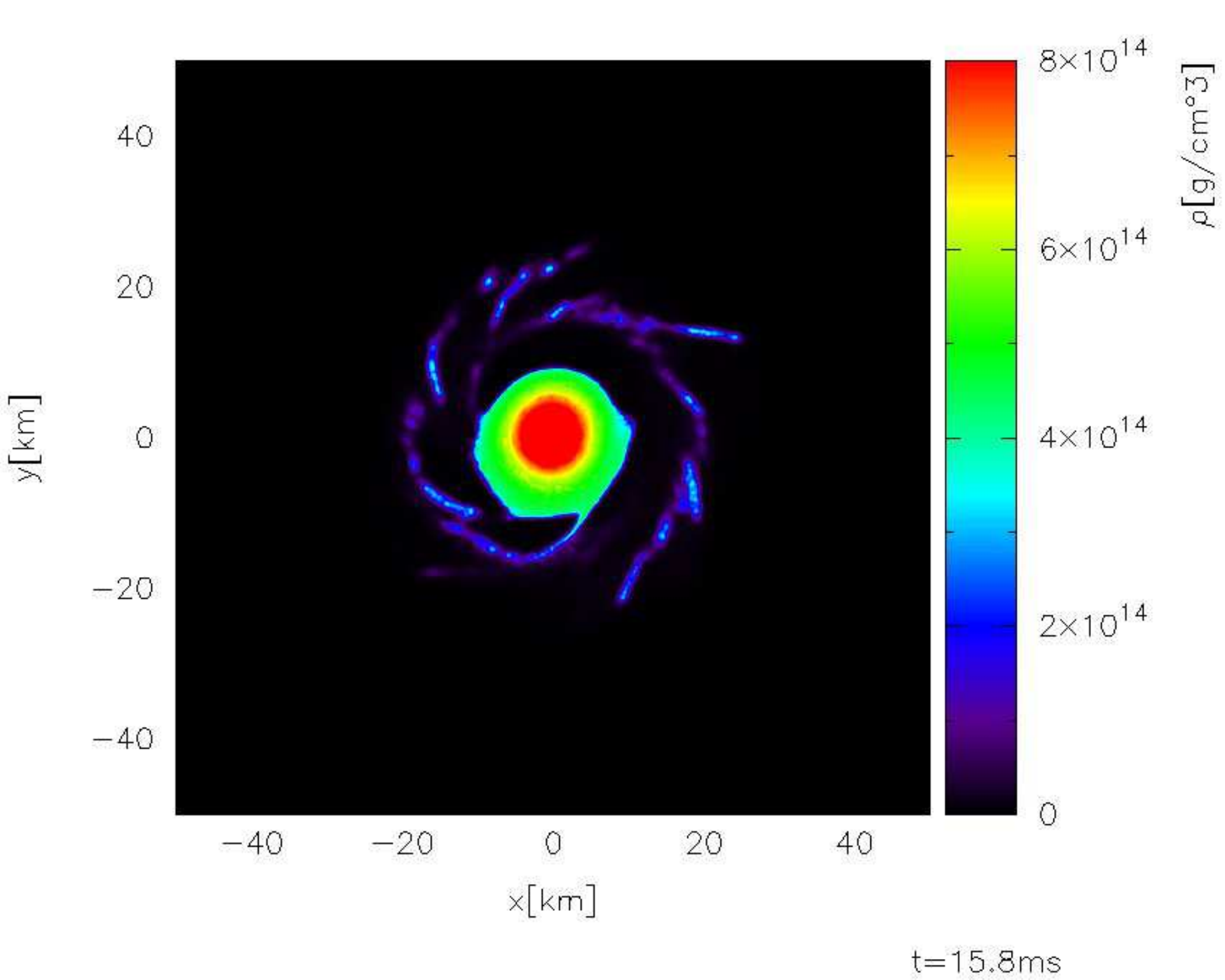}}
    \caption{Merger of two strange stars with masses of 1.2 and 1.35 \msun. Shown is the density in the orbital plane.
  Figure from \cite{bauswein10a}.}
\label{fig:NSNS_strange}
\end{figure}}

This line of work was subsequently continued by Bauswein et al. in a series of papers \citep{bauswein09,bauswein10a,bauswein10b,
bauswein12a,bauswein12b,bauswein13a,bauswein13b}. In their first study, they explored the merger of two strange stars 
\citep{bauswein09,bauswein10a}. If strange quark matter is really the ground state of matter as hypothesized  
\citep{bodmer71,witten84}, compact stars made of strange quark matter might exist. Such stars would differ from neutron 
stars in the sense that they are self-bound, they do not have an overall inverse mass-radius relationship and they can be more
compact. Therefore the gravitational torques close to merger are different and it is more difficult to tidally
tear apart a strange star. 
In their study \citep{bauswein10a} the quark matter EOS was modelled via the MIT bag model \citep{chodos74,farhi84}
for two different bag constants (60 and 80 MeV/fm$^3$) and a large number of binary systems (each time 130K SPH particles) 
was explored.  The coalescence of two strange stars is indeed morphologically different from a neutron star merger:
the result is a differentially rotating hypermassive object with sharp boundary 
layers surrounded by a thin and clumpy strange matter disk, see  Figure~\ref{fig:NSNS_strange} for an example of a strange 
star merger with 1.2 and 1.35 \Msun \citep{bauswein10a}. Moreover, due to the greater compactness, the peak GW 
frequencies were larger during both inspiral and the subsequent ringdown phase. 
If the merger of two strange stars would eject matter in the form of ``strangelets'' these should contribute to the cosmic
ray flux. The ejected mass and therefore the contribution to the cosmic ray flux strongly depends on the chosen bag 
constant and for large values no mass loss could be resolved. For such values neutron stars and 
strange stars could coexist, since neutron stars would not be converted into strange stars by capturing  
cosmic strangelets.
In this and another study \citep{bauswein10b} thermal effects (and their consistent treatment) were shown
to have a substantial impact on the remnant structure.

In subsequent work a very large number of microphysical EOSs was explored 
\citep{bauswein12a,bauswein12b,bauswein13a,bauswein13b}. Here the authors systematically explored which imprint
the nuclear EOS would have on the GW signal. They found that the peak frequency of the post-merger signal correlates well
with the radii of the non-rotating neutron stars \citep{bauswein12a,bauswein12b} and concluded that a GW
detection would allow to constrain the ns radius within a few hundred meters. In a follow-up study \cite{bauswein13b}  explored
the threshold mass beyond which a prompt collapse to a black hole occurs. The study also showed that the ratio
between this threshold mass and the maximum mass is tightly correlated with the compactness of the non-rotating
maximum mass configuration.

In a separate study \citep{bauswein13a} they used their large range of equations of state and several mass ratios to systematically explore 
dynamic mass ejection. According to their study, softer equations of state  with correspondingly smaller radii eject
a larger amount of mass. In the explored cases they found a range from $10^{-3}$ to $2 \times 10^{-2}$ \Msun to
be dynamically ejected. For the arguably most likely case with 1.35 and 1.35 \Msun they found a range of ejecta masses
of about one order of magnitude, determined by the equation of state. Moreover,  consistent with other 
studies, they found  a robust r-process that produces a close-to-solar abundance pattern beyond 
nucleon number of $A= 130$ and they discussed the implications for ``macronovae''
and possibly emerging radio remnants due to the ejecta.

\subsubsection{Neutron star\,--\,black hole mergers}
\label{sec:appl_NSBH}

A number of issues that have complicated the merger dynamics in the WDWD case, such as stability of mass transfer
or the formation of a disk, see Section~\ref{sec:WDWD_MT}, are also very important for  NSBH mergers.\epubtkFootnote{If the neutron star
mass distribution allows for small enough mass ratios, this applies as well to NSNS binaries.} Here, however,
they are further complicated by a poorly known high-density equation of state which determines the mass-radius relationship
and therefore the reaction of the neutron star on mass loss, general-relativistic effects such as the appearance of an
innermost stable circular orbit or effects from the bh spin and the fact that now the GW radiation-reaction time 
scale can become comparable to the dynamical time scale, see Eq.~(\ref{eq:tau_GW_tau_orb}).

Of particular relevance is the question for which binary systems sizeable accretion 
tori form since they are thought to be the crucial ``transformation engines'' that channel available energy into (relativistic)
outflow. The final answer to this question requires 3D numerical simulations with the relevant physics, but a qualitative
idea can be gained form simple estimates (but, based on the experience from WDWD binaries, keep in mind that even plausible 
approximations can yield rather poor results, see Section~\ref{sec:WDWD_MT}). Mass transfer is expected to set in when the Roche volume becomes 
comparable to the volume of the neutron star. By applying Paczynski's estimate for the Roche lobe radius \citep{paczynski71} 
and equating it with the ns radius, one finds that the onset of mass transfer (which we use here as a 
proxy for the tidal disruption radius) can be expected near a separation of
\be
a_{\rm MT}= 2.17 R_{\mathrm{NS}} \left( \frac{1 + q}{q} \right)^{1/3} 
\approx 26\mathrm{\ km} \left( \frac{R_{\mathrm{NS}}}{12\mathrm{\ km}} \right) 
\left( \frac{1 + q}{q} \right)^{1/3} \,. 
\ee
Since $a_{\rm MT}$ grows, in the limit $q \ll 1$, only proportional to $M_{\mathrm{BH}}^{1/3}$, but the ISCO and the event horizon
grow $\propto M_{\mathrm{BH}}$, the onset of mass transfer/disruption can  take place inside the ISCO 
for large bh masses. At the very high end of bh masses, the neutron star is swallowed as whole 
without being disrupted at all. A qualitative illustration (for fiducial neutron star properties, $M_{\mathrm{NS}}= 1.4$ 
\Msun and $R_{\mathrm{NS}}= 12$ km) is shown in Figure~\ref{fig:BH_radii}.
Roughly, already for black holes near $M_{\mathrm{BH}}\approx 8$ \Msun the mass transfer/disruption occurs near the ISCO 
which makes it potentially difficult to form a massive torus from ns debris. So, low mass black holes are clearly preferred as
GRB engines. Numerical simulations \citep{faber06b} have shown,
however,  that even if the disruption occurs deep inside the ISCO this does not necessarily mean that all the matter is 
doomed to fall straight into the hole and  a torus can still possibly form.

When discussing disk formation in a GRB context, it is worth keeping in mind that even seemingly small disk masses
allow, at least in principle, for the extraction of energies,
\be
E_{\rm extr} \sim 1.8\times 10^{51}\mathrm{\ erg} \; \left( \frac{\epsilon}{0.1}\right) \left( \frac{M_{\rm disk}}{0.01\,M_{\odot}}\right),
\ee
that are large enough to accommodate the isotropic gamma-ray energies, $E_{\gamma, \rm iso} \sim 10^{50}$ erg, that have been inferred
for short bursts \citep{berger14a}. If short bursts are collimated into a half-opening angle $\theta$, their 
true energies are substantially lower than this number, $E_{\gamma, \rm true} = \left( E_{\gamma, \rm iso}/65 \right) \left( 
\theta/10^\circ\right)^2$.

\epubtkImage{}{%
  \begin{figure}[htb]
    \centerline{\includegraphics[width=10cm,angle=0]{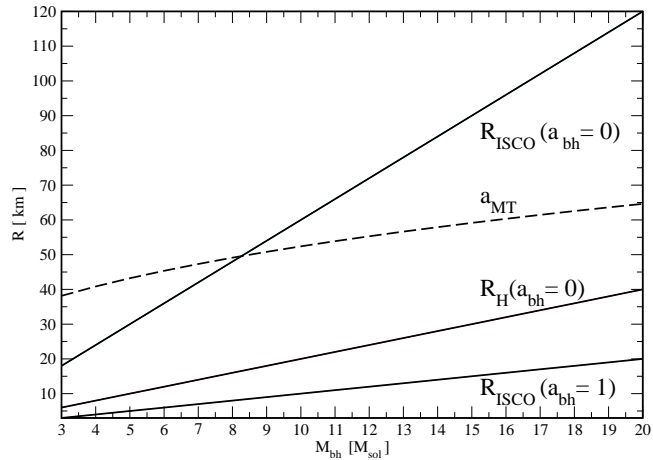}}
    \caption{Illustration of the mass transfer separation $a_{\rm MT}$ with respect to horizon and innermost stable circular orbits based on Newtonian estimates (for fiducial neutron star properties, $M_{\mathrm{NS}}= 1.4$ \Msun and $R_{\mathrm{NS}}= 12$ km).}
    \label{fig:BH_radii}
\end{figure}}

\subsubsection*{Early Newtonian calculations with polytropic equations of state}

The first NSBH simulations were performed by \cite{lee99a,lee99b} and \cite{lee00,lee01a}
using Newtonian physics and polytropic equations of state. Although simple and missing 
some qualitative features of black holes (such as a last stable orbit), these simulations
provided insight and a qualitative understanding of the system dynamics, the impact
of neutron star spin, the mass ratio and the equation of state. The first set of simulations \citep{lee99a}
were carried out with an SPH formulation similar to the one of \cite{hernquist89}, but 
an alternative kernel gradient symmetrization. In this study, the artificial viscosity tensor Eq.~(\ref{eq:basic:PI_AV})
was implemented with fixed dissipation parameters $\alpha$ and $\beta$ and $\sim 10^4$ SPH particles were used. 
The black hole was modelled as a Newtonian point mass with an absorbing boundary 
condition at the Schwarzschild radius $R_{\rm S}$, no backreaction from gravitational wave emission was accounted for.
In the simulations of initially tidally locked binaries with a stiff EOS ($\Gamma=3$; \citealp{lee99a}), 
they found that the neutron star survived the onset of mass transfer and kept orbiting, at a reduced 
mass, for several orbital periods. A similar behavior had been realized in subsequent work with a 
stiff nuclear equation of state \citep{rosswog04b,rosswog07b}. In a follow-up paper \citep{lee99b},
a simple point-mass backreaction force was applied, and, in one case, the Paczynski--Wiita 
\citep{paczynski80} potential was used (now with absorbing
boundary at $1.5\,R_{\rm S}$). But the main focus of this study was to explore the effect of a softer  equation of state 
($\Gamma=5/3$). In all explored cases the system dynamics was very different from the previous 
study, the neutron star was completely disrupted and formed a massive disk of $\sim 0.2 $ \msun,
with $\sim 0.01$ \Msun being dynamically ejected. The sensitivity to the EOS stiffness is not entirely surprising,
since the solutions to the Lane--Emden equations give the mass-radius relationship 
\be
\frac{d\log R}{d \log M}= (\Gamma-2)/(3\Gamma -4)
\label{eq:LE_MR}
\ee 
for a polytropic star, so that the neutron star reacts differently on mass loss: it shrinks for 
$\Gamma > 2$, so mass loss is quenched, and expands for $\Gamma < 2$ and therefore further enhances
mass transfer.

In a second set of calculations ($\approx 80$ K SPH particles), they explored non-rotating neutron 
stars that were modelled as compressible triaxial ellipsoids according to the semi-analytic work of 
\cite{lai93a,lai93b,lai94b}, both with stiff ($\Gamma=2.5$ and 3) \citep{lee00} 
and soft ($\Gamma=5/3$ and 2) \citep{lee01a} polytropic equations of state. They used the same simulation
technology, but also applied a Balsara-limiter, see Eq.~(\ref{eq:Balsara}), in their artificial viscosity treatment
and only purely Newtonian interaction between ns and bh was considered.  For the $\Gamma=3$ case,
the neutron star survived again until the end of the simulation, with $\Gamma=2.5$ it survived
the first mass transfer episode but was subsequently completely disrupted and formed a disk of
nearly 0.2 \msun, about 0.03 \Msun were dynamically ejected.

\cite{lee01b} also simulated mergers between a black hole and a strange star which was modelled
with a simple quark-matter EOS. The dynamical evolution for such systems was quite different from
the  polytropic case: the strange star was stretched into thin matter stream that wound around the 
black hole and was finally swallowed. Although ``starlets'' of $ \approx 0.03$ \Msun formed during the disruption
process, all of them were in the end swallowed by the  hole within milliseconds, no mass loss could be resolved.

\subsubsection*{Studies with focus on microphysics}

The first NSBH studies based on Newtonian gravity, but including detailed microphysics were performed 
by \cite{ruffert99} and \cite{janka99} using a Eulerian PPM code on a Cartesian 
grid.\epubtkFootnote{See Section~\ref{sec:challenges_WDWD} for a discussion of the complications of such an approach.} 
The first Newtonian-gravity-plus-microphysics SPH simulations of NSBH mergers were discussed in
\cite{rosswog04b,rosswog05a}.
Here the black hole was simulated as a Newtonian point mass with an absorbing boundary and a simple GW backreaction
force was applied. For the neutron star the \cite{shen98a,shen98b} temperature-dependent nuclear EOS was 
used and the star was modelled with $3\times 10^5  - 10^6$ SPH particles. In addition, neutrino cooling and electron/positron 
captures were followed with a detailed multi-flavor leakage scheme \citep{rosswog03a}. The initial study 
focussed on systems with low mass black holes ($q=0.5 - 0.1$) since this way there are greater chances to 
disrupt the neutron star outside of the ISCO, see above. Moreover, both (carefully constructed) corotating 
and irrotational neutron stars were studied. In all cases the core of the neutron star ($0.15 - 0.68$ \msun) survived the initial 
mass transfer episodes until the end of the simulations (22 - 64 ms). If disks formed at all during the simulated time, 
they had only moderate masses ($\sim0.005$ \msun). One of the NSBH binary 
($M_{\mathrm{NS}}=1.4$ \msun, $M_{\mathrm{BH}}=3$ \msun) systems was followed throughout the whole mass transfer episode \citep{rosswog07b}
which lasted for 220 ms or 47 orbital revolutions and only ended when the neutron star finally became disrupted and 
resulted the formation of a disk of 0.05 \msun. A set of test calculations with a stiff ($\Gamma=3$) and a softer polytropic EOS ($\Gamma=2$) 
indicated that such episodic mass transfer is related to the stiffness of the ns EOS and only occurs for 
 stiff cases, consistent with the results of  \cite{lee00}. Subsequent studies with better approximations to 
relativistic gravity, e.g., \cite{faber06a}, have seen qualitatively 
similar effects for stiffer EOSs, but after a few orbital periods the neutron was always disrupted. \cite{shibata11} discussed such episodic,
long-lived mass transfer in a GR context  and concluded that while possible, it has so far never been
seen in fully relativistic studies.

A study \citep{rosswog05a} with simulation tools similar to \cite{rosswog04b} focussed on higher mass, non-spinning black holes 
($M_{\mathrm{BH}}= 14 \dots 20$ \msun) that were approximated by  pseudo-relativistic potentials \citep{paczynski80}. While being
very simple to implement, this approach mimics some GR effects quite well and in particular it has an innermost stable
circular particle orbit at the correct location ($6 GM_{\mathrm{BH}}/c^2$), see \cite{tejeda13a} for quantitative assessment of
various properties. In none of these high black hole mass cases was episodic mass transfer observed, the neutron star was 
always completely disrupted shortly after the onset of mass transfer. Although disks  formed for
systems below 18 \msun, a large part of them was inside the ISCO and was  falling practically radially into the hole on a dynamical
time scale. As a result, they were thin and cold and not considered promising GRB engines. It was suggested, however, that 
even black holes at the high end of the mass distribution could possibly be GRB engines, provided they spin rapidly 
enough, since then both ISCO and  horizon move closer to the bh. The investigated systems ejected 
between 0.01 and 0.2 \Msun at large velocities ($\sim 0.5$ c) and analytical estimates suggested that such systems
should produce  bright optical/near-infrared transients (``macronovae'') powered by the radioactive decay of the freshly 
produced r-process elements within the ejecta, as originally suggested by \cite{li98}.

\subsubsection*{Studies with approximate GR gravity around a non-spinning black hole}

\cite{faber06a,faber06b} studied the merger of a non-rotating black hole with a polytropic neutron star
in approximate GR gravity. While the hydrodynamics code was fully relativistic, the self-gravity of the neutron star was
treated within the conformal flatness (CF) approximation. Since the black hole was kept at a fixed position its
mass needed to be substantially larger than the one of the neutron star, therefore a mass ratio of $q=0.1$ was chosen.
The neutron star matter was modelled with two (by nuclear matter standards) relatively soft polytropes ($\Gamma= 1.5$ 
and 2). In the first study \citep{faber06a} they focussed on tidally locked neutron stars and solved the five linked non-linear
field equations of the CF approach by means of the \textsc{Lorene} libraries, the second study \citep{faber06b} used
irrotational neutron stars and solved the CF equations by means of a Fast Fourier transform solver. In a first case
they considered a neutron star of compactness $\mathcal{C}= 0.15$, and, to simulate a case
where the disruption of the neutron star occurs near the ISCO, they also considered a second case where
$\mathcal{C}$ was only 0.09. The first case turned out to be astrophysically unspectacular: the entire
neutron star was swallowed as a whole without leaving matter behind. The second, ``undercompact'' case, however, see 
Figure~\ref{fig:NSBH_faber}, showed some very interesting behavior: the neutron star spiralled towards the black hole,
became tidally stretched and although at some point 98\% of the ns mass were inside of the ISCO, see panel two,
via a rapid redistribution of angular momentum, a substantial fraction of the matter was ejected as a one-armed
spiral. Approximately 12\% of the initial neutron star formed a disk and an additional 13\% of the initial neutron star 
were tidally ejected into unbound orbits. Such systems, they concluded, would be interesting sGRB engines.

\epubtkImage{}{%
  \begin{figure}[htb]
    \centerline{\includegraphics[width=10cm,angle=-90]{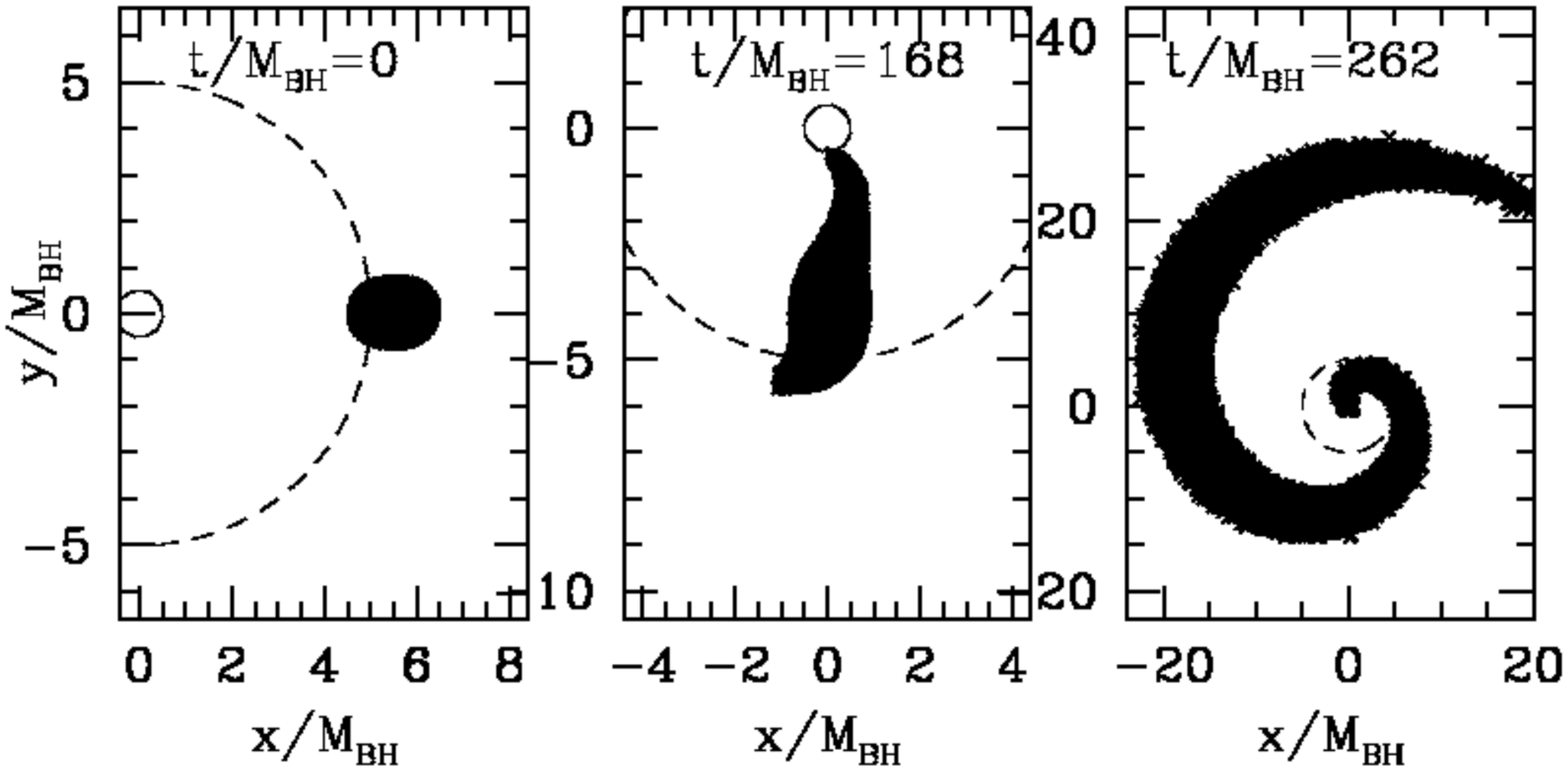}}

\vspace*{-2cm}

    \caption{Disruption of an ``undercompact'' neutron star (compactness $\mathcal{C} = 0.15$),
               simulated within the conformal flatness approximation 
               by a non-rotating black hole ($q=0.1$). Although at some stage 
               98\% of the neutron star are within the innermost stable circular orbit
               (dashed circle),  rapid redistribution of angular momentum
               leads to the ejection of a one-armed spiral. Finally a disk made of 
               $0.12\,M_{\mathrm{NS}}$ forms and $0.13\,M_{\mathrm{NS}}$  are dynamically ejected.
               Figure from \cite{faber06b}.}
    \label{fig:NSBH_faber}
\end{figure}}

\subsubsection*{Studies in the fixed metric of a spinning black hole}

\cite{rantsiou08} explored how the outcome of a neutron star black hole merger depends on
the spin of the black hole and on the inclination angle of the binary orbit with respect to the equatorial plane
of the black hole. They used the relativistic SPH code originally developed by
\cite{laguna93a,laguna93b} to study the tidal disruption of a main sequence star by a massive black hole.
The new code version employed Kerr--Schild coordinates to avoid coordinate singularities at the horizon as
they appear in the frequently used Boyer--Lindquist coordinates. Since the spacetime was kept fixed,
they focused on a small mass ratio $q= 0.1$, where the impact of the neutron star on the spacetime is sub-dominant.
Both the black hole mass and spin were frozen at their initial values during the simulation and the GW backreaction
was implemented via the quadrupole approximation in the point mass limit, similar to the one used by \cite{lee99b}.
The neutron star itself was modelled as a $\Gamma=2.0$ polytrope with Newtonian self-gravity, the artificial
dissipation parameters were fixed to 0.2 (instead of values near unity which are needed to properly deal with 
shocks). Note that $\Gamma=2$ is a special choice, since a Newtonian star does not change its 
radius if mass is added or lost, see Eq.~(\ref{eq:LE_MR}). The bulk of the simulations was calculated
with $10^4$ SPH particles, in one case $10^5$ particles were used to confirm the robustness of the
results.

For the case of a Schwarzschild black hole ($a_{\mathrm{BH}}=0$) they found that neither a disk 
formed nor any material was ejected.  For equatorial mergers with spinning black holes, it 
required a spin parameter of $a_{\mathrm{BH}} > 0.7$ for any mass to form a disk or to become ejected. 
For a rapidly spinning bh  ($a_{\mathrm{BH}}=0.75$) an amount of matter of order
0.01 \Msun became unbound, for a close-to-maximally spinning bh ($a_{\mathrm{BH}}=0.99$) a huge amount of matter 
($> 0.4$ \msun) was ejected. Mergers with inclination angles $> 60^\circ$ lead to the complete swallowing of the neutron
star. An example of close-to-maximally spinning black hole ($a_{\mathrm{BH}}=0.99$) and a neutron star whose orbital plane is
inclined by $45^\circ$ with respect to the black hole spin is shown in Figure~\ref{fig:NSBH_rantsiou}. Here, as much as
25\% of the neutron star, in the shape of a helix, become unbound.

\epubtkImage{}{%
\begin{figure}[htb]
   \centerline{
     \includegraphics[width=0.46\textwidth]{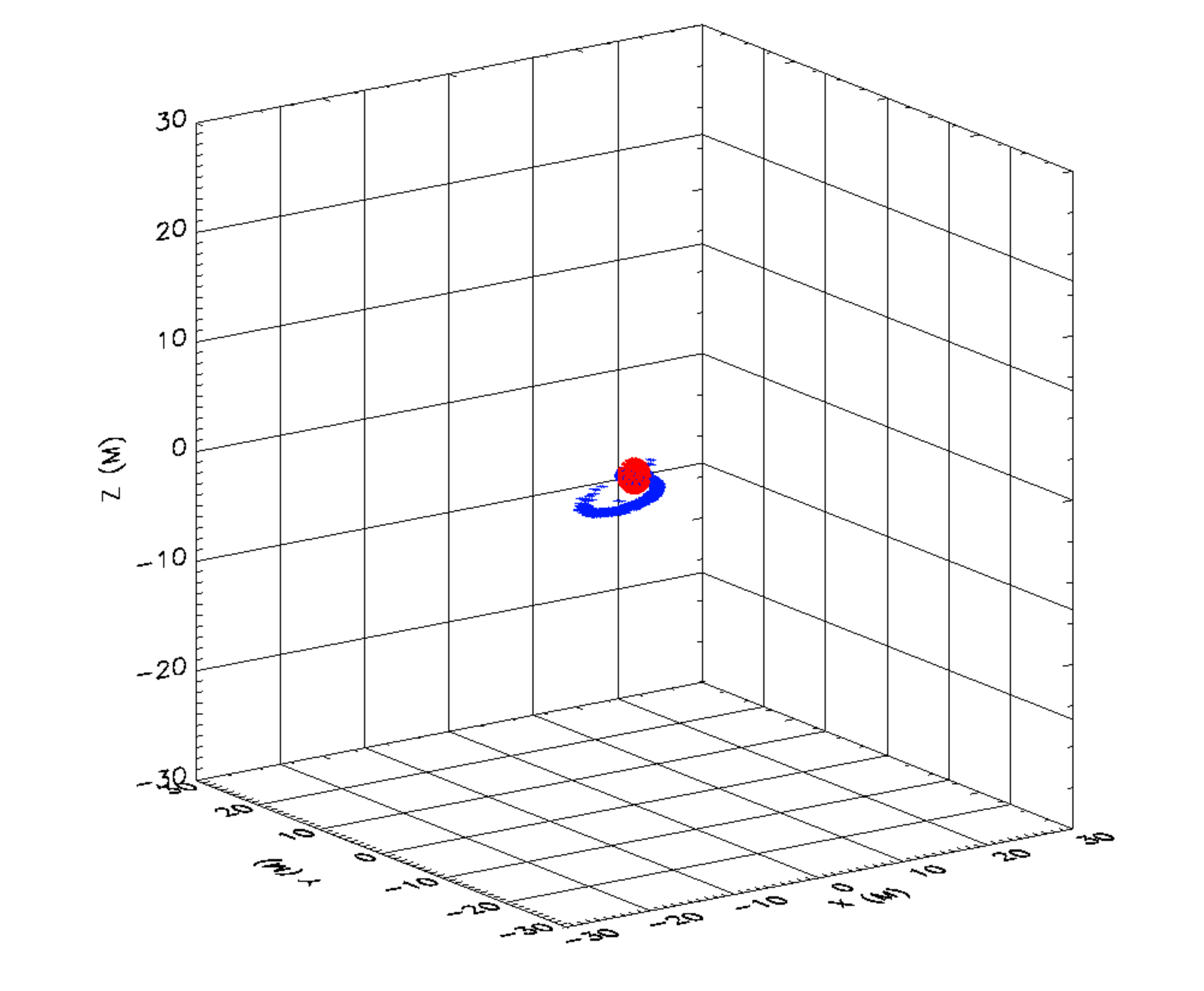}\qquad
     \includegraphics[width=0.46\textwidth]{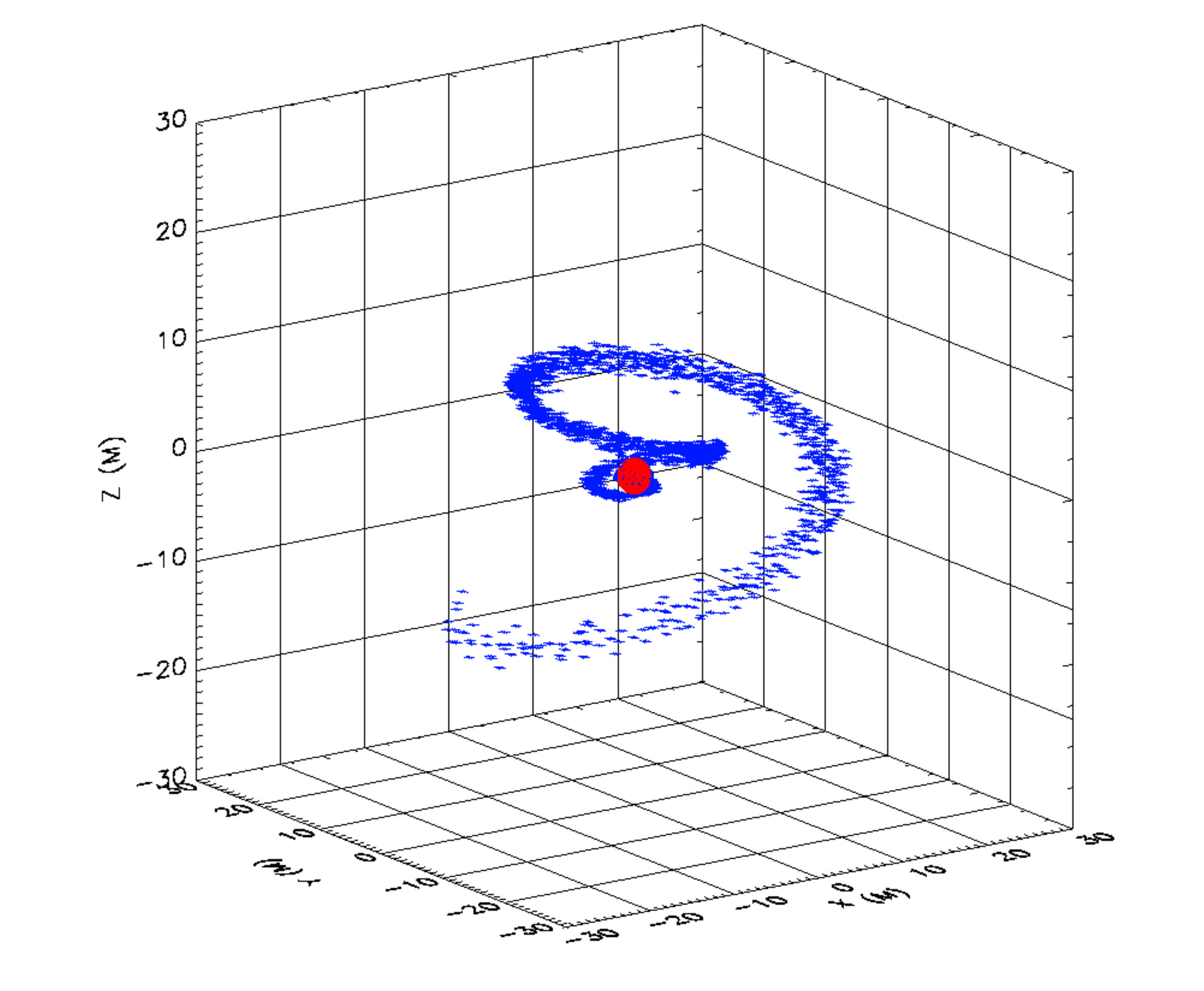}
   }
   \caption{Disruption of a polytropic star ($\Gamma=2.0$) by a close-to-maximally spinning black hole ($a_{\mathrm{BH}}=0.99$) where the binary orbit has an initial inclination of $45^\circ$ with respect to the black hole spin. The ``helix'' of unbound material contains 25\% of the original neutron star mass. Figure from \cite{rantsiou08}.}
   \label{fig:NSBH_rantsiou}
\end{figure}}

\subsubsection{Collisions between two neutron stars and between a neutron star and a black hole}
\label{sec:compact_collisions}

Traditionally, the focus of compact object encounters have been GW-driven binary systems such as the Hulse--Taylor pulsar
\citep{taylor89,weisberg10}. 
More recently, however, also dynamical collisions/high-eccentricity encounters between two compact objects have 
attracted a fair amount of interest \citep{kocsis06a,oleary09,lee10a,east12,kocsis12,gold12,gold13}. Unfortunately, 
their rates are at least as difficult to estimate as those of GW-driven mergers.

Collisions differ from gravitational wave-driven mergers in a number of ways. For example, since gravitational wave emission
of eccentric binaries efficiently removes angular momentum in comparison to energy, primordial binaries will have radiated away
their eccentricity and will finally merge from a nearly circular orbit. On the contrary, binaries that have formed dynamically, 
say in a globular cluster, start from a small orbital separation, but with large eccentricities and may not have had the time 
to circularize until merger. This leads to pronouncedly different gravitational wave signatures, ``chirping'' signals of 
increasing frequency and amplitude for mergers and initially well-separated, repeated GW bursts that continue from 
minutes to days in the case of collisions. Moreover, 
compact binaries are strongly gravitationally bound at the onset of the dynamical merger phase while collisions, in contrast, 
have total orbital energies close to zero and need to get rid of energy and angular momentum via GW emission 
and/or through mass shedding episodes in order to form a single remnant. Due to the strong dependence on the
 impact parameter and the lack of strong constraints on it, one expects a much larger variety of dynamical
behavior for collisions than for mergers.

\cite{lee10a}  provided detailed rate estimates of compact object collisions and concluded that such encounters could 
possibly produce an interesting contribution to the observed GRB rate. They also performed the first SPH simulations 
of such encounters.  Using the SPH code from their earlier studies \citep{lee99a,lee99b,lee00,lee01a}, 
they explored the dynamics and remnant structure of encounters with different strengths between all types of compact stellar objects
(WD/NS/BH; typically with 100K SPH particles). Here polytropic equations of state were used and black holes were treated as 
Newtonian point masses with absorbing boundaries at their Schwarzschild radii. Their calculations indicated in particular that
such encounters would produce interesting GRB engines with massive disks and additional  external
reservoirs (one tidal tail for each close encounter) where a large amounts of matter ($>0.1$ \msun) could be stored to 
possibly prolong the central engine activity, as observed in some bursts. In addition, a substantial amount of mass 
was dynamically ejected (0.03 \Msun for NSNS and up to 0.2 \Msun for NSBH systems).

\epubtkMovie{}{}{%
  \begin{figure}[htbp]
    \centerline{\includegraphics[width=24cm,angle=0]{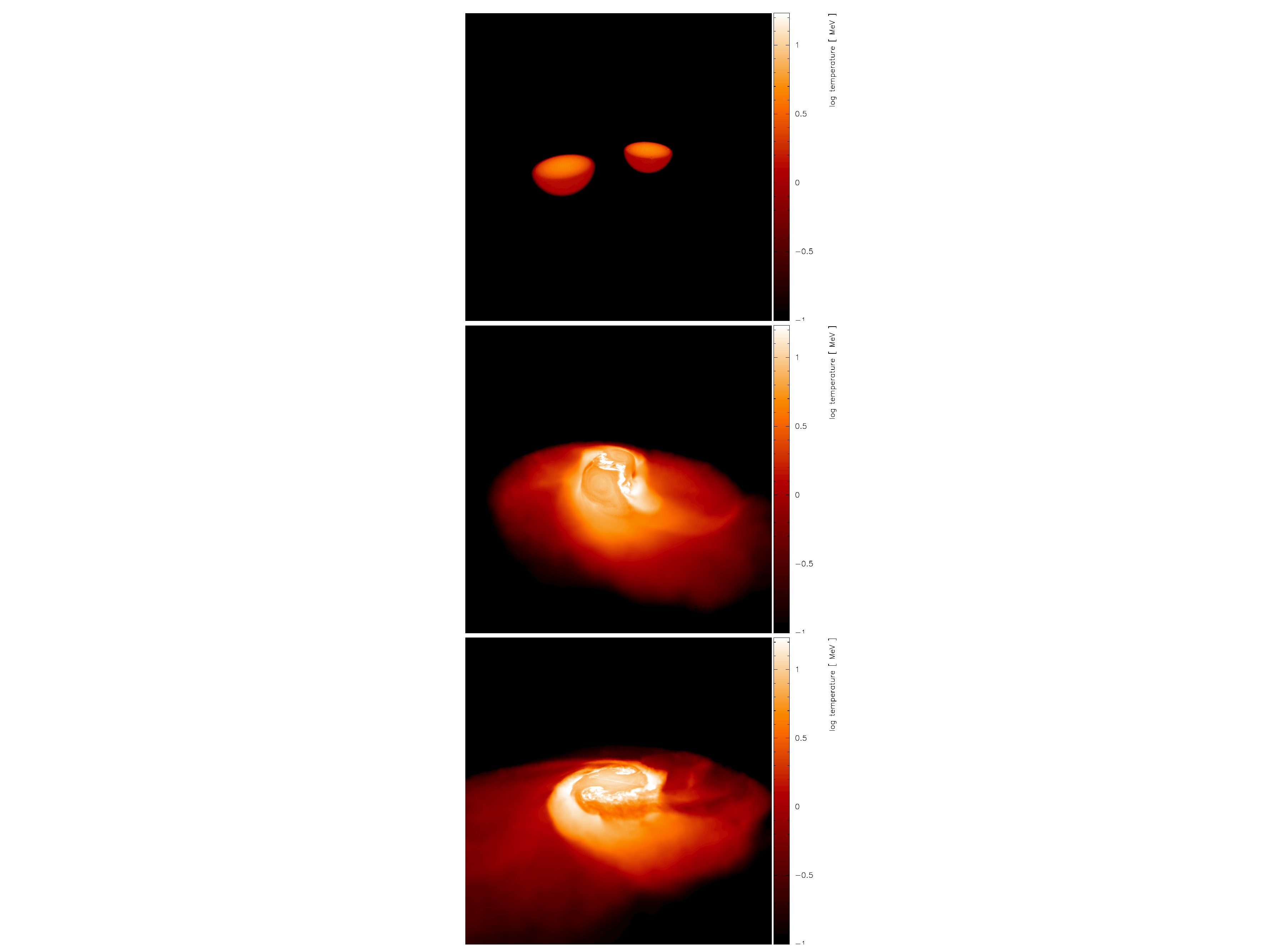}} 
    \caption{Collision between a 1.3 \Msun and a 1.4 \Msun neutron stars (modelled by $8 \times 10^6$ SPH particles; pericentre distance equal to the average of the two neutron star radii). Only matter below the orbital plane is shown, color-coded is temperature in MeV. The corresponding simulations are discussed in detail in \cite{rosswog13a}.}
   \label{fig:NSNS_collision}
\end{figure}}

In \cite{rosswog13a} various signatures of gravitational wave-driven mergers and dynamical collisions were compared,
both for NSNS and NSBH encounters. The study applied Newtonian SPH (with up to $8 \times 10^6$ particles) together with a nuclear equation
of state \citep{shen98a,shen98b} and a detailed neutrino leakage scheme \citep{rosswog03a}. As above, black holes were modelled as Newtonian
point masses with absorbing boundaries at the $R_S$. A simulation result of a strong encounter between a 1.3 and a 1.4 \Msun neutron
star (pericentre distance equal to the average of the two neutron star radii) is shown in Figure~\ref{fig:NSNS_collision}. Due to the strong
shear at their interface, a string of Kelvin--Helmholtz vortices forms in each of the  close encounters before a final central 
remnant is formed. Such conditions offer plenty of opportunity for magnetic field amplification 
\citep{price06,anderson08b,obergaulinger10,rezzolla11,zrake13}. In all explored cases the neutrino luminosity was at least comparable to
the merger case, $L_\nu \approx 10^{53}$ erg/s, but for the more extreme cases exceeded this value by an order of magnitude. Thus, if
neutrino annihilation should be the main agent driving a GRB outflow, chances for collisions should be at least as good as in the
merger case. But both scenarios also share the same caveat: neutrino-driven, baryonic pollution could prevent in at least a fraction of cases
the emergence of relativistic outflows. In NSBH collisions the neutron star took usually several encounters before being completely
disrupted. In some cases its core survived several encounters and was finally ejected  with a mass of $\sim 0.1$ \msun. Of course,
this offers a number of interesting possibilities (production of low-mass neutron stars, explosion of the NS core at the minimal mass etc.). But first
of all, such events may be very rare and it needs to be seen  whether such behavior can occur at all in the general-relativistic case.

Generally, both NSNS and NSBH collisions ejected  large quantities of unbound debris. Collisions between neutron stars ejected a few percent
(dependent on the impact strength) of a solar mass, while all investigated NSBH collisions ejected $\sim 0.15$ \msun, consistent with the
findings of \cite{lee10a}. Since NSBH encounters should dominate the rates \citep{lee10a}, it was concluded in \cite{rosswog13a} 
that collisions must be (possibly much) less frequent than 10\% of the NSNS merger rate to avoid  a conflict with constraints from the
chemical evolution of galaxies.

Since here the ejecta velocities and masses are substantially larger ($v_{\rm ej}\sim 0.2$c and $m_{\rm ej}\sim 0.1$ \msun) than in the neutron  
star merger case ($v_{\rm ej} \sim 0.1$c and $m_{\rm ej}\sim 0.01$ \msun) simple scaling relations \citep{grossman14a}
suggest that a resulting radioactively powered macronova should peak after 
\be
t_{\rm P}= 11 \; {\rm days} \; \left( \frac{\kappa}{10\mathrm{\ cm^2/g}}  \;  \frac{m_{\rm ej}}{0.1\,M_{\odot}} \; \frac{0.2 {\rm c}}{v_{\rm ej}}\right)^{1/2}
\ee
with a luminosity of
\be
 L_{\rm P}= 8.8 \times 10^{40}\mathrm{\ erg/s} \left( \frac{v_{\rm ej}}{0.2{\rm c}}  \;  \frac{10\mathrm{\ cm^2/g}}{\kappa} \right)^{0.65}
 \left( \frac{m_{\rm ej}}{0.1\,M_{\odot}} \right)^{0.35}.
\ee
Here $\kappa$ is the r-process material opacity \citep{kasen13a} and a radioactive heating rate 
$\dot{\epsilon} \propto t^{-1.3}$ \citep{metzger10b,korobkin12a} has been assumed. 

\epubtkMovie{}{}{%
\begin{figure}[htbp]
  \centerline{\includegraphics[width=24cm,angle=0]{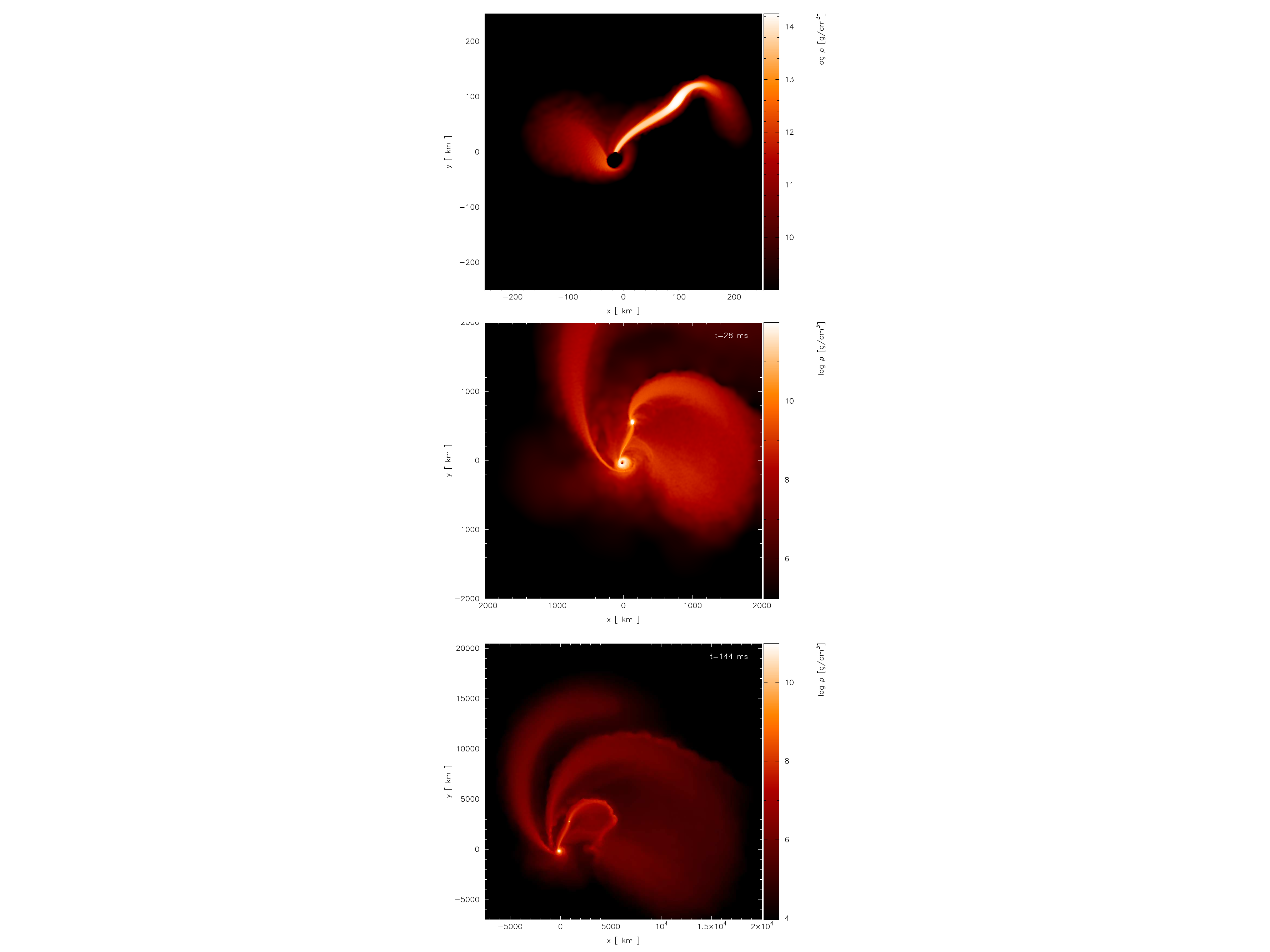}}
  \caption{Grazing collision between a 5 \Msun black hole and a 1.3 \Msun neutron star. Shown are the densities in the orbital plane after the first, second and third close encounter. In this particular case the neutron star's core survives all three encounters, each time producing a tidal tail, and is finally ejected at $\sim 0.1$ \msun. The corresponding simulations are discussed in detail in \cite{rosswog13a}.}
  \label{fig:NSBH_collision}
\end{figure}}

\subsubsection{Post-merger disk evolution}

SPH simulations were also applied to study the long-term evolution of accretion disks that 
have formed during a CBM. Lee and collaborators \citep{lee02,lee04,lee05b,lee09} 
started from their NSBH merger simulations, see Section~\ref{sec:appl_NSBH}, and followed the fate of the 
resulting disks. Since the viscous disk time scale, see Eq.~(\ref{eq:tau_visc}), by far exceeds the numerical time step
allowed by the CFL condition, Eq.~(\ref{eq:CFL}), the previous results were mapped into a 2D version
of their code and they followed the evolution, driven by  a Shakura-Sunyaev ``$\alpha$-viscosity'' 
prescription \citep{shakura73}, for hundreds of milliseconds. Consistent with their NSBH merger simulations, the 
black hole was treated as a Newtonian point mass with an absorbing boundary at 
$R_{\rm S}= 2 G M_{\mathrm{BH}}/c^2$, the disk self-gravity was neglected. In the first study \citep{lee02}
the disk matter was modelled with a polytropic EOS ($\Gamma=4/3$) and locally dissipated
energy was assumed to be emitted via neutrinos. Subsequent studies \citep{lee04,lee05b}
applied increasingly more sophisticated microphysics, the latter study accounted for pressure contributions 
from relativistic electrons with an arbitrary degree of degeneracy and an ideal gas of nucleons and
alpha particles. These latter studies  accounted for opacity-dependent neutrino cooling
and also considered trapped neutrinos as a source of pressure,  but no distinction 
between different neutrino flavors was made. It turned out that at the transition between 
the inner, neutrino-opaque and the outer, transparent regions an inversion of the lepton number 
gradient builds up,  with minimum values $Y_e \approx 0.05$ close to the transition radius 
($\sim 10^7$ cm), values close to 0.1 near the BH and proton-rich conditions ($Y_e>0.5$) 
at large radii. Such lepton number gradients drive strong convective motions that shape
the inner disk regions. Overall, neutrino luminosities around $\approx 10^{53}$ erg/s were found
and around $10^{52}$ erg were emitted in neutrinos over the lifetime of the disk ($\sim 0.4$ s).

\subsubsection{Summary encounters between neutron stars an black holes}

Compact binary mergers are  related to a number of vibrant astrophysical topics. 
They are likely the first sources whose gravitational waves are detected directly, they
probably power short Gamma-ray Bursts and they may be the astrophysical sites where  
a large fraction of the heaviest elements in the Universe are forged. With these high promises for 
different fields comes also the necessity to reliably include a broad range of 
physical processes. Here, much progress has been achieved in the last one and a half decades. 
The task for the future will be to bring the different facets of compact binary encounters into 
a coherent, bigger astrophysical picture. Very likely, our understanding will be challenged 
once, say, the first Gamma-ray burst is also observed as a gravitational wave event. 
If the interpretation of the recent nIR transient related to GRB~130603B as a ``macronova''  is 
correct, this is the first link between two suspected, but previously unproven aspects of the compact 
binary merger phenomenon: Gamma-ray bursts and heavy element nucleosynthesis. This
may just have been one of the first heralds of the beginning era of multi-messenger astronomy.

SPH has played a major role in achieving our current understanding of the astrophysics of
compact binary mergers. The main reasons for its use were its geometrical 
flexibility, its excellent numerical conservation properties and the ease with which new 
physics ingredients can be implemented. A broad range of physical 
ingredients has been implemented into the simulations that exist to date. These include a large 
number of different equations of state, weak interactions/neutrino emission and
magnetic fields. In terms of gravitational physics, mergers have been 
simulated in Newtonian, post-Newtonian and in conformal flatness approximations 
to GR. An important milestone, however, that at the time of writing (June 2014) still has 
to be achieved,  is an SPH simulation where the spacetime is self-consistently 
evolved in dynamic GR.


\section*{Acknowledgements}
\label{sec:acknowledgements}

This work has benefited from the discussions with many teachers, students, colleagues and friends and I want to collectively
thank all of them. Particular thanks goes to W.~Benz, M.~Dan, M.B.~Davies, W.~Dehnen, O.~Korobkin, W.H.~Lee, J.J.~Monaghan, D.J.~Price,
E.~Ramirez-Ruiz,  E.~Tejeda and F.K.~Thielemann.
This work has been supported by the Swedish Research Council (VR) under grant 621-2012-4870,
the CompStar network, COST Action MP1304,
and in part by the National Science Foundation Grant No. PHYS-1066293. The hospitality of the 
Aspen Center for Physics is gratefully acknowledged.
Some of the figures in this article have been produced by the visualization software 
\textsc{Splash} \citep{price07d}.

\newpage



\bibliography{refs}

\end{document}